\documentclass[11pt,document,nofootinbib,superscriptaddress,onecolumn,preprintnumbers,balancelastpage]{article}
\pdfoutput=1
\usepackage{jheppub}
\usepackage{graphicx}
\usepackage{epstopdf}
\usepackage{dcolumn}
\usepackage{bm}
\usepackage{hyperref}
\usepackage{booktabs}
\usepackage[dvipsnames]{xcolor}
\usepackage{amsmath}
\usepackage{cancel}
\usepackage{xpatch}
\usepackage{mathtools}
\definecolor{c1}{rgb}{0.368417, 0.506779, 0.709798}
\definecolor{c2}{rgb}{0.880722, 0.611041, 0.142051}
\definecolor{c3}{rgb}{0.560181, 0.691569, 0.194885}
\definecolor{c4}{rgb}{0.922526, 0.385626, 0.209179}
\definecolor{c5}{rgb}{0.528488, 0.470624, 0.701351}
\definecolor{c6}{rgb}{0.772079, 0.431554, 0.102387}
\definecolor{c7}{rgb}{0.363898, 0.618501, 0.782349}
\definecolor{turq}{rgb}{0.181,0.638,0.594}
\definecolor{pink}{rgb}{1.000,0.54,0.8}
\definecolor{purple}{RGB}{155,100,155}
\definecolor{gray}{RGB}{128,128,128}
\definecolor{lightBlue}{RGB}{148,179,229}
\definecolor{lightRed}{RGB}{213,157,131}
\definecolor{violet}{RGB}{130,121,173}
\definecolor{gold}{RGB}{255,191,0}

\def\beq{\begin{align}}
\def\eeq{\end{align}}

\newcommand{\vev}[1]{ \left\langle {#1} \right\rangle }

\newcommand{\EV}{ {\rm eV} }
\newcommand{\KEV}{ {\rm keV} }
\newcommand{\MEV}{ {\rm MeV} }
\newcommand{\GEV}{ {\rm GeV} }

\def\mpl{M_{\rm Pl}}

\def\GeV{{\rm GeV}}

\title{
Sterile Neutrino Dark Matter and Leptogenesis \\
in Left-Right Higgs Parity
}

\author[1,2]{David Dunsky}
\author[1,2]{Lawrence J. Hall}
\author[3]{Keisuke Harigaya}
\affiliation[1]{Department of Physics, University of California, Berkeley, California 94720, USA}
\affiliation[2]{Theoretical Physics Group, Lawrence Berkeley National Laboratory, Berkeley, California 94720, USA}
\affiliation[3]{School of Natural Sciences, Institute for Advanced Study, Princeton, New Jersey 08540, USA}

\abstract{
The standard model Higgs quartic coupling vanishes at $(10^{9}-10^{13})$ GeV. We study $SU(2)_L \times SU(2)_R \times U(1)_{B-L}$ theories that incorporate the Higgs Parity mechanism, where this becomes the scale of Left-Right symmetry breaking, $v_R$. Furthermore, these theories solve the strong CP problem and predict three right-handed neutrinos. We introduce cosmologies where $SU(2)_R \times U(1)_{B-L}$ gauge interactions produce right-handed neutrinos via the freeze-out or freeze-in mechanisms. In both cases, we find the parameter space where the lightest right-handed neutrino is dark matter and the decay of a heavier one creates the baryon asymmetry of the universe via leptogenesis.  A theory of flavor is constructed that naturally accounts for the lightness and stability of the right-handed neutrino dark matter, while maintaining sufficient baryon asymmetry. The dark matter abundance and successful natural leptogenesis require $v_R$ to be in the range $(10^{10}-10^{13})$ GeV for freeze-out, in remarkable agreement with the scale where the Higgs quartic coupling vanishes, whereas freeze-in requires $v_R \gtrsim 10^9$ GeV. The allowed parameter space can be probed by the warmness of dark matter, precise determinations of the top quark mass and QCD coupling by future colliders and lattice computations, and measurement of the neutrino mass hierarchy.
}

\date{\today}

\begin{document}
\maketitle
\flushbottom

\newpage

\section{Introduction}
The discovery at the Large Hadron Collider of a Higgs boson with mass 125 GeV~\cite{Aad:2012tfa,Chatrchyan:2012ufa} suggests a new paradigm for particle physics: the mass scale of new physics beyond the Standard Model (SM) is the scale where the Higgs quartic coupling vanishes, $\Lambda_{NP} \sim (10^{9} - 10^{13})$ GeV, and not the weak scale. In this case, a variety of precision measurements at colliders, searches for rare processes, and cosmological observations could reveal this new physics.  $\Lambda_{NP}$ may be the scale where new symmetries emerge, for example Peccei-Quinn symmetry~\cite{Redi:2012ad} or supersymmetry~\cite{Hall:2013eko,Ibe:2013rpa,Hall:2014vga,Fox:2014moa}.

In this paper we study a Higgs Parity extension of the SM \cite{Hall:2018let}. The $SU(2)$ gauge group is extended to $SU(2) \times SU(2)'$ and the Higgs sector is extended to $H(2,1) + H'(1,2)$, with a parity interchanging these Higgs multiplets, $H(2,1) \leftrightarrow H'(1,2)$.  This Higgs Parity is spontaneously broken at $\Lambda_{NP}$ by $\vev{H'}$, yielding the SM as the low energy effective theory. Remarkably, in the limit that the weak scale is far below $\Lambda_{NP}$, the Higgs quartic coupling is found to vanish at $\Lambda_{NP}$. One possibility is that $SU(2)'$ is part of a mirror sector, with mirror matter heavier than ordinary matter by a factor $\vev{H'}/\vev{H}$. This yields a highly predictive scheme for dark matter composed of mirror electrons~\cite{Dunsky:2019api,Dunsky:2019upk}. 

The most economical version of Higgs Parity, which we study in this paper and review in Sec.~\ref{sec:HP}, is based on the simple extension of the SM electroweak gauge group to $SU(2)_L \times SU(2)_R \times U(1)_{B-L}$, first introduced in the 1970s \cite{Pati:1974yy, Mohapatra:1974gc,  Senjanovic:1975rk}. We introduce Higgs doublet multiplets, $H_L(2,1) + H_R(1,2)$, rather than the conventional case of weak triplets and a $(2,2)$ multiplet. Higgs Parity is imposed, $H_L(2,1) \leftrightarrow H_R(1,2)$, and spontaneously broken by $\vev{H_R} =v_R$, so that the SM Higgs quartic coupling vanishing at this Left-Right (LR) symmetry breaking scale $v_R$.  This theory has the same number of gauge couplings and charged fermion Yukawa couplings as the SM.  The Higgs potential has three parameters rather than two; but one of these is irrelevant as it only determines the mass of the right-handed Higgs boson.  Another determines the electroweak scale $\vev{H_L} =v$, while the third provides a correlation between the Higgs boson mass, the top quark mass, the QCD coupling and $v_R$. For this theory, precision measurements at future colliders will play a key roll in sharpening this prediction for $v_R$, which is presently highly uncertain
\begin{align}
v_R \sim (10^{9} - 10^{13}) \; \GEV.
\label{eq:vR}
\end{align}
This will test whether precision gauge coupling unification in $SO(10)$ can be realized, and whether proton decay is within reach of future searches~\cite{Hall:2019qwx}. 

It has been known for many years that spacetime parity can solve the strong CP problem, in particular in the context of the gauge group $SU(2)_L \times SU(2)_R \times U(1)_{B-L}$ broken solely by doublets $H_{L,R}$ \cite{Babu:1989rb}. Indeed, the Higgs Parity theory we study actually has one less relevant parameter than the SM, since $\bar{\theta}=0$ at tree-level.  Non-zero contributions arise at the two-loop level and are estimated to typically generate the neutron electric dipole moment of order $10^{-27}$ ecm~\cite{Hall:2018let}, and may be within the reach of current searches. Given the simplicity of the parity solution of the strong CP problem proposed in \cite{Babu:1989rb}, why does the solution involving an anomalous Peccei-Quinn symmetry \cite{Peccei:1977hh, Peccei:1977ur} dominate the literature? The answer may be that it requires an axion \cite{Weinberg:1977ma, Wilczek:1977pj}; a candidate for the cosmological dark matter with plausible production mechanisms~\cite{Preskill:1982cy, Abbott:1982af,Dine:1982ah,Sikivie:1982qv,Davis:1986xc,Kawasaki:2014sqa,Co:2017mop,Co:2019jts}. Furthermore, the axion can be searched for in many ways and will be probed in the coming decade over much of its parameter range. In Secs.~\ref{sec:RHnuDM} and \ref{sec:N1prod} of this paper, we show that the LR Higgs Parity theory also contains a dark matter candidate that can be produced in the early universe, leading to constraints and tests on the theory.

The minimal description of neutrino masses is to add the dimension 5 operator $\ell_i \ell_j HH$ to the SM, where $\ell_i$ are the lepton doublets and $H$ the Higgs doublet.  Alternatively, right-handed neutrinos $N_i$ can be added to the theory together with the two operators
\begin{align}
{\cal L}_{ \rm SM +N}\; \supset   \; y_{ij} \,\ell_i  N_j \, H \, + \, \frac{M_{ij}}{2} \, N_i N_j, \hspace{1in}    (+\,\ell_i \ell_j \, HH )
\label{eq:SMnu}
\end{align}
involving two flavor flavor matrices. (The $\ell_i \ell_j HH$ operator could also be present, but in the seesaw mechanism~\cite{Yanagida:1979as,GellMann:1980vs,Minkowski:1977sc, Mohapatra:1979ia} it is taken to be subdominant.) A virtue of adding the right-handed neutrinos is that, if they are produced in the early universe, their decays can lead to the cosmological baryon asymmetry via leptogenesis~\cite{Fukugita:1986hr}.

Theories containing $SU(2)_L \times SU(2)_R$ gauge symmetry necessarily contain $N_i$ as the neutral member of the $SU(2)_R$ doublets $\bar{\ell}_i$. In the effective theory below the scale $v_R$, the generic structure of the operators  leading to neutrino masses is
\begin{align}
{\cal L} _{ \rm LR}\; \supset   \;  y_{ij} \, \ell_i \,  N_j \, H_L    +  \frac{M_{ij}}{2} \, N_i  \, N_j   +   c \, \frac{M_{ij}}{2 v_R^2} \, \ell_i  \, \ell_j \, H_LH_L. 
\label{eq:LRnu}
\end{align}
Even though there are three operators, the flavor matrices for the $\ell_i \ell_j$ and $N_i N_j$ terms are identical, although there is a model dependent coefficient $c$ in the relative strengths of these two terms.  If the lightest right-handed neutrino $N_1$ has a very small mass $M_1$, it could be dark matter, produced in the early universe via $SU(2)_R\times U(1)_{B-L}$ gauge interactions~\cite{Khalil:2008kp,Bezrukov:2009th,Dror:2020jzy}. With an abundance set by freeze-out (and subsequent dilution by the decay of a heavier right-handed neutrino, $N_2$) the allowed range of the  $(M_1,v_R)$ parameter space was found to be restricted to a triangle, with a location that depended on $c$~\cite{Dror:2020jzy}.  With $c=1$, the allowed ranges within the triangle were roughly  $M_1 \sim 2-300$ keV and $v_R \sim 10^{10\pm2}$ GeV. Lowering $c$ led to a lowering of $v_R$ and a reduction in the range for $M_1$, with no parameter space for $v_R< 10^6$ GeV.  Increasing $c$ above unity requires fine-tuning in the theory, but opens up regions to larger values of $M_1$ and $v_R$. Large values of these parameters were also consistent with $N_1$ dark matter produced via freeze-in.

In the LR Higgs Parity theory, neutrino masses are generated by the operators of (\ref{eq:LRnu}) with $c=1$.  As noted above, without interactions for neutrino masses the LR Higgs Parity theory has one fewer relevant parameter than the SM; adding the neutrino mass interactions, (\ref{eq:SMnu}) for the SM and (\ref{eq:LRnu}) with $c=1$ for Higgs Parity, does not alter this. Thus $N_1$ dark matter can arise as in \cite{Dror:2020jzy} and, remarkably, in the case that its abundance is determined by freeze-out, the required scale $v_R \sim 10^{10\pm2}$ GeV lies inside the range (\ref{eq:vR}) determined by the Higgs mass. $N_1$ dark matter can be probed by future precision collider data that tightens the range of (\ref{eq:vR}). 

In Sec.~\ref{sec:lepto} we show that leptogenesis from the decay of $N_2$ is possible in this theory, at the same time that $N_1$ provides the dark matter, and we investigate the extent to which the resulting reduced range for $M_1$ can be probed using 21cm cosmology.

In theories of sterile neutrino dark matter, there are naturalness issues for the small mass and long lifetime of the sterile neutrino.  This is especially true in the LR symmetric theory, as the interactions of $N_i$ are either determined by symmetry or constrained by the observed neutrino masses and mixings.  In Sec.~\ref{sec:natural_EFT} we study radiative corrections to the mass and lifetime in the effective theory where quark and lepton masses arise from dimension 5 operators. These lead to significant naturalness constraints on the parameter space for dark matter. In Sec.~\ref{sec:model} we introduce UV completions of these operators that greatly improve the naturalness of the long-lived, light right-handed neutrino dark matter.  In Sec.~\ref{sec:natlepto} we study the naturalness of leptogenesis in these theories and find highly restricted ranges for the LR symmetry breaking scale, $v_R$, and the dark matter mass, $M_1$.  Conclusions are drawn in Sec.~\ref{sec:conclusion}.

\section{Higgs Parity}
\label{sec:HP}
We begin with a brief review of Higgs Parity, first introduced in~\cite{Hall:2018let}, as a model that simultaneously predicts a nearly vanishing Higgs quartic coupling at a scale $10^{9-13} \GEV$ and solves the strong CP problem. 

\subsection{Vanishing quartic}
Higgs Parity is a $Z_2$ symmetry that exchanges the $SU(2)_L$ gauge interaction with a new $SU(2)'$ interaction. The SM Higgs field $H(2,1)$ is exchanged with its $Z_2$ partner $H'(1,2)$, where the brackets show the $(SU(2)_L,SU(2)')$ charges. The scalar potential of $H$ and $H'$ is
\begin{align}
\label{eq:potential}
V(H,H') = - m  ^2 \left( \left| H \right|^2  + \left| H' \right|^2 \right) + \frac{\lambda}{2} \left( \left| H \right|^2  + \left| H' \right|^2 \right) ^2 + \lambda' \left| H \right|^2 \left| H' \right|^2.
\end{align}
We assume that the mass scale $m$ is much larger than the electroweak scale, $v$.

With positive $m^2$, $H'$ obtains a large vacuum expectation value $\vev{H'} = m  / \lambda^{1/2} \equiv v'$ and Higgs Parity is spontaneously broken. After integrating out $H'$ at tree-level, the low energy effective potential of $H$ is 
\begin{align}
V_{\rm LE}(H) = \lambda' \; v'^2  \;  \left| H \right|^2 - \lambda' \left(1  + \frac{\lambda'}{2 \lambda} \right) \left| H \right|^4 .
\end{align}
The hierarchy $v \ll v'$ is obtained only if the quadratic term is small, which requires a small value of $\lambda' \sim - v^2/v^{'2}$. The quartic coupling of the Higgs $H$, $\lambda_{\rm SM}$, is then very small at the symmetry breaking scale $v'$.
The nearly vanishing quartic coupling can be understood by an approximate global $SU(4)$ symmetry under which $(H,H')$ forms a fundamental representation. For $|\lambda'| \ll 1$ the potential in Eq.~(\ref{eq:potential}) becomes $SU(4)$ symmetric. The $SU(4)$ symmetry is spontaneously broken by $\vev{H'}$ and the SM Higgs is understood as a Nambu-Goldstone boson with vanishing potential.

At tree-level the potential still leads to $\vev{H} = \vev{H'} = v'/\sqrt{2}$ because of the small quartic coupling. However, for extremely small $\lambda'$, vacuum alignment in the $SU(4)$ space is fixed by quantum corrections which violate the $SU(4)$ symmetry. The dominant effect is renormalization group running from energy scale $v'$ down to $v$. The top contribution dominates over the gauge contribution and generates a positive quartic coupling $\lambda_{\rm SM}(v) \simeq 0.1$, and creates the minimum of the potential at $v \ll v'$. From the perspective of running from low to high energy scales, the scale at which the SM Higgs quartic coupling nearly vanishes is the scale $v'$. Threshold corrections to $\lambda_{\rm SM}(v')$ are computed in~\cite{Dunsky:2019api,Hall:2019qwx} and are typically $O(10^{-3})$.

The vacuum alignment can be also understood in the following way.
For $\lambda' > 0$, the minima of the potential are $\left( \vev{H}, \vev{H'} \right)= \left( v',0 \right)$ and $\left( 0,v' \right)$, where $v'\equiv m / \lambda^{1/2}$, and the mass of Higgses are as large as $m$. For $\lambda' < 0$, the minima are $\vev{H} = \vev{H'} \sim v'$. None of the minima for $\lambda>0$ and $\lambda'<0$ has a non-zero but small $v$.
To obtain a viable vacuum, we need $\lambda' \simeq 0$, for which the potential has an accidental $SU(4)$ symmetry and nearly degenerate vacua with $\vev{H^2} + \vev{H'}^2 = {v'}^2$. In this case, quantum corrections must be taken into account to determine the minimum. The dominant effect is given by the top quark Yukawa coupling. The Colemann-Weinberg potential given by the top Yukawa makes $\left( \vev{H}, \vev{H'} \right)= \left( v',0 \right)$ and $\left( 0,v' \right)$ minima. By switching-on small negative $\lambda'$, the vacuum  $\left( \vev{H}, \vev{H'} \right)= \left( 0,v' \right)$ is slightly destabilized and we may obtain $\left( \vev{H}, \vev{H'} \right)= \left( v ,v' \right)$ with $v \ll v'$. There also is a physically equivalent minimum connected to this by Higgs Parity, $\left( \vev{H}, \vev{H'} \right)= \left( v' ,v \right)$.

\subsection{Left-right Higgs Parity}
In this work, we consider the case where only the right-handed (SM) fermions are charged under $SU(2)'$, i.e., $SU(2)_R$, and we accordingly relabel $(H,H')$ as $(H_L, H_R)$. The gauge group of the theory is $SU(3)_c \times SU(2)_L\times SU(2)_R \times U(1)_{B-L}$ and the matter content is listed in Table~\ref{tab:charge}. The presence of the right-handed neutrinos is now required by the gauge symmetry. Higgs Parity maps $SU(2)_L \leftrightarrow SU(2)_R$, and hence $\ell \leftrightarrow \bar{\ell}^\dagger$, $q \leftrightarrow \bar{q}^\dagger$, and $H_L \leftrightarrow H_R^\dag$.%
\footnote{If the $Z_2$ does not include spacetime parity, $\ell \leftrightarrow \bar{\ell}$, $q \leftrightarrow \bar{q}$ and $H_L \leftrightarrow H_R$}
The symmetry breaking pattern is,
\begin{align} 
& SU(3)_c \times SU(2)_L\times SU(2)_R \times U(1)_{B-L} \times Z_2 \nonumber \\
& \hspace{1cm}\xrightarrow{\left\langle H _R \right\rangle } SU(3)_c \times SU(2)_L \times U(1)_{Y} \xrightarrow{\left\langle H _L \right\rangle} SU(3)_c \times U(1)_{\rm EM}.
\end{align} 

\begin{table}[htp]
\begin{center}
\setlength{\tabcolsep}{3pt}
\begin{tabular}{ccccccc}
\toprule[1.5pt]
 & $q$ & $\ell$  & $\bar{q} = (\bar{u}, \bar{d}) $ &  $\bar{\ell} \equiv (N,\bar{e}) $ & $H_L$ & $H_R$ \\ \midrule[.5pt] 
$SU(3)_c$ & $3$ & 1  & $\bar{3}$ & 1 & 1& 1 \\
$SU(2)_L$ & $2$ & $2$ & $1$  & $1$ & $2$  & $1$ \\
$SU(2)_R$ & $1$ & $1$ & $2$ &  $2$ & $1$  & $2$ \\ 
$U(1)_{B-L}$ & $1/6$ & $-1/2$ & $-1/6$ &  $1/2$ & $1/2$  & $-1/2$ \\ 
\bottomrule[1.5pt]
\end{tabular}
\end{center}
\caption{The gauge charges of quarks, leptons, $H_L$, and $H_R$.}
\label{tab:charge}
\end{table}%

In contrast to conventional Left-Right symmetric models, we do not introduce scalar multiplets in $(2,2)$, $(3,1)$ or $(1,3)$ representations of $SU(2)_L \times SU(2)_R$ around the scale $v_R$; the Higgs Parity explanation for the vanishing quartic coupling holds only if $SU(2)_R$ and $SU(2)_L$ symmetry are dominantly broken by $H_R$ and $H_L$. Thus, Yukawa couplings are forbidden at the renormalizable level, and arise from dimension-5 operators,
\begin{align}
\label{eq:yukawa}
-{\cal L} _{\rm e,u,d} &=  \frac{c_{ij}^u}{M} \, q_i \bar{q}_j H_L H_R +    \frac{c_{ij}^d}{M} \, q_i \bar{q}_j H_L^\dag H_R^\dag +   \frac{c_{ij}^e}{M} \, \ell_i \bar{\ell}_j H_L^\dag H_R^\dag + {\rm h.c.}, &\hspace{0.5in} y^f_{ij} \equiv c^f_{ij} \frac{v_R}{M} \\
- {\cal L} _{\nu, \rm N}  &= \frac{c_{ij}}{2M} \left( \ell_i \ell_j H_L H_L  +   \bar{\ell}_i \bar{\ell}_j H_R H_R \right)  -  \frac{b_{ij}}{M} \,  \ell_i \bar{\ell}_j  H_L H_R + {\rm h.c.} &\hspace{0.5in} y_{ij} \equiv b_{ij} \frac{v_R}{M}
\label{eq:yukawaNu}
\end{align}
These can arise, e.g., from exchanges of massive Dirac fermions (as considered in~\cite{Hall:2018let,Hall:2019qwx}) or from the exchange of a massive scalar with a charge $(1,2,2,0)$.%
\footnote{To obtain the up and down quark masses solely from the exchange of $(1,2,2,0)$, it must be a complex scalar rather than a pseudo-real scalar. In this case, the strong CP problem cannot be solved by parity because of the complex vacuum expectation value of the complex scalar, unless extra symmetries, such as supersymmetry, are imposed~\cite{Beg:1978mt,Mohapatra:1978fy,Kuchimanchi:1995rp,Mohapatra:1995xd}.}
In Sec.~\ref{sec:model}, we take some of the masses of the Dirac fermions to be small. In this case, the corresponding SM right-handed fermions dominantly come from the Dirac fermions rather than the $SU(2)_R$ doublets.
The origin of the neutrino masses are discussed in Sec.~\ref{sec:RHnuDM}.

\subsection{Strong CP problem}
Higgs Parity can also solve the strong CP problem if $SU(3)_c$ is $Z_2$ neutral and the $Z_2$ symmetry includes space-time parity~\cite{Hall:2018let}. Then spacetime parity forbids the QCD $\theta$ parameter at tree-level and requires the quark mass matrices $y^f_{ij}v$ in Eq.~(\ref{eq:yukawa}) to be Hermitian and thus enjoy real eigenvalues. The determinant of the quark mass matrix is then real and hence $\bar{\theta}$ is absent at both tree-level and at one-loop. Two-loop corrections to the quark mass matrix give non-zero $\bar{\theta}$~\cite{Hall:2018let}, but can be below the experimental upper bound from the neutron electric dipole moment.

Solving the strong CP problem by restoring space-time parity was first pointed out in ~\cite{Beg:1978mt,Mohapatra:1978fy}.
The first realistic model was proposed in~\cite{Babu:1988mw,Babu:1989rb}, which used $(2,1) + (1,2)$ Higgses and Dirac fermions to generate the Yukawa coupling in Eq.~(\ref{eq:yukawa}). In their model, space-time parity is assumed to be softly broken in the Higgs potential to obtain the hierarchy $v \ll v_R$. In the setup of~\cite{Hall:2018let}, Higgs Parity including space-time parity is spontaneously broken without soft breaking and predicts vanishing $\lambda_{\rm SM}(v_R)$. The embedding of the theory into $SO(10)$ unification is achieved in~\cite{Hall:2018let,Hall:2019qwx}, with Higgs Parity arising from a $Z_2$ subgroup of $SO(10)$.

\subsection{Prediction for the Higgs Parity symmetry breaking scale}
\begin{figure}[tb]
    \centering
    \begin{minipage}{0.5\textwidth}
        \centering
        \includegraphics[width=.95\textwidth]{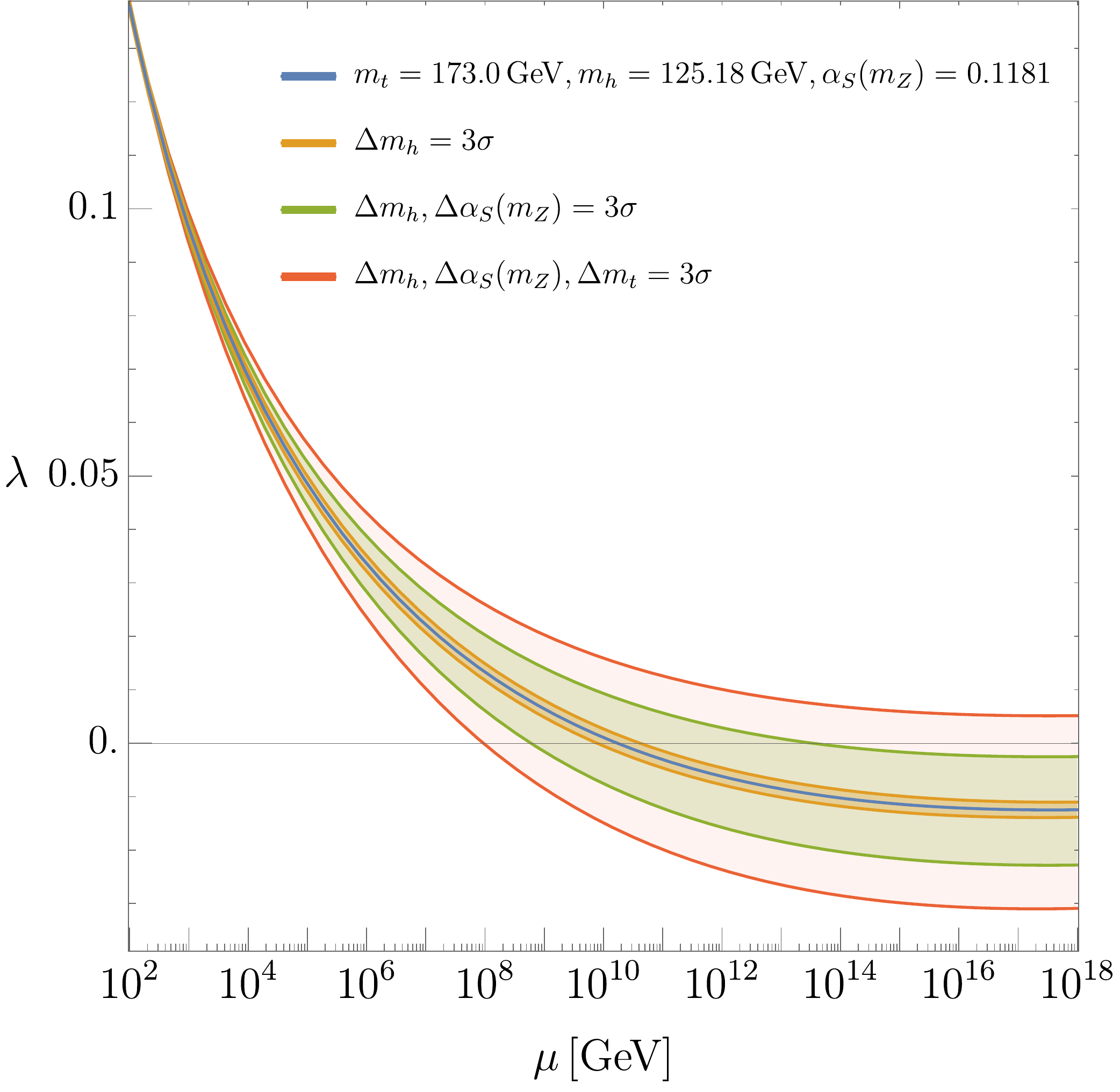} 
    \end{minipage}\hfill
    \begin{minipage}{0.5\textwidth}
        \centering
        \includegraphics[width=.95\textwidth]{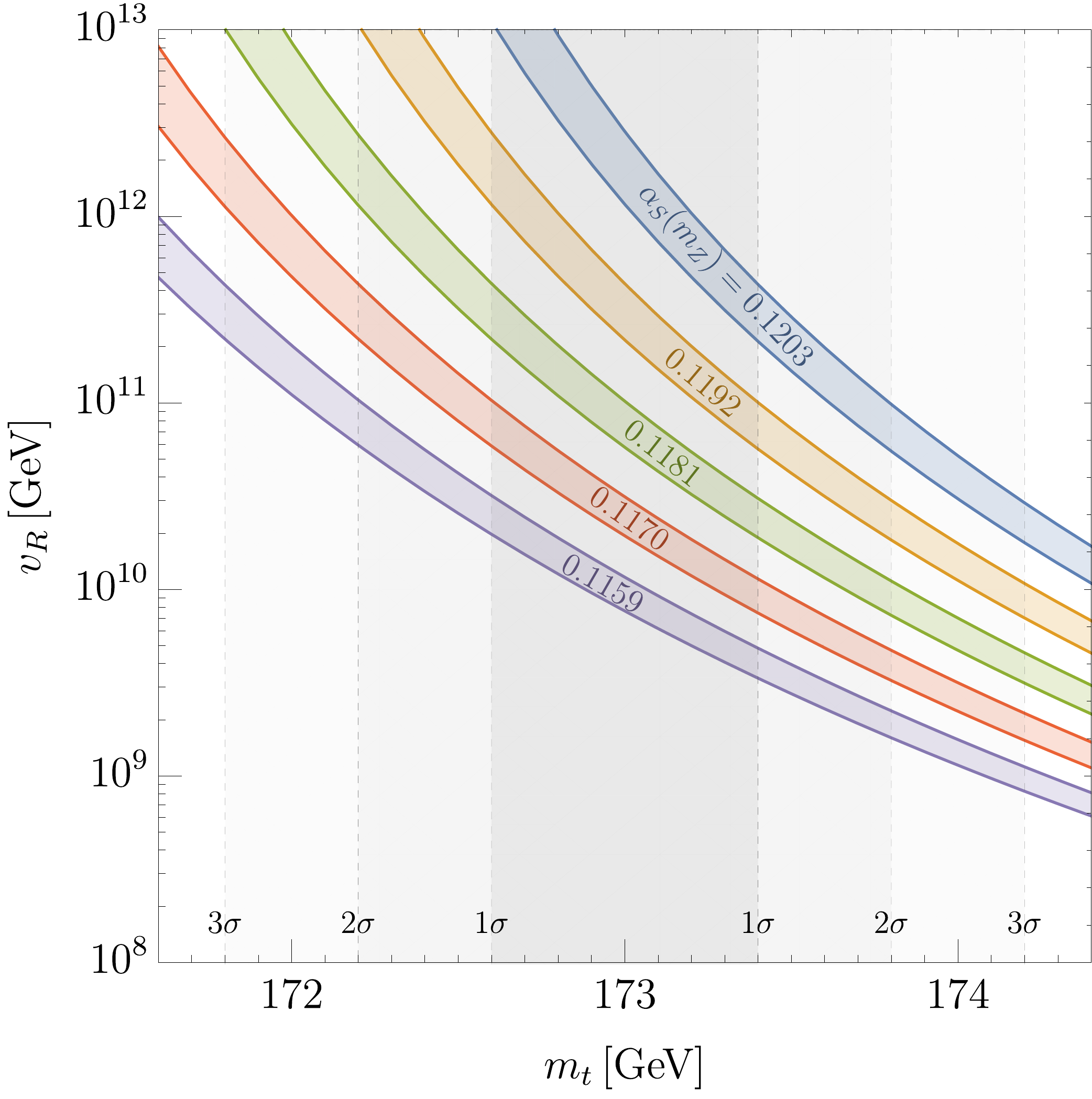} 
    \end{minipage}
    \caption{\small (\textbf{Left}) Running of the SM quartic coupling. (\textbf{Right}) Predictions for the scale $v_R$ as a function of the top quark mass, $m_t$. Contours of $\alpha_S(M_Z)$ show how the prediction changes with the uncertainty in the QCD couping constant. The thickness of each countour corresponds to $\pm 1\sigma$ deviation in $m_h$.}
    \label{fig:vpPrediction}
\end{figure}

Between the electroweak scale and the Left-Right scale $v_R$, the running of the Higgs quartic coupling $\lambda_{\rm SM}$ is exactly the same as in the SM. We follow the computation in~\cite{Buttazzo:2013uya} and show the running in the left panel of Fig.~\ref{fig:vpPrediction} for a range of values for the top quark mass $m_t = (173.0\pm 0.4)$ GeV, QCD coupling constant at the $Z$ boson mass $\alpha_S(m_Z) = (0.1181 \pm 0.0011)$, and Higgs mass $m_h = (125.18 \pm 0.16)$ GeV.

The value of the SM quartic coupling at the scale $v_R$ is not exactly zero because of the threshold correction~\cite{Dunsky:2019api},
\begin{align}
\lambda_{\rm SM}(v_R) \simeq - \frac{3}{8\pi^2} y_t^4 \, {\rm ln} \frac{e}{y_t} + \frac{3}{128\pi^2} (g^2 + {g'}^2)^2 \left( {\rm ln} \frac{e\sqrt{2}}{\sqrt{g^2 + {g'}^2 }} -  {\rm ln} \frac{g^2}{\sqrt{g^4 - g'^4}} \right) + \frac{3}{64\pi^2} g^4 \, {\rm ln} \frac{e\sqrt{2}}{ g},
\end{align}
where the $\overline{\rm MS}$ scheme is assumed.
The prediction for the scale $v_R$ is shown in the right panel of Fig.~\ref{fig:vpPrediction} as a function of $m_t$. Colored contours show how the prediction in $v_R$ changes when the QCD coupling constant varies by $\pm 2$ deviations about its mean, $\alpha_S(M_Z) = 0.1181 \pm 0.0011$. The thickness of each curve corresponds to the 1-sigma uncertainty in the measured Higgs mass, $m_h = (125.18 \pm 0.16)$ GeV. With 2$\sigma$ uncertainties, $v_R$ can be as low as $10^9$ GeV. Future measurements of SM parameters can pin down the scale $v_R$ with an accuracy of a few tens of percent~\cite{Dunsky:2019api}.

\section{Right-handed neutrino dark matter}
\label{sec:RHnuDM}
In this section, we review the results of \cite{Dror:2020jzy} on the general properties and constraints of right-handed neutrino dark matter in LR theories.
\subsection{Neutrino masses}
The effective Lagrangian of \eqref{eq:yukawaNu} leads to a $6 \times 6$ neutrino mass matrix,
\begin{align}
\begin{array}{c} \big( \begin{array}{cc}\nu  _i   & N _i \end{array} \big) \\ {} \end{array}
  \begin{pmatrix}
 M_{ij} \,v ^2 /v_R ^2  & y_{ij} v \\
 y_{ji} v & M_{ij}^{(*)}
\end{pmatrix} \bigg( \begin{array}{c} 
 \nu _j \\  
 N _j 
\end{array} \bigg)
  \,,
\end{align}
where $M_{ij} = c_{ij} v_R^2/M$.  Without loss of generality, we can work in a basis where $c_{ij}$ is diagonal such that
\begin{align}
	M_{ij} &= M_i \, \delta_{ij},
\end{align}
with all $M_i$ real and positive.  Upon integrating out the three heavy states, we obtain a mass matrix for the three light neutrinos:
\begin{align}
	m_{ij} \, &= \, \delta_{ij} \frac{v^2}{v_R^2} M_i - y_{ik} v \; \frac{1}{M_k} \; y_{jk} v \,\equiv \, \delta_{ij} \, m_{i}^{(5)} - m_{ij}^{(ss)}.
	\label{eq:numassmatrix}
\end{align}
In this basis, and in the limit that $y _{ij}$ is diagonal, the lepton flavor mixing arises entirely from the charged lepton mass matrix.

\subsection{The lightest right-handed neutrino as dark matter}
We define $N_1$ as the right-handed neutrino responsible for the dark matter (DM) density of the universe.%
\footnote{Note that our numbering of SM neutrinos does not necessarily coincide with the neutrino numbering commonly found in the literature. }
Even though there is no symmetry that stabilizes $N_1$, it may be sufficiently long-lived to be a DM candidate.

$ N _1 $ decays via $ N _1 - \nu $ mixing controlled by $ y _{i1} $.
The $N_1-\nu$ mixing angle is given by
\begin{align}
	\sin 2 \theta_1 &\equiv \frac{ v}{M_1} \sqrt{\Sigma_i \; |y_{i1}|^2},
	\label{eq:sin2thetaDef}
\end{align}
where $v\simeq 174$ GeV.
The experimental constraints on $ \sin 2 \theta_1 $ arise from two different processes:
1) $N_1$ DM may be overproduced via the Dodelson-Widrow mechanism~\cite{Dodelson:1993je}.
2) $N_1$ DM decays into $\nu \gamma$ and may overproduce photons relative to observed diffuse photon backgrounds and galaxy fluxes~\cite{Adhikari:2016bei}. This decay rate is given by:
\begin{align}
\Gamma_{N_1\rightarrow \nu\gamma}  \simeq \frac{9 \alpha}{8192 \pi^4}  \; \frac{M_1^5}{v^4} \; \sin^2 2 \theta_1 \,
 \simeq  \left(1.5 \times 10^{30} \sec \right)^{-1} \left(\frac{M_1}{1 ~\KEV}\right)^5 \left( \frac{ \sin^2 2 \theta_1 }{5 \times 10 ^{ - 9} } \right) \,.
\label{eq:mixing}
\end{align}
These two constraints are summarized by the experimental limit on the mixing angle~\cite{Adhikari:2016bei},
\begin{equation}
\frac{ v^2}{M_1^2} \Sigma_i \; |y_{i1}|^2 \leq
\sin ^2 2 {\theta_1}_{\rm exp} \simeq 5 \times 10^{-9} 
\begin{cases}
\left(\dfrac{M_1}{3 ~\KEV}\right)^{-1.8} \times D &  
\quad \text{(Overproduction)}\\
\left(\dfrac{M_1}{3 ~\KEV}\right)^{-5} & 
\quad \text{(Decay)}.
\end{cases}
\label{eq:yi1}
\end{equation}
Here $D$ is a possible dilution factor after $N_1$ is produced by the Dodelson-Widrow mechanism.  The higher photometric sensitivities of next generation x-ray and gamma-ray telescopes such as ATHENA~\cite{Nandra:2013shg} and  e-ASTROGAM~\cite{Tatischeff:2016ykb} may probe an order of magnitude smaller decay rate~\cite{Caputo:2019djj}. For $M_1 > 1$ MeV, the tree-level decay $N_1 \rightarrow e^+ e^- \nu$ is open and the resultant constraint on $y_{i1}$ is similar to (\ref{eq:yi1}).

Regardless of how small $y_{i1}$ is, constraints arise from $N_1$ decays mediated by gauge exchange. For example, $N_1$ decays into $\ell^\pm +$ hadron(s) via $W_R$ exchange when kinematically allowed. In addition, $W_R$ and $W_L$ mix with each other by a top-bottom-loop, and $N_1$ may decay into $\ell^+ \ell^- \nu$. The experimental upper bounds on these decay rates are about $10^{-25}\,{\rm sec}^{-1}$~\cite{Essig:2013goa}. Furthermore, the $W_R-W_L$ mixing also generates a radiative decay of $N_1$ into $\nu\gamma$ \cite{Bezrukov:2009th,Lavoura:2003xp,Greljo:2018ogz}, which has a stronger experimental upper limit of about $10^{-27}\, {\rm sec}^{-1}$ due to the emission of a hard photon \cite{Adhikari:2016bei}. The parameter region with large $M_1$ and/or small $v_R$ is excluded by these gauge-induced decays as discussed more in~\cite{Dror:2020jzy} and shown graphically in Fig.~\ref{fig:thermal_FI}.

\section{Cosmological production of right-handed neutrino dark matter}
\label{sec:N1prod}
In this section, we review the two production mechanisms of $N_1$ DM considered in this paper~\cite{Dror:2020jzy}:
\begin{itemize}
\item 
At sufficiently high reheating temperatures $T_{\rm RH}^{\rm inf}$ after inflation, $N_i$ have a thermal abundances from $W_R$ exchange. The $N_1$ abundance is reduced by an appropriate amount to the DM abundance by making $N_2$ long-lived so that entropy is produced upon decaying.
\item
At low reheating temperatures $T_{\rm RH}^{\rm inf}$ after inflation, the $N_1$ DM abundance is produced by freeze-in via $W_R$ exchange.  $N_{2}$ are also produced by freeze-in, via $W_R$ exchange or via the Yukawa couplings with $\ell H$.
\end{itemize}
In these two scenarios, $N_1$ DM can be obtained over a wide range of parameter space. 

\subsection{Relativistic freeze-out and dilution}
The right-handed neutrinos couple to the SM bath via $W_R$ exchange. If the reheat temperature of the universe after inflation is sufficiently high,
\begin{align}
\label{eq:eqtemp}
T_{\rm{RH}} ^{ {\rm inf}} \gtrsim 10^8 ~\GEV \left(\frac{v_R}{10^{10} ~\GEV}\right)^{4/3},
\end{align}
the right-handed neutrinos reach thermal equilibrium and subsequently decouple with a thermal yield $Y_{\rm therm} \simeq 0.004$.%
\footnote{The analysis is this section is also applicable to lower $T_{\rm RH}^{\rm inf}$ 
as long as $N_1$ and $N_2$ are frozen-in from $W_R$ exchange, and $N_1$ is overproduced as DM (see Eq.~\eqref{eq:FIAbundance}). In such a scenario, the required dilution to realize $N_1$ DM is diminished, and hence the warmness constraints on $N_1$ slightly increase above $2 \, \KEV$. See Fig.~\ref{fig:thermal_FI} for the warmness constraints on a pure freeze-in cosmology without any dilution.}
For $N_1$ to have the observed DM abundance requires $m_{N_1} \simeq 100$~eV. Such light sterile neutrino DM, however, is excluded by the Tremaine-Gunn~\cite{Tremaine:1979we,Boyarsky:2008ju,Gorbunov:2008ka} and warmness~\cite{Narayanan:2000tp,Irsic:2017ixq,Yeche:2017upn,Seljak:2006qw} bounds; see~\cite{Adhikari:2016bei} for a recent review.

$N_1$ may be DM  
if their abundance is diluted. If another right-handed neutrino, $N_2$, is sufficiently long-lived such that it comes to dominate the energy density of the universe and produces entropy when it decays, it can dilute the DM abundance and cool $N_1$ below warmness bounds~\cite{Asaka:2006ek,Bezrukov:2009th}. The relic density of $N_1$ is
\begin{align}
\frac{\rho_{N_1}}{s} & =  1.6 \; \frac{3}{4} \; \frac{ M_1}{M_2} \; T_{\rm RH} \,, \nonumber \\ 
\Rightarrow \frac{ \Omega_{N_1} }{ \Omega _{ {\rm DM}}} & \simeq \left( \frac{ M _1 }{ 10 \, {\rm keV} } \right) \left( \frac{ 300 \,{\rm GeV} }{ M _2 } \right) \left( \frac{ T _{\rm RH}}{ 10 \, {\rm MeV}  } \right) \,,
\label{eq:DMabund}
\end{align}
where the numerical factor $1.6$ is taken from~\cite{Harigaya:2018ooc}, $\rho _{ N _1 } $ is the energy density, $ s $ is the entropy density, $ \Omega _{ {\rm DM}} \simeq 0.25 $ is the observed cosmic relic abundance, and $T_{\rm RH}$ is the decay temperature of $N_2$, as set by its total decay rate $\Gamma_{N _2 }$
\begin{align}
T_{\rm RH} = \left(\frac{10}{\pi^2 g_*}\right)^{1/4}\sqrt{\Gamma_{ N _2 } \mpl}.
\label{eq:TRH}
\end{align}

The reheating bound from hadronic decays of $N_2$ during BBN ($T_{\rm RH} > 4 \,\MEV$)~\cite{Kawasaki:1999na,Kawasaki:2000en,Hasegawa:2019jsa},
requires that $N_2$ is heavy enough,
\begin{align}
	M_2 \gtrsim 24 \, \GeV \frac{M_1}{2~\KEV}.
	\label{eq:m2bound}
\end{align}
Low reheating temperatures can also affect the CMB since some decays occur after neutrinos decouple and heat up only electrons and photons, relatively cooling neutrinos and reducing the effective number of neutrinos~\cite{Kawasaki:1999na,Kawasaki:2000en,Ichikawa:2005vw}. In our case, $N_2$ also decays into neutrinos and the bound from the CMB, $T_{\rm RH}> 4$ MeV~\cite{deSalas:2015glj}, may be relaxed.

To achieve the dilution of $N_1$ dark matter, $N_2$ must be long-lived enough. $N_2$ can always beta decay through $W_R$ exchange into right-handed fermions, $N_2 \rightarrow (\ell^+ \bar{u} d,\, \ell^- u \bar{d})  $ and $N_2 \rightarrow N_1 \ell^+ \ell^- $. These decay channels are unavoidable as they are independent of the free-parameter $y_{i2}$, and prevent $N_2$ from efficiently diluting $N_1$ for large $M_2$ and/or small $v_R$. In addition, $N_2$ can decay through the couplings $y_{i2}$.
When $M_2 \gtrsim v$, $N_2$ can decay at tree-level via $ N _2 \rightarrow \nu  h ,  \nu Z , \ell ^\pm  W_L ^\mp $ while for $M_2 \lesssim v$, $N_2$ can beta decay through $W_L/Z$ exchange and active-sterile mixing to SM fermions, $N_2 \rightarrow \ell ud, \ell^+ \ell^- \nu, \nu \nu \bar{\nu}$. As discussed in more detail in Ref.~\cite{Dror:2020jzy}, these decays require $y_{i2}$ to be sufficiently small.

In Ref.~\cite{Dror:2020jzy}, we used the above results, together with the radiative stability bound on $N_1$, to derive constrains on the neutrino mass matrix of \eqref{eq:numassmatrix}. 
We considered the cases with $M_3 \gtrsim M_2$ and $M_3 \ll M_2$.
As we will see later, efficient leptogenesis require that $M_3\gtrsim M_2$. For this case, Ref.~\cite{Dror:2020jzy} shows that the lightest neutrino mass eigenstate is closely aligned with $\nu_1$ and has a mass $m_1 \ll \sqrt{\smash[b]{ \Delta m_{\rm sol}  ^2}}$. The other two mass eigenstates are very close to $\nu_2$ and $\nu_3$ and have masses $m_2 = (v^2/v_R^2) M_2$ and $m_3 = \,(v^2/v_R^2) M_3 - y_{33}^2 v^2/M_3$.
The mass of $N_2$ is thus fixed as
\begin{align}
M_2 &\simeq m_2 \left(\frac{v_R}{v}\right)^2.
\label{eq:M2m2}
\end{align}

\label{sec:signals}
\begin{figure}[]
    \centering
    \begin{minipage}{0.5\textwidth}
        \centering
        \includegraphics[width=1\textwidth]{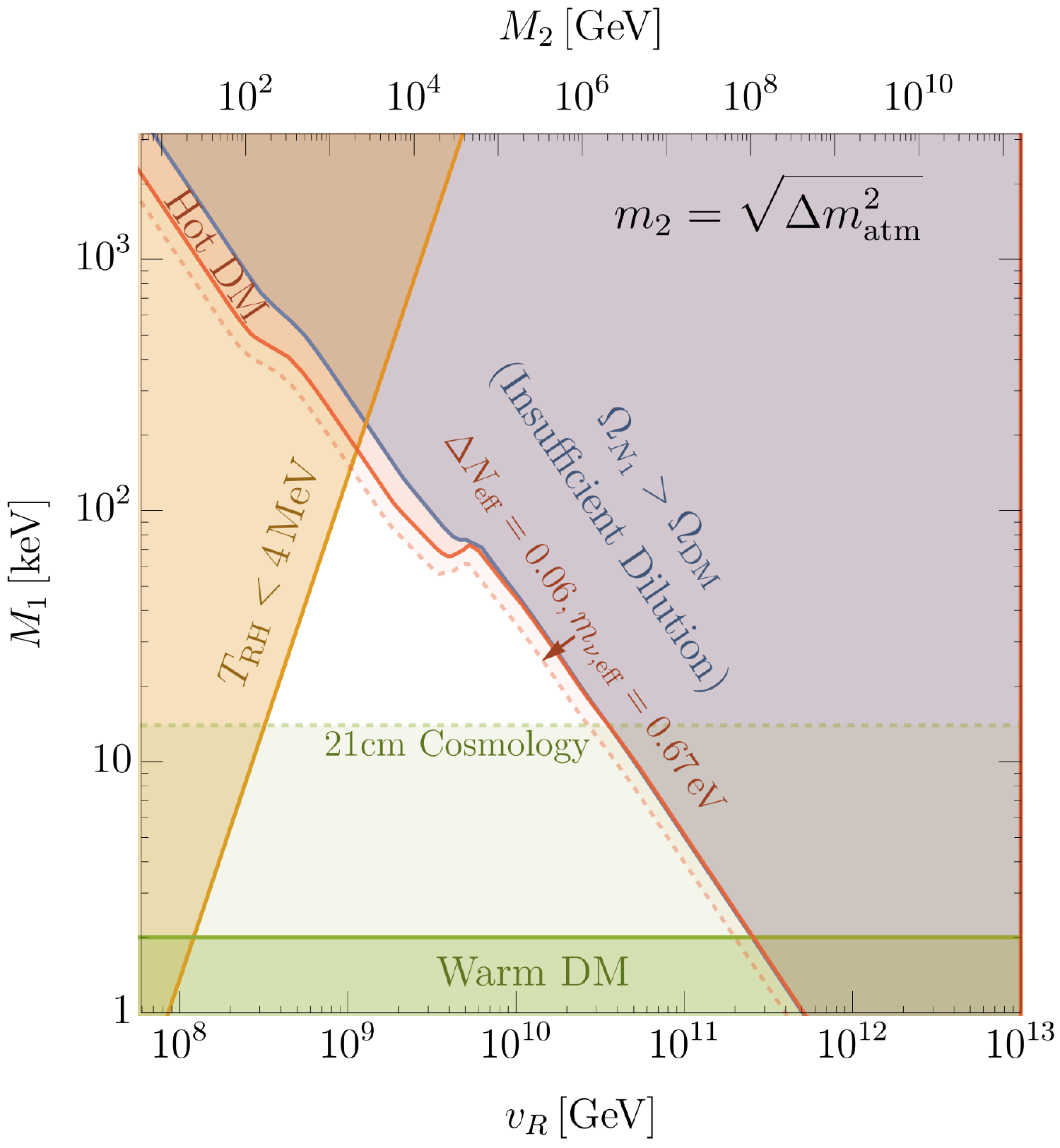} 
    \end{minipage}\hfill
    \begin{minipage}{0.5\textwidth}
        \centering
        \includegraphics[width=1\textwidth]{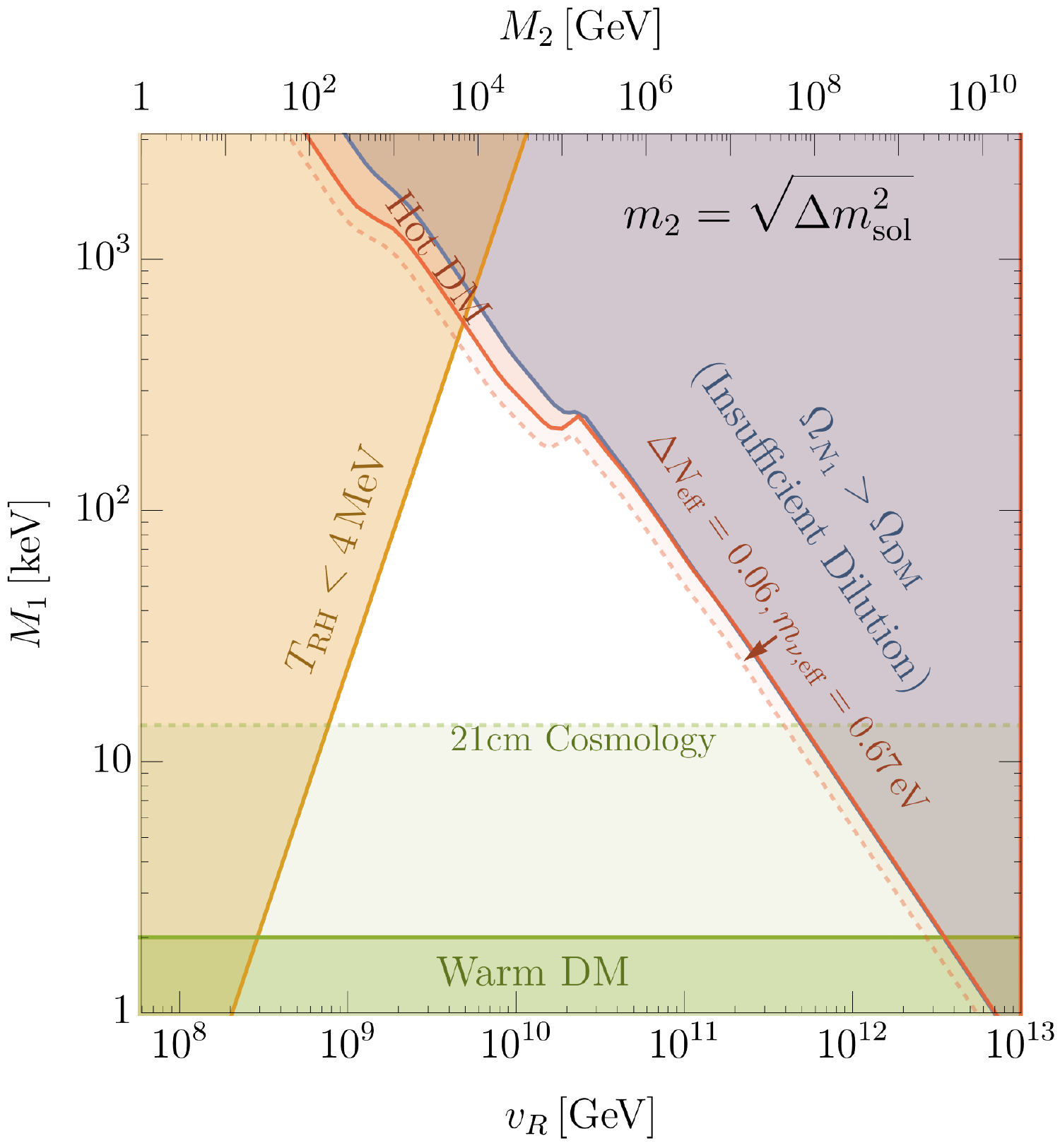} 
    \end{minipage}
    \caption{The parameter space of $N_1$ DM produced by relativistic freeze-out and dilution from $N_2$ decay in terms of the Left-Right symmetry breaking scale, $ v _R $, and the mass of $ N _1 $, $ M _{ 1 } $. We show constraints from $ N _2 $ decaying after Big Bang Nucleosynthesis ({\color{c2} \bf orange}), decaying too early to provide sufficient $ N _1 $ dilution ({\color{c1} \bf blue}), warm DM bounds ({\color{c3} \bf green}), and hot DM bounds ({\color{c4} \bf red}). In addition we show prospects of improved searches for hot DM from CMB telescopes (dashed {\color{c4} \bf red}), and warm DM from $21$-cm cosmology (dashed {\color{c3} \bf green}). 
    We fix the $ \nu _2 $ mass with the atmospheric neutrino mass difference, $m_2 = \sqrt{ \smash[b]{ \Delta m_{\rm atm}  ^2}}$, {\bf left}, and the solar neutrino mass difference, $m_2 = \sqrt{ \smash[b]{ \Delta m_{\rm sol}  ^2}}$, {\bf  right}.}
    \label{fig:thermal_FO}
\end{figure}

In Fig.~\ref{fig:thermal_FO}, we show the constraints on $(v_R,M_1)$ when $m_2 = \sqrt{ \smash[b]{ \Delta m_{\rm atm}  ^2}} $ $\,$({\bf left}) and $m_2 = \sqrt{\smash[b]{ \Delta m_{\rm sol}  ^2}}$ $\,$({\bf right}). 
In the orange shaded region, the required $T_{\rm RH}$ is below 4 MeV, which is excluded by hadronic decays of $N_2$ during BBN~\cite{Kawasaki:1999na,Kawasaki:2000en}. 
The green-shaded region is excluded due to the warmness of $N_1$ affecting large scale structure~\cite{Narayanan:2000tp,Irsic:2017ixq,Yeche:2017upn,Seljak:2006qw}. The light green-shaded region shows the sensitivity of future observations of 21cm lines~\cite{Munoz:2019hjh}. In the blue-shaded region, $ N _2 $ decays too quickly through $W_R$ exchange to efficiently dilute the $ N _1 $ energy density. 
The non-trivial shape of the blue-shaded region is due to the $T_{\rm RH}$ dependent effective degrees of freedom.

The blue line itself is an interesting region of parameter space, which does not require any tuning but simply corresponds to the limit where the dominant decay is set entirely by $ W _R $ exchange. In this limit, the $ N _1 $ abundance has two contributions:~from $ N _2 $ decay through $ N _2 \rightarrow  N _1 \ell^+ \ell^- $ as well as the prior thermal abundance from relativistic decoupling. The former contribution makes up $10\%$ of DM and is hot. The red-shaded region is excluded by the effect of the hot component on the CMB and  structure formation, as set by current limits of $\Delta N_{\rm eff}$ and $m_{\nu, \rm{eff}}$~\cite{Feng:2017nss}. The low $v_R$ part of the blue line is already excluded, and high $v_R$ is in tension. CMB Stage IV experiments~\cite{Abazajian:2016yjj,Abazajian:2019eic} can cover the light red-shaded region and probe the limit where $N_2$ dominantly decays via the $W_R$ exchange.

In sum, as can be seen from Fig.~\ref{fig:thermal_FO}, the allowed region of $N_1$ DM from freeze-out in LR theories forms a bounded triangle in the $v_R-M_1$ plane.

\subsection{Freeze-in}
\label{sec:freezeInDM}

\begin{figure}[h]
	\begin{center}\includegraphics[width=0.8\linewidth]{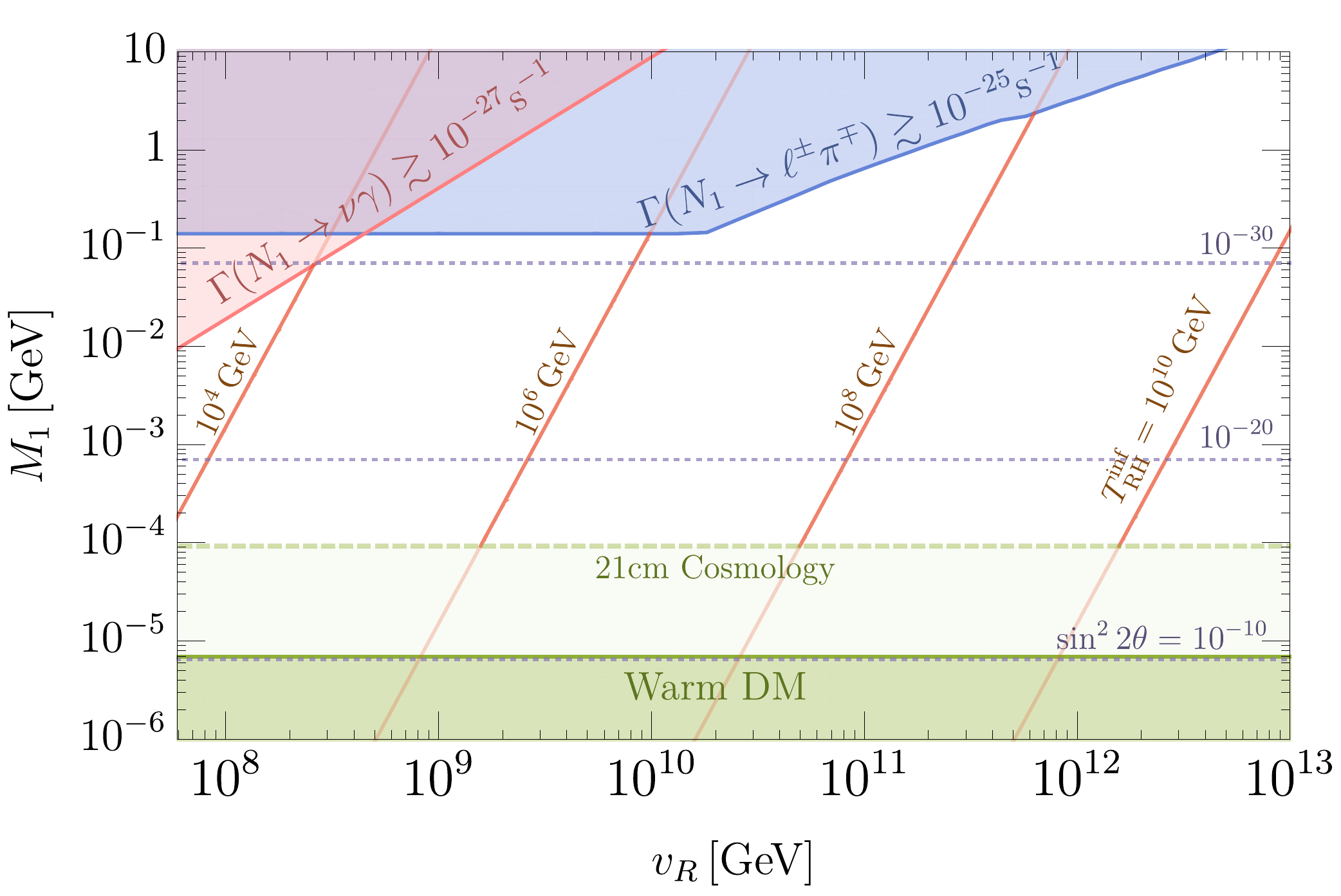}\end{center}
	\caption{The parameter space for $N_1$ DM produced by freeze-in. The observed relic abundance occurs in the unshaded region for values of $T^{\rm inf}_{\rm{RH}}$ shown by the {\color{c4} \bf red} contours. Constraints from small scale structure are shown in {\color{c3} \bf green}, with projections from future probes of small scale structure using the 21cm line in dashed {\color{c3} \bf green}.  In the {\color{c1} \bf blue} region $N_1$ decays too rapidly via $W_R$ to $\ell^\pm \pi^\mp$ and in the {\color{pink} \bf pink} region $N_1$ decays too rapidly via $W_R-W_L$ mixing to $\nu \gamma$. The horizontal dashed {\color{c1} \bf  blue} lines show the limit (\ref{eq:yi1}) on the mixing angle of $N_1$ with active neutrinos.}
	\label{fig:thermal_FI}	
\end{figure}

When the reheat temperature of the universe is below the thermalization temperature of the right-handed neutrinos (see~\eqref{eq:eqtemp}), neither $N_1$ nor $ N _2 $ has a thermal abundance. Instead, the $ N _1 $ abundance is determined by scattering via heavy $W_R$ and $Z_R$ exchange, which, being UV-dominated, depends on the reheating temperature after inflation,
\begin{align}
\frac{\rho_{N_1}}{s} & \simeq 1 \times 10 ^{ - 5 } \left( \frac{M _1 \left(T^{\rm inf} _{ {\rm RH}}\right) ^3 M_{\rm pl}  }{ v _R ^4 } \right)\,,\\ 
\Rightarrow \;\;\;\;  \frac{ \Omega }{ \Omega _{ {\rm DM}}} & \simeq \left(\frac{M_1}{150 \, \KEV}\right)\left(\frac{10^{10} ~\GEV}{v_R}\right)^4  \left(\frac{T^{\rm inf}_{\rm RH}}{10^{7} ~\GEV}\right)^3\,.
\label{eq:FIAbundance}
\end{align}
The production of sterile neutrino DM by $B-L$ gauge boson exchange is considered in~\cite{Khalil:2008kp}.
Freeze-in production from other sources, such as $ \ell H \rightarrow N _1  $, are subdominant since $ y _{i1} \ll 1 $ is needed to ensure that $ N _1 $ is long-lived. 
$N_1$ may be also produced from beta decays of $N_2$ and $N_3$. These contributions, however, are always subdominant to the direct freeze-in production of $N_1$, whether $N_{2,3}$ are produced by the $W_R$ interaction or the $\ell N H$ interaction. 

The contours of Fig.~\ref{fig:thermal_FI} show the reheat temperature after inflation for $N_1$ DM to arise from freeze-in, in the $(v_R,M_1)$ plane. In the green region, the warmness of $N_1$ affects large scale structure. Since $N_1$ from freeze-in are not diluted, they are warmer than $N_1$ from freeze-out and dilution, for a fixed $M_1$. More concretely, the free-streaming length is larger by a factor of approximately $(4/3.2)(Y_{\rm therm} M_1 s/ \rho_{\rm DM})^{1/3}$, giving a commensurately stronger warm DM bound compared to Fig.~\ref{fig:thermal_FO}. Here, the factor of $4/3.2$ comes from the difference in $\langle p/T \rangle$ between the non-thermal freeze-in and the thermal freeze-out distributions, as discussed in~\cite{Heeck:2017xbu}. In the blue and pink regions, the decay of $N_1$ mediated by $W_R$ or $W_R-W_L$-mixing overproduces the observed amount of galactic gamma-rays, respectively~\cite{Essig:2013goa}. Similarly, the decay of $N_1$ via active-sterile mixing overproduces the observed galactic x-rays and gamma-rays for the mixing angle $\sin^2 2\theta_1$ labeling the purple dotted contours. Unlike the $W_R$-mediated decay, which is fixed by $v_R$, the decay via $N_1-\nu$ mixing is set by the free parameters $y_{i1}$.

Fig.~\ref{fig:thermal_FI} shows that the parameter space for $N_1$ DM from freeze-in is weakly constrained compared to that of $N_1$ DM from freeze-out and dilution, shown in Fig.~\ref{fig:thermal_FO}. For example, $v_R$ could be as low as about 100 TeV, with the reheat temperature after inflation below $100 \, \GEV$. Likewise, bounds on $M_1$ are weak; although, as $M_1$ increases, $\sin^2 2\theta_1$ is constrained to become extremely small to keep $N_1$ sufficiently long-lived. In the next section we find that, if leptogenesis via $N_2$ decay is incorporated into the $N_1$ DM freeze-in cosmology, the $(M_1,v_R)$ parameter space becomes more tightly constrained.

\section{Leptogenesis from heavy right-handed neutrino decay}
\label{sec:lepto}

In both the freeze-out and freeze-in cosmologies, where $N_1$ makes up DM, the decays of $N_2$ can produce a baryon asymmetry through leptogenesis. Producing a large enough lepton asymmetry requires $N_3$ to have a sizable Yukawa coupling $y_{33}$ or $y_{23}$; $y_{13}= y_{31}^*$ is small due to the longevity of $N_1$. $N_3$ is therefore short-lived.  

The lepton asymmetry yield from $N_2$ decay is
\begin{align}
Y_L & \;=\;  \epsilon \eta Y_{\rm therm} B
\label{eq:YL}
\end{align}
where $\epsilon$ is the asymmetry created per $N_2$ decay into $\ell H_L$ or $\ell^\dag H_L^\dag$, $\eta$ is the efficiency factor, and $B \equiv  {\rm Br}(N_2 \rightarrow \ell  H_L)  + {\rm Br}(N_2 \rightarrow \ell ^\dagger H_L^\dagger$). In the next two sub-sections we discuss the abundance of $N_2$, which differs in the two cosmologies, and the quantities $\epsilon$ and $\eta$.

\subsection{The baryon asymmetry in freeze-out and freeze-in cosmologies}
\label{sec:leptoN2}
\begin{figure}[]
    \centering
    \begin{minipage}{0.5\textwidth}
        \centering
        \includegraphics[width=1\textwidth]{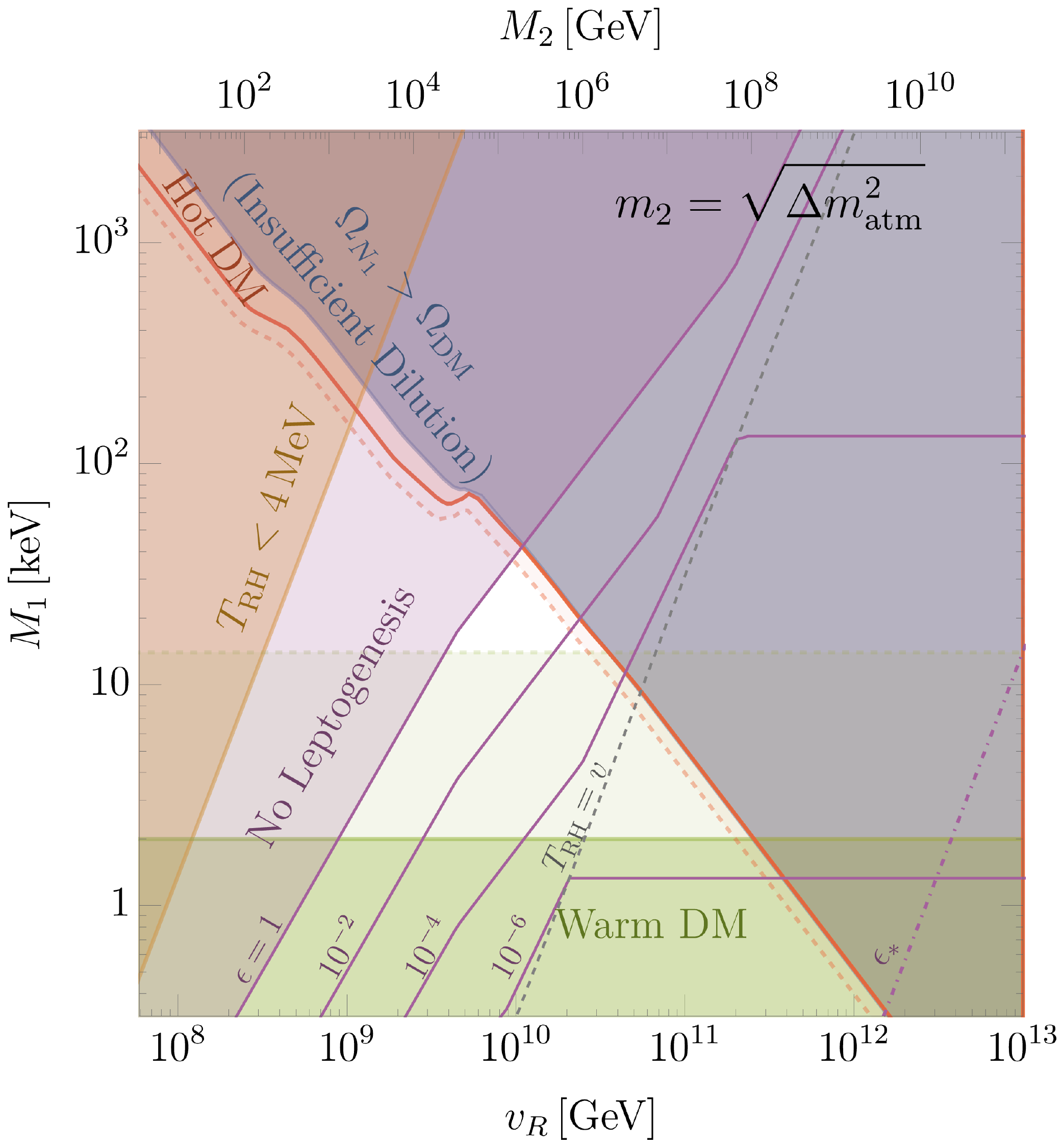} 
    \end{minipage}\hfill
    \begin{minipage}{0.5\textwidth}
        \centering
        \includegraphics[width=1\textwidth]{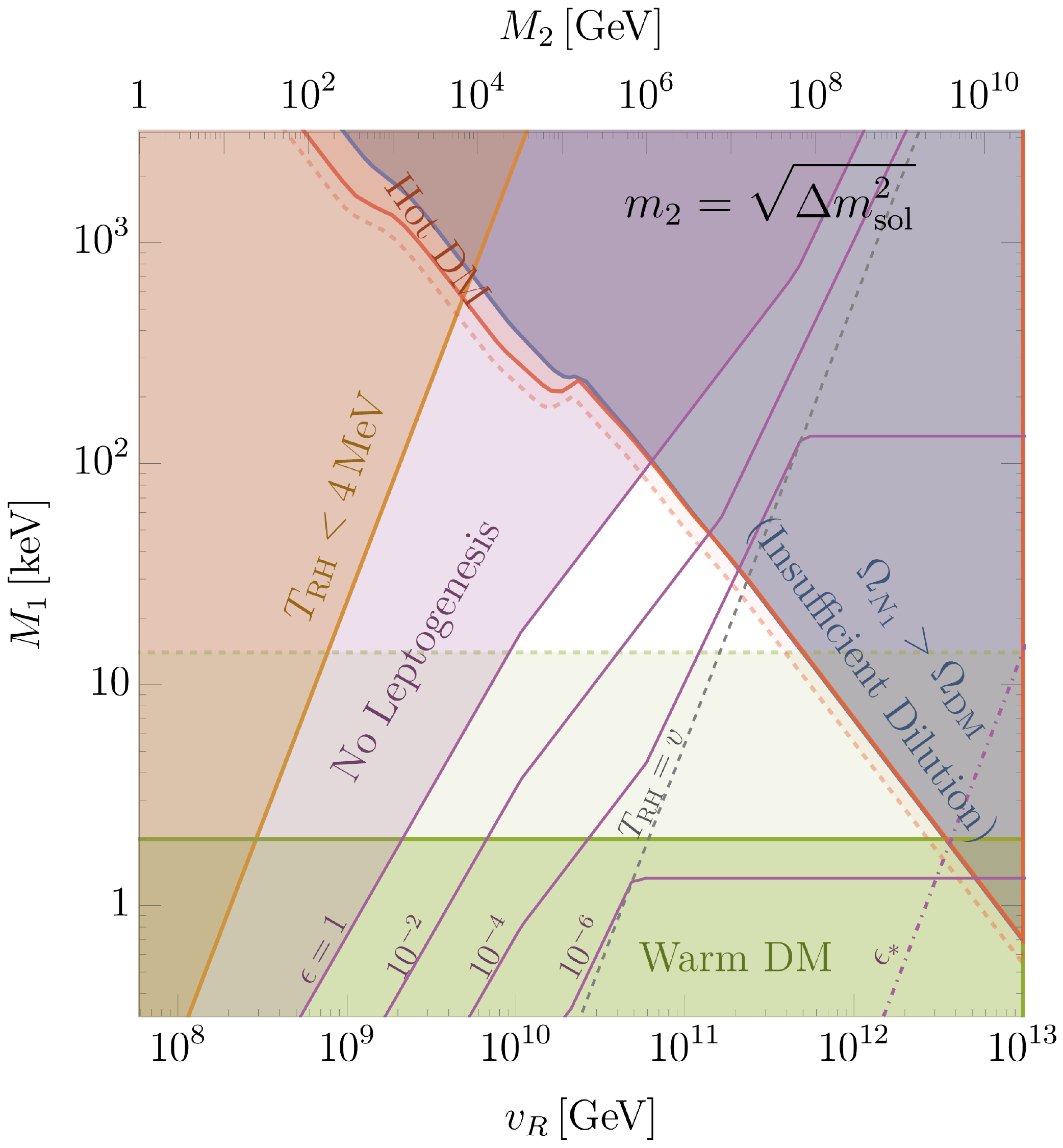} 
    \end{minipage}
    \caption{{\color{purple} \bf Purple} contours of the asymmetry parameter, $\epsilon$, required to produce the observed baryon asymmetry, $Y_B \simeq 8 \times 10^{-11}$ in the freeze-out cosmology.  Larger values of $\epsilon$ are required as  $M_1$ increases due to the greater dilution necessary to realize $N_1$ dark matter. Likewise, larger values of $\epsilon$ are required at low $v_R$ when $T_{\rm RH}$ is below the weak scale, as indicated by the dashed {\color{gray} \bf gray} line. In this regime, the baryon asymmetry is generated only by $N_2$ that decay at temperatures above the weak scale, where electroweak sphalerons are operative. To the left of the dot-dashed {\color{purple} \bf purple}  contour, the baryon asymmetry can only be realized when $\epsilon$ is greater than its natural maximum, $\epsilon_*$.}
    \label{fig:epsilonFO}
\end{figure}
When the reheat temperatures after inflation, $T_{\rm RH}^{\rm inf}$, is high,  $N_1$ DM is produced by freeze-out and subsequent dilution from $ N _2 $ decay. Although the initial $N_2$ abundance is thermal, the efficiency $\eta$ is reduced by the dilution produced from $N_2$.  Also, if the reheat temperature after the $N_2$ MD-era, $T_{\rm RH}$, is below the weak scale, the baryon asymmetry is reduced because only the lepton number produced above the weak scale is converted to baryons by sphaleron processes. The $N_2$ decays yield a baryon asymmetry 
\begin{align}
Y_B &= \frac{28}{79}  \epsilon \; \left( \frac{3T_{\rm RH}}{4M_2} \right)f \, B  
= \frac{28}{79}  \epsilon \; \left( \frac{\rho_{{\rm DM}}/s}{M_1} \right)f \,B, \, \hspace{1cm} & \text{(Freeze-Out + Dilution)} 
\label{eq:YBFO}
\end{align}
where the factor of $28/79$ accounts for the conversion of the lepton asymmetry into the baryon asymmetry via sphaleron processes~\cite{Harvey:1990qw}. $f$ is the fraction of decays that occur when the temperature of the universe is above the weak scale where sphalerons convert the lepton asymmetry into a baryon asymmetry. The fraction depends on whether the temperature of the universe falls below $v$ during a radiation-dominated or $N_2$ matter-dominated era:
\begin{align}
	f = \Gamma_{N_2}t(T = v) \simeq
	\label{eq:fractionDecay}
	\begin{dcases}
	(T_{\rm RH}/v)^2 						\hspace{1cm} & T_{\rm MD} < v \\
	(T_{\rm RH}/v)^2 (v/T_{\rm MD})^{1/2} 	\hspace{1cm} & T_{\rm NA} < v <T_{\rm MD} \\
	(T_{\rm RH}/v)^4						\hspace{1cm} & T_{\rm RH} < v <T_{\rm NA} \\
	1 										\hspace{1cm} & v < T_{\rm RH}.
	\end{dcases}
\end{align}
Here, $T_{\rm MD} = \frac{4}{3} M_2 Y_{\rm therm}$ is the temperature at the start of the adiabatic matter-dominated era, and $T_{\rm NA} = (T_{\rm MD} T_{\rm RH}^4)^{1/5}$ is the temperature at the start of the non-adiabatic matter-dominated era~\cite{Kolb:1990vq,Co:2015pka}. Fig.~\ref{fig:epsilonFO} shows contours of $\epsilon$ required to produce the observed baryon asymmetry, $Y_B \simeq 8 \times 10^{-11}$, in the $(v_R, M_1)$ plane. The contours zig-zag through the plane due to the era-dependent change in $f$, according to Eq.~\eqref{eq:fractionDecay}. For large $v_R$, the reheat temperature is high and $N_2$ always decays before the electroweak phase transition so that $f = 1$ and the required $\epsilon$ depends solely on $M_1$. As $T_{\rm RH}$ drops below $v$, as indicated by the dashed gray line, $f$ falls below unity and $\epsilon$ is suppressed. 

In addition, there is no efficiency lost due to cancellations between the lepton asymmetry generated during production with the lepton asymmetry generated during decay, since the production of $N_2$ through $W_R$ exchange does not generate any lepton asymmetry. Since $y_{i2}$ are small, the wash-out effect is negligible. Finally, we use the DM abundance from (\ref{eq:DMabund}) to obtain the final result. 

Conversely, in the limit when the reheat temperature after inflation, $T_{\rm RH}^{\rm inf}$, is low, $ N _{ 1,2} $ abundances are frozen-in and the resultant baryon asymmetry is 
\begin{align}
Y_B &= \frac{28}{79}  \epsilon \eta Y_{\rm therm} B.  \hspace{2cm} \text{(Freeze-In)}
\label{eq:FIYB}
\end{align}
Note that without a thermal abundance, the freeze-in yield of $N_2$ is too low to induce a matter-dominated era, so that no entropy is produced when $N_2$ decays; this accounts for the difference between Eq.~\eqref{eq:FIYB} and \eqref{eq:YBFO}. The efficiency factor, $\eta$, of $N_2$ is \cite{Giudice:2003jh}
\begin{align}
\label{eq:etaY2FI}
\eta Y_{\rm therm} \simeq 
\begin{dcases}
	Y_{W_R} + 0.03 \left(\frac{\widetilde{m}_2}{10^{-4} \,\EV} \right)Y_{\rm therm} & :\quad \widetilde{m}_2 <10^{-3}~{\rm eV}~\text{(Weak Washout)}\\
	0.03 \, Y_{\rm therm} \left(\frac{\widetilde{m}_2 }{10^{-2} \,\EV} \right)^{-1.16} & :\quad \widetilde{m}_2  >10^{-3}~{\rm eV}~ \text{(Strong Washout)}\\
\end{dcases}
\end{align}
where
\begin{align}
	\widetilde{m}_2 \equiv \sum_i |y_{i2}|^2 v^2 /M_2.
	\label{eq:m2tilde}
\end{align}
In the weak washout regime, when $\widetilde{m}_2 < 10^{-3} \, \EV$, $N_2$ decays out-of-equilibrium. $Y_{W_R}$ is the freeze-in yield of $N_2$ from $W_R$ exchange, where we have set $\eta \simeq 1$ again for this production mechanism. Since the freeze-in abundance of $N_1$ and $N_2$ via $W_R$ exchange is identical, $Y_{W_R}$ is simply
\begin{align}
	Y_{W_R} = \frac{\rho_{N_1}/s}{M_1} = \frac{\rho_{\rm DM}/s}{M_1}.
\end{align}
In the strong washout regime, where $Y_2$ reaches $Y_{\rm therm}$ by the Yukawa coupling $y_{i2}$, $N_2$ is in thermal equilibrium when $T\sim M_2$, and the lepton asymmetry is washed-out until the Yukawa interection is out-of-equilibrium, strongly reducing the efficiency of leptogenesis. The maximum possible $\eta Y_{\rm therm}$ for freeze-in is about $0.1 Y_{\rm therm}$, which occurs when $\widetilde{m}_2  \simeq 10^{-3} ~\EV$ at the transition between the weak and strong washout regimes \cite{Giudice:2003jh}. 
The leptogenesis CP asymmetry parameter, defined by the difference between the branching ratio of $N_2$ into a lepton and an anti-lepton~\cite{Covi:1996wh}, is given in the limit that $ y _{i1} \ll 1 $ by
\begin{align}
\epsilon & \;=\;  \frac{(y_{33} + y_{22})^2}{8\pi} \; \frac { {\rm Im} ( y_{23}^2 ) }{ y_{22}^2 + |y_{32}|^2 } \; g(x) \; = \; \frac{ (y_{33} + y_{22})^2}{8\pi} \; g(x) \; \sin^2\alpha \sin 2 \beta,  \hspace{0.4in} x= \frac{M_3^2}{M_2^2}.
\label{eq:asymParameter}
\end{align}
Since the Higgs Parity solution to the strong CP problem requires $y_{ij}$ to be Hermitian, the heavy $2 \times 2$ space contains a single phase $y_{23} \equiv  |y_{23}| e^{i \beta}$. Furthermore, we introduce an angle $\alpha$ defined by $|y_{23}|/y_{22} \equiv  \tan \alpha$. The function $g(x)$ is \cite{Fukugita:1986hr,Davidson:2008bu}
\begin{align}
g(x) &\equiv \sqrt{x}\left(\frac{1}{1-x} + 1 - (1+x) \log\left(\frac{1}{x}+1\right) \right),
\label{eq:g(x)}
\end{align}
and is much less than unity when $M_3$ and $M_2$ are disparate, near unity when $M_3$ and $M_2$ are comparable, and much greater than unity as $M_3$ and $M_2$ become degenerate.

It is possible to choose $y_{33}, M_3/M_2, \alpha$ and $\beta$ to achieve a sufficiently large asymmetry per decay (\ref{eq:asymParameter}) for successful $N_2$ leptogenesis in both the freeze-out and freeze-in $N_1$ DM cosmologies. For freeze-out, the baryon asymmetry generated by $N_2$, \eqref{eq:YBFO}, can match the observed baryon asymmetry $Y_B \simeq 8 \times 10^{-11}$ everywhere in the unshaded region of Fig.~\ref{fig:epsilonFO}. At larger values of $v_R$, $\epsilon \sim 10^{-5}$ is sufficient. However, at lower $v_R$ as $T_{\rm RH}$ drops below the weak scale, larger values are needed, as shown by the purple contours, as only the fraction of $N_2$ decaying above $v$ result in baryogenesis. 
At the lowest values of $v_R$ that give $N_1$ dark matter, an insufficient baryon asymmetry is generated even if $y_{33}$ becomes non-perturbative and $\epsilon =1$, as shown by the shaded purple region of Fig.~\ref{fig:epsilonFO}.
In the case of freeze-in cosmology there is no dilution, so that the baryon asymmetry of \eqref{eq:FIYB} can successfully yield the observed asymmetry everywhere in Fig.~\ref{fig:thermal_FI}, except in the region not shown at very low $v_R$ where $T_{\rm RH}^{\rm inf} \ll v$.

\subsection{Enhancing the lepton asymmetry parameter}
\label{sec:asymPerDecay}

For comparable $M_2$ and $M_3$, $g(x) \sim 1$, and for large angles $\alpha,\beta \sim 1$, the asymmetry parameter is of order $ (y_{33} + y_{22})^2/8 \pi$. For the freeze-out cosmology, $y_{22}$ is negligible, while for the freeze-in cosmology $y_{22}$ is subject to the similar constraints as $y_{33}$. We thus focus on $y_{33}$ in this subsection. The coupling $y_{33}$ determines the size of the seesaw contribution to the $\nu_3$ mass via 
\begin{align}
	m_{33} &= m_3^{(5)} - m_{33}^{(ss)} = \frac{v^2}{v_R^2} M_3 - \frac{y_{33}^2 v^2}{M_3} - \underbrace{\frac{y_{32}^2 v^2}{M_2}}_{< 10^{-3} \, \EV }.
	\label{eq:m3}
\end{align}
In the freeze-out cosmology the last term is negligible due to the long lifetime of $N_2$. Moreover, $m_{33}$ is aligned with the neutrino mass eigenstate $m_3$ \cite{Dror:2020jzy}. In the freeze-in cosmology, we assume that the last term is less than $10^{-3} \, \EV$, since otherwise $Y_B$, \eqref{eq:FIYB}, is strongly suppressed from strong washout effects.%
\footnote{If $y_{23}^2 v^2 /M_2$ is taken much greater than $\mathcal{O}(0.1 \,\EV)$, it is possible that $y_{33}^2$ commensurately grows to ensure $m_{33}$ remains $\mathcal{O}(0.1 \,\EV)$. Although this appears to enhance $\epsilon$ by increasing $y_{33}^2$, the strong washout reduces $Y_B$ by a slightly higher power, so the net effect is a decrease in $Y_B$. We avoid this route.}
Unlike the freeze-out cosmology, $m_{33}$ is not necessarily $m_3$, but $\mathcal{O}(0-0.1 \, \EV)$, since $m_{23}$ may be non-negligible.

Avoiding a finely tuned cancellation between the two terms, $y_{33}$ is maximized when the two terms are comparable, giving $y_{33} \sim m_{33} v_R/v^2$. This leads to a maximal natural value for the asymmetry parameter
\begin{align}
	\epsilon_*  & \; \equiv  \frac{m_3^2 \, v_R^2}{8 \pi \, v^4} \sim 10^{-11} \left( \frac{v_R}{10^{10} \, \GEV} \right)^2  \left( \frac{m_{33}
	}{ 0.05 \, \EV} \right)^2. 
	\label{eq:eps*}
\end{align}
Using this value for $\epsilon$, the baryon asymmetry in the freeze-out plus dilution cosmology (\ref{eq:YBFO}) is too small, except for the very highest values of $v_R \sim 3 \times 10^{12}$ GeV as shown by the dot-dashed contour labeled $\epsilon_*$ in Fig.~\ref{fig:epsilonFO}. 
Hence, except for a very small region near $v_R \sim 3 \times 10^{12}$ GeV, simultaneous $N_1$ dark matter and $N_2$ leptogenesis requires an enhancement of $\epsilon$ above $\epsilon_*$.  By comparing (\ref{eq:eps*}) with the contours of required $\epsilon$ in Fig.~\ref{fig:epsilonFO}, it is apparent that the enhancement must be very significant at lower values of $v_R$.
A similar conclusion applies to leptogenesis with freeze-in dark matter, (\ref{eq:FIYB}).

There are two possibilities for this enhancement. One is to take $y_{33} \gg m_{33} v_R/v^2$ by having $m_3^{(5)}, m_{33}^{(ss)} \gg |m_{33}|$ so that a cancellation between the two terms of (\ref{eq:m3}) occurs. Alternatively, $ g ( x ) $ may be large when $M_2$ and $M_3$ are nearly degenerate (i.e.~$x \simeq 1)$.
It is useful to introduce
\begin{align}
	\chi & \equiv \frac{m_3^{(5)} - m_{33}^{(ss)}}{m_{3}^{(5)}+m_{33}^{(ss)}} = \frac{m_{33}}{m_{3}^{(5)}+m_{33}^{(ss)}}\,. 
    \label{eq:chi}
\end{align}
As $\chi$ goes to zero, the fine-tuning between the dimension-five and see-saw masses increases since each becomes larger than $m_{33}$ and hence increasingly degenerate so as to keep their difference equal to $m_{33}$. That is, as $\chi \rightarrow 0$, $M_3$ grows (so that $m_3^{(5)}$ increases) and $y_{33}^2$ grows (even faster than $M_3$, so that $m_{33}^{(ss)}$ increases) in the following manner: 
\begin{align}
	M_3 &= m_{33}\frac{v_R^2}{v^2} \frac{1+\chi}{2\chi} \\
	y_{33}^2 &= m_{33}^2 \frac{v_R^2}{v^4} \frac{(1+\chi)(1-\chi)}{4\chi^2} \label{eq:y33chi}.
\end{align}
Note that $-1 < \chi < 1$ and that the sign of $m_{33}$ is the same as the sign of $\chi$. For the freeze-out cosmology, $m_{22}=m_2 =m_2^{(5)}$ is always positive.
In terms of $\chi$ and $\epsilon_*$ of (\ref{eq:eps*}), the lepton asymmetry parameter can be written as
\begin{align}
	\epsilon &= 
	\epsilon_* \; \frac{(1+\chi)(1-\chi)}{4\chi^2} \; g(x)  \sin^2\alpha \sin 2 \beta .
\label{eq:epsilon}
\end{align}
The observed baryon asymmetry can be explained by the enhancement from small $\chi$ and/or $x \simeq 1$.

We focus on the freeze-out cosmology for the rest of this subsection and identify $m_{22}$ and $m_{33}$ with $m_2$ and $m_3$, respectively.
Combining (\ref{eq:YBFO}), (\ref{eq:eps*}), and (\ref{eq:epsilon}), the baryon asymmetry produced by $N_2$ decays is
\begin{align}
\frac{Y_B}{8 \times 10^{-11}} &= 10^{-5} \; \frac{2 \, \KEV}{M_1} \left( \frac{v_R}{10^{10} \, \GEV} \right)^2  \left( \frac{m_{3}
	}{ 0.05 \, \EV} \right)^2  \frac{(1+\chi)(1-\chi)}{4\chi^2} \; g(x) \, f B \sin^2\alpha \sin 2\beta .
\label{eq:YBFOfinal}
\end{align}

Since $m_2$ is dominated by the dimension 5 contribution to its mass,
\begin{equation} 
x = \frac{ M _3 ^2  }{ M _2 ^2  } = \frac{m_3^2}{m_2^2}\frac{(1+\chi)^2}{4 \chi^2}.
\label{eq:xchi}
\end{equation} 
This is an important result since it shows that $x$ and $\chi$ are not independent; they are related by the neutrino spectrum. The two choices for enhancing $\epsilon$, $x$ near unity and small $\chi$, are seen to be mutually exclusive: if $\chi \ll 0.1$ then $x\gg 1$ for any realistic neutrino spectrum. Thus $N_1$ freeze-out dark matter and leptogenesis from $N_2$ decay requires either $x$ near unity {\it or} small $\chi$.

For the case of $x$ very close to unity, $\chi$ is fixed from (\ref{eq:xchi}), giving
\begin{equation} 
\chi \; \simeq \; \left( \frac{-1}{1+2\sqrt{r}} , \;\; -\frac{1}{1\pm2/\sqrt{r}}; \;\;
1- r,  \;\; -\frac{1}{3} (1 \pm \frac{r}{3}) \right),
\hspace{0.5in} r = \frac{\Delta m_{\rm sol}^2 }{\Delta m_{\rm atm}^2}
\label{eq:enhancementx2}
\end{equation} 
where the first two cases are for a normal hierarchy, with $|m_3| > m_2$ and $|m_3| < m_2$, respectively, while the last two cases are for the inverse hierarchy with $m_3$ positive (and $m_2 > m_3$), and negative. These give values for the enhancement factor of 
\begin{equation} 
\frac{(1+\chi)(1-\chi)}{4\chi^2} \; g(x) \; \simeq \; \left(0.20, \; 30; \;
0.015, 1.96 \right) \frac{1}{1-x}.
\label{eq:enhancementx}
\end{equation} 
We see that the inverse hierarchy requires $g(x)$ to be larger than in the normal hierarchy.
Using this result, for the normal hierarchy with  $|m_3| < m_2$, we find the observed baryon asymmetry results for
\begin{align}
	x-1 \simeq 2 \,\frac{|M_2 - M_3|}{M_{2,3}} \simeq 1\times 10^{-5}  \left(\frac{2 \, \KEV}{M_1} \right) 
	\left( \frac{v_R}{10^{10} \, \GEV} \right)^2  \left( \frac{m_3
	}{ 0.01 \, \EV} \right)^2  f B \sin^2 \alpha \sin 2\beta.
\end{align}

For the case of a cancellation of large contributions to the neutrino mass $m_3$, with $\chi$ very small, we find that (\ref{eq:xchi}) gives $g(x) \sim 3 \chi (m_2/m_3) \ll 1$, so that the observed baryon asymmetry requires
\begin{align}
	\chi \simeq 0.75 \times 10^{-5}  \left(\frac{2 \, \KEV}{M_1} \right) 
	\left( \frac{v_R}{10^{10} \, \GEV} \right)^2  \left( \frac{m_3 m_2
	}{ (0.05 \, \EV)^2} \right)  f B \sin^2 \alpha \sin 2\beta.
\end{align}

We conclude that $N_1$ DM from freeze-out and leptogenesis from $N_2$ decay can occur simultaneously throughout the large unshaded region of Fig.~\ref{fig:epsilonFO}. 
Enhancements in $\epsilon$ are required and can arise in two ways: near degeneracy of $M_{2,3}$ or large $y_{33}$ with $m_3$ resulting from a cancellation between seesaw and dimension 5 contributions.
In the next section we study whether leptogenesis can be obtained naturally, considering both the origin in the enhancement for $\epsilon$ and the effects of radiative corrections from $y_{33}$ on the $N_1$ lifetime.

\subsection{Restriction on neutrino masses in freeze-in cosmology}
\label{sec:numass_FI}
In the freeze-in cosmology without leptogenesis, discussed in Sec.~\ref{sec:freezeInDM}, $y_{i2}$ is not necessarily small since $N_2$ need not be long-lived. Consequently, $m_{22}$ may possess a substantial contribution from $m_{22}^{(ss)}$, spoiling the direct relationship between $M_2$ and $v_R$ of Eq.~\eqref{eq:M2m2} required for the freeze-out cosmology. However, requiring efficient leptogenesis in the freeze-in $N_1$ DM cosmology puts restrictions on the neutrino mass matrix.

To avoid the strong wash-out and maximize the allowed parameter space, the see-saw contribution from $N_2$ is required to be negligible. Then the SM neutrino masses are determined by the see-saw contribution from $N_3$, $m_2^{(5)}$, and $m_3^{(5)}$.

The enhancement of the asymmetry requires $M_3 \gtrsim M_2$ for the following reasons. For enhancement by degeneracy, $M_3 = M_2$. For enhancement by tuning in $m_{33}$, if $M_2 > M_3$, $m_{2}^{(5)}$ must be also cancelled by $m_{22}^{(ss)}$ from $N_3$, giving $y_{33}^2 \simeq  M_3^2 / v_R^2$ and $y_{23}^2 \simeq M_2 M_3 / v_R^2$. However, $m_{23}^{(ss)}\simeq y_{23} y_{33} v^2 / M_3 \simeq \sqrt{M_2 M_3 }v^2 / v_{R}^2$ becomes much larger than the observed SM neutrino masses.

Since $M_3 \gtrsim M_2$, the see-saw contribution from $N_3$ to $m_{22}$, $y_{23}^2 v^2 / M_3$ is also negligible. We obtain a relation similar to Eq.~\eqref{eq:M2m2},
\begin{align}
\label{eq:M2m22}
    M_{2} \simeq m_{22} \left(\frac{v_R}{v}\right)^2.
\end{align}
Moreover, $m_2^{(5)}$ must be as large as the observed neutrino masses. Suppose that it is negligible. To obtain the two observed non-zero neutrino mass eigenvalues, $m_{23}$ must be non-negligible. Since $y_{23}$ is required to be small to avoid strong wash-out, $y_{33}$ must compensate it. Then $m_{33}^{(ss)}$ is large, requiring the cancellation with $m_3^{(5)}$ and hence $y_{33}^2 \simeq  M_3^2 / v_R^2$. However,
\begin{align}
    m_{23}^{(ss)} \simeq \frac{y_{23} y_{33}v^2}{M_3} \simeq y_{23} \frac{v^2}{v_R^2} < \frac{(0.001{\rm eV})^{1/2} M_2^{1/2} v}{v_R} < (0.001{\rm eV})^{1/2} (0.1{\rm eV})^{1/2} =0.01{\rm eV},
\end{align}
which is not large enough to explain the SM neutrino masses.
We conclude that $m_{22}$ in Eq.~\eqref{eq:M2m22} must be $0.01-0.05$ eV.

\section{Naturalness and radiative corrections in the effective field theory}
\label{sec:natural_EFT}

For $N_1$ to be dark matter, whether in the context of (SM+N) or of Left-Right symmetry, small parameters must be introduced to limit its mass and decay rate,
$M_1/M_{2,3}, \, y_{i1} \ll 1$. For sufficient cosmological stability, (\ref{eq:yi1}) can be approximated by
\begin{align}
y_{i1} \; \lesssim \; 3 \times 10^{-13} \left( \frac{3 \, \KEV}{M_1}\right)^{3/2}.
\label{eq:y1iapprox}
\end{align}
The value of $M_1/M_{2,3}$ is model-dependent. In LR Higgs Parity, taking the examples of (\ref{eq:M2m2}) or (\ref{eq:y33chi}) with $|\chi|$ not tuned to be small,   
\begin{align}
\frac{M_1}{M_{2,3}} \; \simeq \; (10^{-12} - 10^{-13}) \left( \frac{M_1}{3 \, \KEV}\right) \left( \frac{10^{11} \GEV}{v_R}\right)^2.
\label{eq:M1/M3approx}
\end{align}
Quite generally, light sterile neutrino dark matter has a small numbers problem.

In (SM+N), with the $N$ interactions of (\ref{eq:SMnu}), the smallness of $y_{i1}$ and $M_1$ can result from an approximate global symmetry under which only $N_1$ transforms. However, since freeze-in production of $N_1$ via $y_{i1}$ violates (\ref{eq:y1iapprox}), the only available production mechanism is via neutrino oscillations, and this also violates (\ref{eq:y1iapprox}) unless it is enhanced by a very high lepton asymmetry \cite{Shi:1998km}.

In LR symmetric theories, $N_1$ may be produced by the $SU(2)_R\times U(1)_{B-L}$ gauge interactions. However, the smallness of the coupling $y_{i1}$ seems to be hard to understand. We need a hierarchy $y_{i1} \ll y_{jk}^e$, despite the right-handed neutrinos and the right-handed charged leptons coming from the same $SU(2)_R$ doublets $\bar{\ell}$. A similar problem arises from the hierarchies $y_{i1} \ll y_{i2}, y_{i3}$ and $M_1 \ll M_{2,3}$. The observed large neutrino mixing angles imply no large symmetry distinction between the $\ell_i$, and the LR symmetry then implies there are none between the $\bar{\ell_i}$. Then no symmetry can distinguish $y_{i1}$ from $y_{i2}, y_{i3}$, nor $M_1$ from $M_{2,3}$.

While one can simply choose $y_{i1}$ and $M_1$ to be small, in this and the next section we seek an explanation for their suppression. At the tree-level, it is possible to obtain the desired hierarchies of parameters by breaking $U(3)_q\times U(3)_{\bar{q}} \times U(3)_{\ell} \times U(3)_{\bar{\ell}} \times U(1)_{H_L}\times U(1)_{H_R}$ by appropriate symmetry breaking fields. However, because of the absence of symmetry protection mentioned above, quantum corrections may destabilize the hierarchies.

To make a comparison, we first examine the conventional LR symmetric theory with an $SU(2)_L\times SU(2)_R$ bi-fundamental and point out the difficulty in guaranteeing the stability of $N_1$. We then argue why the problem can be avoided in Left-Right Higgs Parity, deferring the presentation of a UV completion to the next section. We show that the lepton sector of (\ref{eq:yukawa}) and (\ref{eq:yukawaNu}) has a naturalness problem if the cut-off scale of those interactions are far above $v_R$: in certain regions of parameter space, radiative contributions to $y_{i1}$ and $M_1$ violate (\ref{eq:y1iapprox}) and (\ref{eq:M1/M3approx}).  This gives significant naturalness constraints on $N_1$ dark matter and on leptogenesis from $N_2$ decay. The UV completion discussed in the next section will also solve this problem.

\subsection{Conventional LR symmetric theories}
In the conventional LR symmetric theories, the SM Higgs is embedded into an $SU(2)_L\times SU(2)_R$ bi-fundamental scalar $\Phi$, which can be decomposed under $SU(2)_L\times U(1)_Y$ as
\begin{align}
    \Phi = \left( H_u, H_d\right),~~H_u:(2,\frac{1}{2}),~~H_d:(2,-\frac{1}{2}).
\end{align}
In order for $N_1$ to be stable, the SM Higgs must almost exclusively come from only one of $H_u$ or $H_d$. In fact, the charged lepton Yukawa coupling arises from
\begin{align}
    {\cal L}= y^e_{ij} \ell_i \Phi \bar{\ell}_j = y^e_{ij} \ell_i H_d \bar{e}_j + y^e_{ij} \ell_i H_u N_j
\end{align}
with the SM Higgs $H$ containing $H_d$.
In the basis where the $N_i$ mass matrix is diagonal, $y^e_{i1}$ is as large as $y_{\tau} \sim 10^{-2}$. To satisfy (\ref{eq:y1iapprox}), the fraction of $H_u$ in the SM Higgs must be very small. This can be achieved by coupling $\Phi \Phi^\dag$ to an $SU(2)_R$ triplet that spontaneously breaks $SU(2)_R$, thereby splitting the masses of $H_u$ and $H_d$. Also, the operators $\Phi^2$ and $\ell \Phi^\dag \bar{\ell}$ must be suppressed, since the former introduces $H_u-H_d$ mixing and the latter introduces the Yukawa coupling of $N$ to $\ell H_d^\dag$. This can be achieved by a non-zero charge of $\Phi$ under some symmetry.

We must also introduce up and down quark Yukawa couplings,
\begin{align}
    {\cal L}= y^u q \Phi^\dag \bar{q} + y^d q \Phi \bar{q}. 
\end{align}
These terms necessarily break the aforementioned symmetry of $\Phi$. The dominant effect comes from the quantum correction to the mass of $\Phi$ from the quark loop,
\begin{align}
    \Delta {\cal L} \; \sim \; \frac{y^{t*} y^{b}}{16\pi^2} \Lambda^2 \; \Phi^2  + {\rm h.c.} \; \sim \; 10^{-4} \Lambda^2 \; \Phi^2 + {\rm h.c.},
\end{align}
where $\Lambda$ is the cut-off of the theory. This introduces $H_u-H_d$ mixing and the Yukawa coupling of $N$,
\begin{align}
    {\cal L} = y_{ij} \ell_i H N_j,~~y_{ij}\sim  10^{-4} \frac{\Lambda^2}{m_{H_u}^2} y^e_{ij} \sim 10^{-6} \frac{\Lambda^2}{m_{H_d}^2} > 10^{-6},
\end{align}
violating the bound (\ref{eq:y1iapprox}).

This problem can be avoided by using different $\Phi$s for quark and lepton Yukawa couplings and/or introducing supersymmetry, but we do not pursue this direction further.

\subsection{Left-right Higgs Parity}
\label{sec:radcor}

\begin{figure}
	\center
	\includegraphics[width=0.6\linewidth]{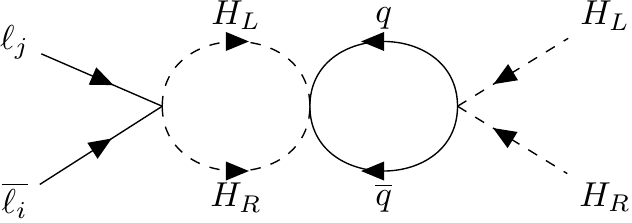}
	\caption{Radiative corrections to $y_{i1}$ from charged leptons and quarks in the EFT.  Loop momenta near the quark EFT cutoff scale, $\Lambda_c$, lead to \eqref{eq:y1radQuarks}.}
	\label{fig:eltonJohn}	
\end{figure}

The coupling $y_{ij}$ receives quantum correction also in Left-Right Higgs Parity.
The quantum correction from the quark and charged lepton Yukawa couplings is given by the Feynman diagram in Fig.~\ref{fig:eltonJohn}. We estimate this radiative correction to $y_{i1}$ to be
\begin{align}
	\Delta y_{i1} &\sim \frac{1}{(16\pi^2)^2} \; 3y_{t}y_{b}y_{\tau} \, U_{\tau I_i} U^*_{\tau I_1}  \,\left(\frac{\Lambda_c}{v_R}\right)^2 \simeq  10^{-9}\left(\frac{\Lambda_c}{v_R}\right)^2,
	\label{eq:y1radQuarks}
\end{align}
where the PMNS matrix $U$ appears in the charged current $\bar{e} \, U \gamma^\mu \nu$, and $I_{i}$ is the standard PDG numbering for the $LR$ partner of $N_{i}$. In the following we take $U_{\tau I} \sim 0.5$.
This correction is quadratically divergent, for loop momenta above $v_R$ up to $\Lambda_c$, the cutoff of the effective theory with the dimension-five operators for the charged fermion masses of \eqref{eq:yukawa}.
The stability of $N_1$, \eqref{eq:yi1}, requires $y_{i1} \lesssim 10^{-13}$ for any $M_1$, which is violated for $\Lambda_c > v_R$.
The dimension-five operators may be, however, UV-completed by introduction of particles with masses below $v_R$. In the next section, we present such a setup and show that the quantum correction to $y_{i1}$ can be suppressed.

\begin{figure}
	\center
	\includegraphics[width=0.6\linewidth]{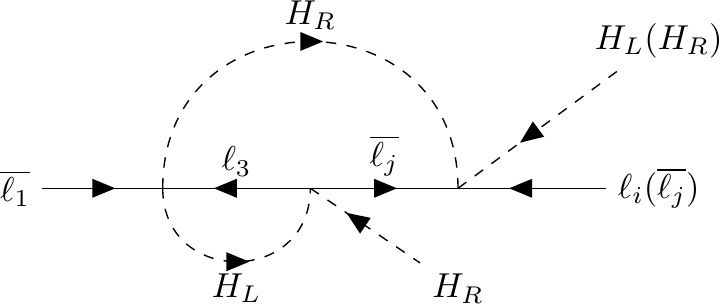}
	\caption{Radiative corrections to $y_{i1}$ and $M_{1j}$ (parenthesis) in the EFT.  Loop momenta near the EFT cutoff scale  lead to \eqref{eq:y1rad} and \eqref{eq:M1jrad}.}
	\label{fig:octopus}	
\end{figure}

Successful leptogenesis from $N_2$ decay requires $y_{33}$ to be sufficiently large. Since the flavor symmetry that distinguishes $N_1$ from $N_{2,3}$ is broken by the charged lepton Yukawa couplings, quantum corrections involving $y_{33}$ and the charged lepton Yukawas generate non-zero $y_{i1}$. Similarly, $M_1$ should also receive quantum corrections from $M_{2,3}$ and charged lepton Yukawa couplings.

The Feynman diagrams for quantum corrections to $y_{i1}$ and $M_{1j}$ from the lepton sector are shown in Fig.~\ref{fig:octopus}. Two further diagrams involve the same vertices with different connections of the Higgs lines. They are quadratically divergent for loop momenta above $v_R$ up to $\Lambda$, the cutoff of the effective field theory described by the Lagrangian \eqref{eq:yukawaNu} and the third term of \eqref{eq:yukawa}. 
We estimate this radiative correction to $y_{i1}$ to be
\begin{align}
\label{eq:y1rad}
\Delta y_{i1} \sim  \frac{1}{(16\pi^2)^2} \sum_{j=2,3} y_{ij} \, y_{\tau}^2 \; U_{\tau I_j} U_{\tau I_1}^*  \left(\frac{\Lambda}{v_R}\right)^2.
\end{align}
Requiring this radiative correction to $y_{i1}$ not exceed the limit of (\ref{eq:y1iapprox}) from the radiative decay of $N_1$ bounds $y_{ij}$ ($i,j = 2,3$),
\begin{align}
	y_{ij} \; \lesssim \; y_{\rm max} &= \frac{M_1 \sin 2\theta_{1 \rm exp}}{v} \; \frac{(16\pi^2)^2}{0.25 \, y_{\tau}^2} \; \left(\frac{v_R}{\Lambda}\right)^2 
		\;	\lesssim \; 10^{-5} \left(\frac{3 \, \KEV}{M_1}\right)^{3/2} \left(\frac{10}{\Lambda/v_R}\right)^2,
	\label{eq:ysqMax}
\end{align}
where we used $U_{\tau I_1}^* U_{\tau I_j} \sim 0.25$
and assumed no cancellation in (\ref{eq:y1rad}) between $j=2$ and $j=3$ contributions.
For $N_1$ dark matter, whether by freeze-out or freeze-in, $y_{ij}$ may be chosen small enough to satisfy this bound.  However, leptogenesis requires a significant $y_{33}$ and we discuss this below.

Similarly, diagrams such as the one in  Fig.~\ref{fig:octopus} lead to radiative corrections to the $\bar{\ell}_1 \bar{\ell}_j H_R H_R$ operator
\begin{align}
\label{eq:M1jrad}
\Delta M_{j1} \sim  \frac{1}{(16\pi^2)^2} \, M_j \, y_{\tau}^2 \; U_{\tau I_j} U_{\tau I_1}^*   \left(\frac{\Lambda}{v_R}\right)^2.
\end{align}
Diagonalizing the $N$ mass matrix leads to a radiative correction to $M_1$ from $M_{2,3}$
\begin{align}
\label{eq:M1rad}
\Delta M_1 \sim  \frac{1}{(16\pi^2)^4} \;   M_{2,3} (0.25 \, y_{\tau}^2)^2         \left(\frac{\Lambda}{v_R}\right)^4.
\end{align}
For this not to exceed the value of $M_1/M_{2,3}$ given in
(\ref{eq:M1/M3approx}) requires
\begin{align}
\label{eq:M1bound}
M_1 \; \gtrsim \; 3 \, \KEV  
\left(\frac{v_R}{10^{12} \, \GEV}\right)^2
\left(\frac{\Lambda/v_R}{10}\right)^4,
\end{align}
where we assumed no cancellation between $j=2,3$ contributions.
Thus, for $N_1$ dark matter, a cutoff $\Lambda = 10 \, v_R$ just allows the entire triangular regions of Fig.~\ref{fig:thermal_FO} for the freeze-out cosmology but limits very large $v_R$ in Fig.~\ref{fig:thermal_FI} for the freeze-in cosmology.

\begin{figure}[!]
    \centering
    \begin{minipage}{0.49\textwidth}
        \centering
        \includegraphics[width=1\textwidth]{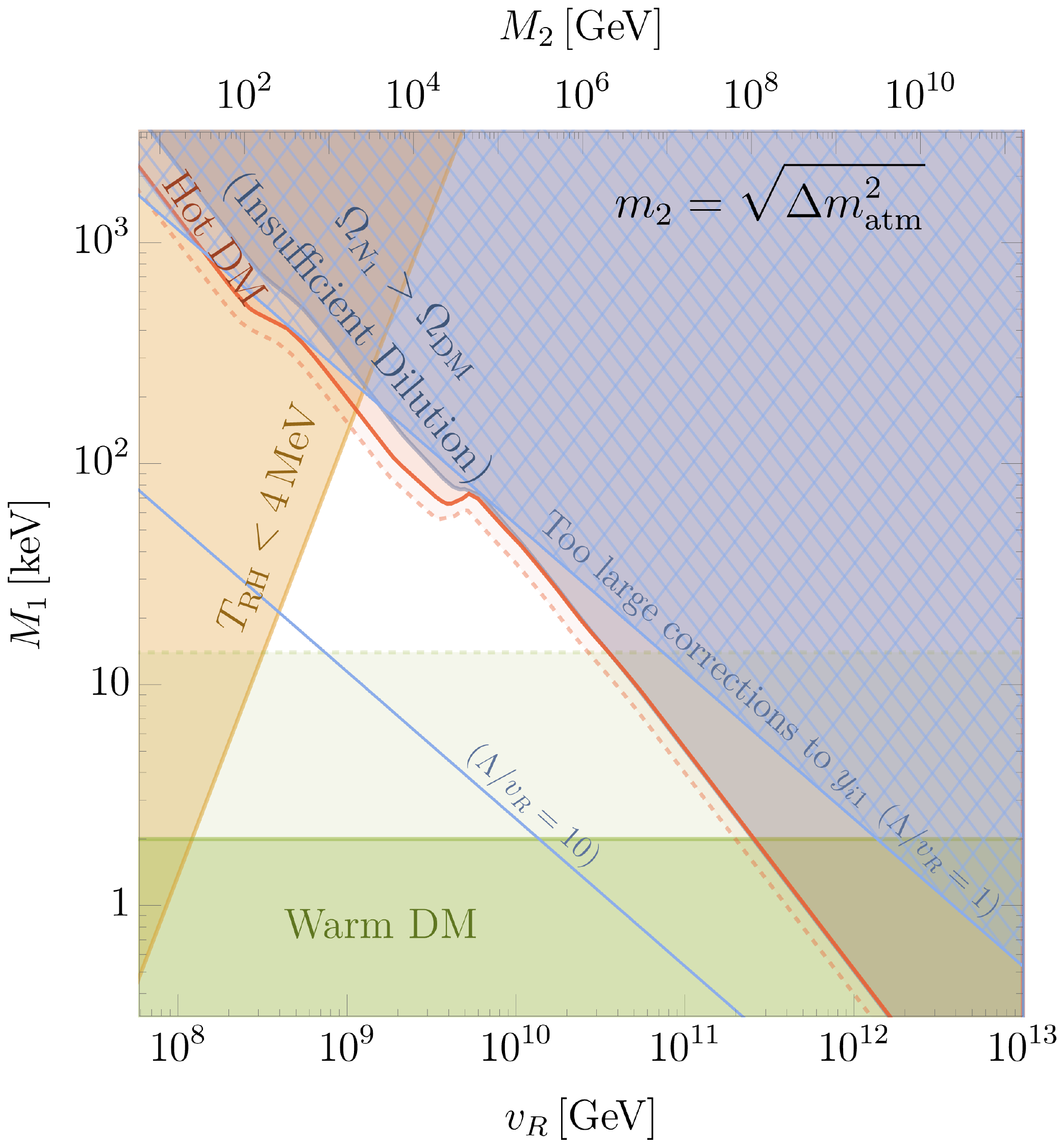}
    \end{minipage}
    \begin{minipage}{0.49\textwidth}
        \centering
        \includegraphics[width=1\textwidth]{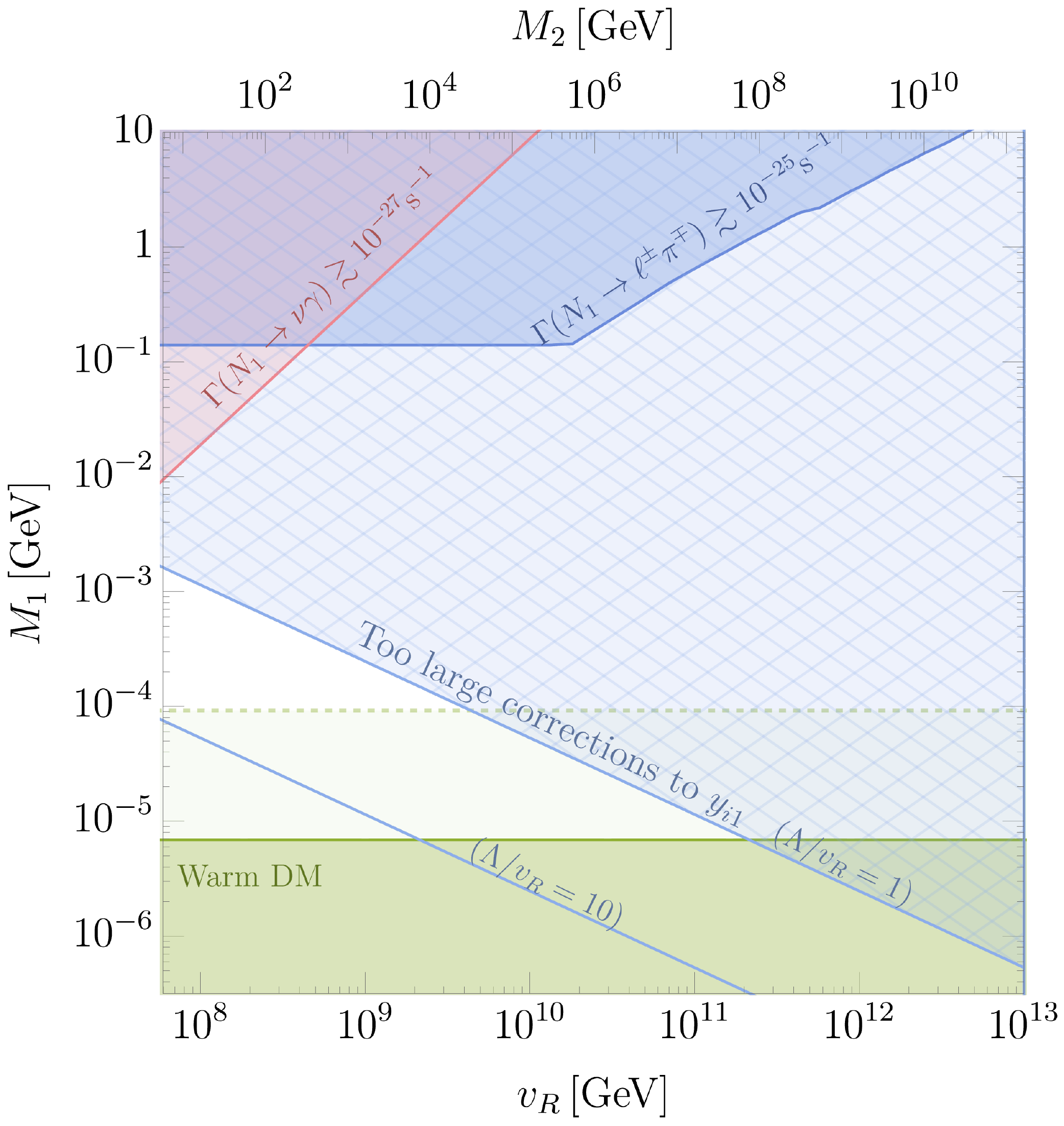}
    \end{minipage}
    \caption{The parameter space where the mass and stability of $N_1$ DM can be realized without fine tuning in the effective theory  $\ell \ell H_L H_L + \bar{\ell} \bar{\ell} H_R H_R + \ell \bar{\ell} H_L H_R$. The charged fermion masses are UV completed below $v_R$ to avoid the radiative correction of Fig.~\ref{fig:eltonJohn}.  In the hatched {\color{lightBlue} \bf blue} region, the value of $y_{33}$ required to set $x \equiv (M_3/M_2)^2 \simeq 1$ for leptogenesis, approximately $m_{33} v_R/v^2$, is sufficiently large that the tree and loop contributions to $y_{i1}$ must be unnaturally tuned to keep $N_1$ stable when $\Lambda/v_R = 1$. $\Lambda$ is the UV cutoff.  The lower blue contour shows the same region if $\Lambda/v_R = 10$.
    The unhatched shaded regions are constraints solely on $N_1$ DM in the freeze-out (left) and freeze-in (right) cosmologies, as in Figs. \ref{fig:thermal_FO} and \ref{fig:thermal_FI}. }
   \label{fig:y33Nat}
\end{figure}

The quadratically divergent correction to $y_{i1}$ (\ref{eq:y1rad}) places a naturalness constraint on $y_{33}$ and therefore, via (\ref{eq:asymParameter}), on leptogenesis
\begin{align}
	\epsilon \; \lesssim \; 3 \times 10^{-12} \left(\frac{3 \, \KEV}{M_1}\right)^3 \left(\frac{10}{\Lambda/v_R}\right)^4  g(x) \, \sin^2 \alpha \, \sin 2 \beta.
	\label{eq:epsnatlimit}
\end{align}
This is far below the required values of $\epsilon$ shown in Fig.~\ref{fig:epsilonFO} for freeze-out dark matter and given in (\ref{eq:FIYB}) for freeze-in cosmology, unless $g(x) \gg 1$.\footnote{We will discuss a natural origin for $g(x) \gg 1$ in Sec.~\ref{sec:natlepto}.}  
This requires $x$ near unity and, from (\ref{eq:y33chi}) and (\ref{eq:xchi}), $y_{33} \sim m_3 v_R/v^2$.  Requiring this value of $y_{33}$ to satisfy the bound of (\ref{eq:ysqMax}) leads to the naturalness constraint
\begin{align}
	\left(\frac{M_1}{3 \, \KEV}\right)^{3/2} 
	\left(\frac{v_R}{10^{10} \, \GEV}\right)
	\; \lesssim \;   \left(\frac{10}{\Lambda/v_R}\right)^2 
\end{align}
shown by blue lines in Fig.~\ref{fig:y33Nat}.  Thus, in the EFT the quadratic divergence of $y_{i1}$ greatly limits the range of $(M_1,v_R)$ that naturally allows successful leptogenesis.

In the next section we give a UV completion of the lepton and quark sector. This is important for two reasons:  first it provides an understanding for why $N_1$ is very light and long-lived, and second it allows a very large reduction in the radiative corrections for $y_{i1}$ and $M_1$, reopening large regions of the $(M_1,v_R)$ plane to natural leptogenesis.

\section{A UV Completion yielding a light, long-lived $N_1$}
\label{sec:model}

As we have seen in the previous section, to naturally protect the stability of $N_1$ against quantum corrections, the UV completion of the dimension-5 operators~(\ref{eq:yukawa}) and ~(\ref{eq:yukawaNu}) should occur at a mass scale below $v_R$ for the correction from Fig.~\ref{fig:eltonJohn}, and at the most, not far above $v_R$ for the correction from Fig.~\ref{fig:octopus}. In this section, we present a UV completion and show that the quantum corrections can be sufficiently suppressed.

\subsection{The UV completion: tree-level}
The operators $\ell \ell H_L H_L$, $\bar{\ell} \bar{\ell} H_R H_R$ and $\ell\bar{\ell} H_L H_R$ can be obtained by introducing singlet fields $S_a$ and $\bar{S}_a$ with the following couplings and masses,
\begin{align}
{\cal L} = \lambda_{i a} \ell_i \bar{S}_a H_L +  \bar{\lambda}_{i a} \bar{\ell}_i S_a H_R +  \frac{1}{2} M_{\bar{S},a} \bar{S}_a \bar{S}_a + \frac{1}{2} M_{S,a} S_a S_a + { M_{S\bar{S},ab}}S_a \bar{S}_b + {\rm h.c.}, \nonumber \\
\bar{\lambda}_{i a} = {\lambda}_{i a}^*,~ M_{\bar{S},a} = M_{S,a},~M_{S\bar{S},ab}^* = M_{S\bar{S},ba},
\label{eq:Smodel}
\end{align}
and integrating out $S$ and $\bar{S}$. With three pairs of $S$ and $\bar{S}$, the neutrino sector has $U(3)_\ell \times U(3)_{\bar{\ell}}\times U(3)_S \times U(3)_{\bar{S}}\times U(1)_{H_L}\times U(1)_{H_R}$ flavor symmetry. Hierarchical breaking of the symmetry can explain the hierarchy $y_{i1} \ll y_{i2}, y_{i3}$ and $M_1 \ll M_{2,3}$. We assume flavor symmetry breaking such that among three pairs of $S$ and $\bar{S}$, only two pairs have significant coupling $\lambda$ and/or small masses $M_S$; we may instead start from the theory where only two pairs of $S$ and $\bar{S}$ are present. This suppresses the quantum correction to $y_{i1}$ and $M_1$ for the following reason. Although the vertex corrections to $\lambda$ from the tau Yukawa may couple $\bar{\ell}_1$ to $S$, one linear combination of $\bar{\ell}_i$ does not couple to $\bar{S}$. We may redefine the linear combination as $\bar{\ell}_1$, which is light. The operator $\ell\bar{\ell} H_L H_R$ is obtained from the mass term $M_{S\bar{S}}S\bar{S}$. This gives rise to Yukawa couplings between the massive linear combinations of $\ell_i$ and of $\bar{\ell}_i$, but the massless combinations, which do not couple to $S$ and $\bar{S}$, do not obtain Yukawa couplings.

If there are (effectively) only two pairs of $S$ and $\bar{S}$, the $U(3)_\ell\times U(3)_{\bar{\ell}}$ symmetry may be anarchically broken in the neutrino sector. This model explains why $N_1$ is much lighter and has a smaller Yukawa coupling than $N_{2,3}$. However, to show that $N_1$ is sufficiently light and stable, we must study higher-dimensional operators from the cutoff scale of the theory $M_{\rm cut}$, e.g.~the Planck scale (and, in the next subsection, from radiative corrections).
If the $U(3)_{\ell} \times U(3)_{\bar{\ell}}$ symmetry is anarchically broken, the following higher-dimensional operators are allowed:
\begin{align}
\label{eq:UV correction}
{\cal L} \; \sim \; \frac{\bar{\lambda}^2 M_S^*}{M_{\rm cut}^2} \; \bar{\ell}\bar{\ell} H_R H_R + \frac{\lambda \bar{\lambda} M_{S\bar{S}}^*}{M_{\rm cut}^2} \; \ell\bar{\ell} H_L H_R,
\end{align}
with $\lambda$ and $\bar{\lambda}$ being typical entries in the matrices $\lambda_{ia}$ and $\bar{\lambda}_{ia}$.\footnote{Although $M_S$ is a real parameter, we put the superscript $*$ to clarify the charge structure.}
These operators give $N_1$ a mass and a coupling to $\ell H$ with values
\begin{align}
\label{eq:UV correction result}
\Delta M_1 \simeq & \frac{\lambda^2 M_S v_R^2}{M_{\rm cut}^2}
\simeq \left(\frac{M_S}{M_{\rm cut}}\right)^2M_3 
\simeq {\rm keV} \frac{M_3}{m_\nu (v_R/v)^2} \left( \frac{v_R}{3\times 10^{11} {\rm GeV}} \right)^4 \left( \frac{M_S}{v_R} \right)^2 \left( \frac{\mpl}{M_{\rm cut}} \right)^2, \nonumber \\
\Delta y_{i1} \simeq &  \frac{\lambda^2 M_{S\bar{S}}v_R}{M_{\rm cut}^2} \simeq \frac{M_S^2}{M_{\rm cut}^2} y_{33},
\end{align}
where we take the largest $M_i$ and $y_{ij}$, i.e.~$M_3$ and $y_{33}$. 
It is possible to reduce the size of these corrections by taking $M_S$ smaller than $v_R$, when $H_R = v_R$ in (\ref{eq:Smodel}) and (\ref{eq:UV correction}).  In this case the effective theory below $M_S$ takes the form of Eq.~\eqref{eq:yukawaNu} with $H_R$ replaced with $v_R$. It is clear that (\ref{eq:UV correction result}) can satisfy (\ref{eq:y1iapprox}) and (\ref{eq:M1/M3approx}) for the range of $v_R$ of interest.\footnote{In fact, further suppression results if supersymmetry exists in the UV, since holomorphy of the superpotential can forbid the operators in Eq.~(\ref{eq:UV correction}).} We delay a discussion of the implications of these results as the quantum corrections to $y_{i1}$ are larger than the tree result of (\ref{eq:UV correction result}), unless $v_R > 10^{-4} M_{\rm cut}$. 

In the model without $\bar{S}$, shown in Eq.~(\ref{eq:model_cancel}), $M_{S\bar{S}}$ in Eq.~(\ref{eq:UV correction}) is replaced by $M_S$, but the corrections to $M_1$ and $y_{i1}$ are still given by Eq.~(\ref{eq:UV correction result}).

\subsection{The UV completion: quantum corrections}
\label{sec:appQCModels}

\subsubsection{Corrections from lepton Yukawas}

We first discuss the quantum corrections from $y_{i2}$, $y_{i3}$ and charged lepton Yukawa couplings. All three $\bar{\ell}_{i}$ have Yukawa interactions in Eq.~(\ref{eq:yukawa}), among which the tau Yukawa is the largest. The tau Yukawa necessarily breaks the approximate or accidental symmetry  of (\ref{eq:Smodel}) that discriminates $\bar{\ell}_1$ from $\bar{\ell}_{2,3}$, and gives quantum contributions to $M_1$ and $y_{i1}$.

The quantum corrections depend on the UV model that generates the dimension-5 interactions in Eq.~(\ref{eq:yukawa}). Let us first consider the case where the charged lepton Yukawas arise from the exchange of a heavy scalar $\Phi$ with charge $(1,2,2,0)$,
\footnote{$\Phi$ couples exclusively to leptons, not quarks, so that potential CP violating phases of $\Phi$ do not enter into the quark sector. Consequently, the strong CP problem remains solved when introducing $\Phi$.}
\begin{align}
\label{eq:yukawa_charge}
{\cal L } =    - m_{\Phi}^2 |\Phi|^2 + (x_{ij} \Phi \ell_i \bar{\ell}_j  - A \Phi^\dag H_L^\dag H_R^\dag + {\rm h.c.}).
\end{align}
After integrating out $\Phi$ and inserting the vev of $H_R$, we obtain the Yukawa coupling
\begin{align}
y^e_{ij} = \frac{A v_R}{m_{\Phi}^2} \; x_{ij}.
\end{align}
\begin{figure}
\centering
	 \includegraphics[width=0.48\linewidth]{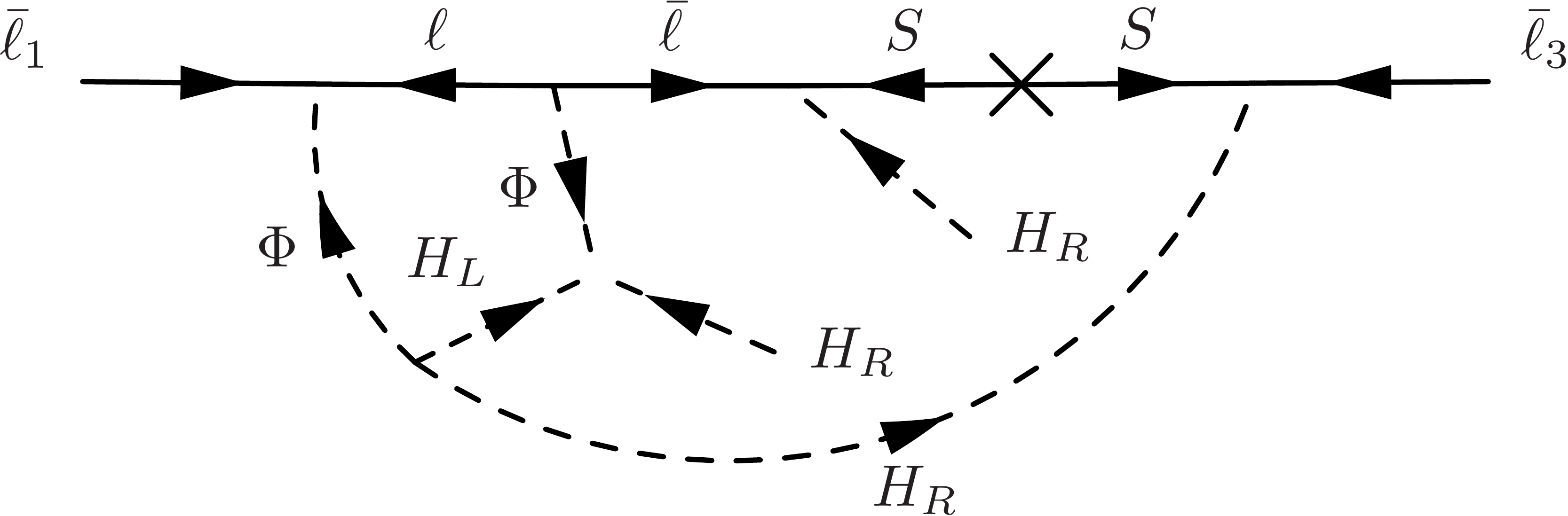} \hfill
	 \includegraphics[width=0.48\linewidth]{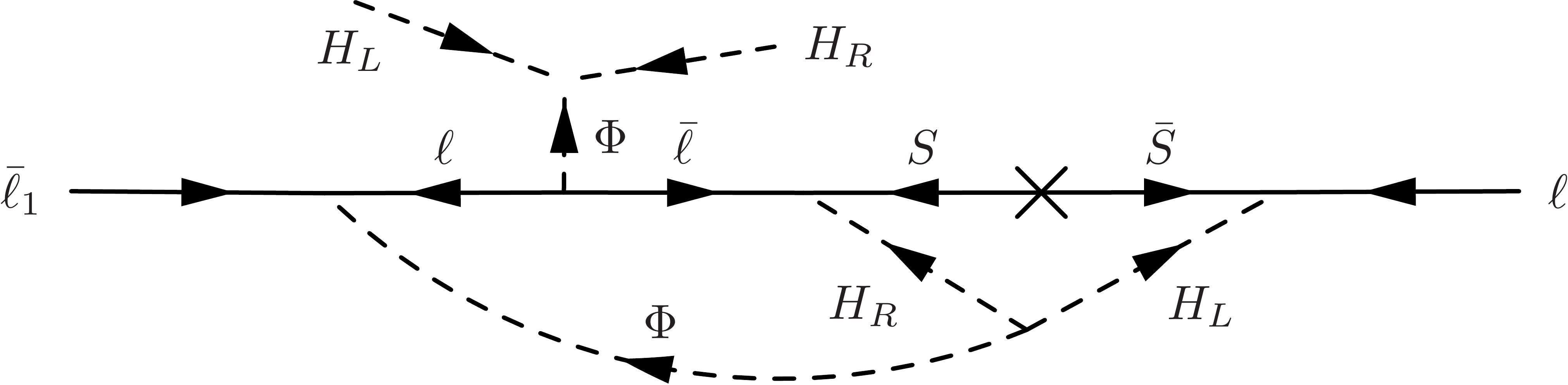}
	\caption{Two-loop diagrams correcting the mass and decay rate of the dark matter, $N_1$, when the neutrino masses are generated by the exchange of a heavy singlet $S$, and the charged lepton masses are generated by the exchange of a heavy scalar, $\Phi$. The diagrams are UV completions to the EFT diagrams of Fig.~\ref{fig:octopus}.}
	\label{fig:correction_mass_coupling}	
\end{figure}

The quantum correction above the scale $M_S$ renormalizes $\lambda$ and $M_S$ but, by the approximate (accidental) symmetry, one linear combination of the $N_i$ still has a small (zero) mass and coupling to $\ell H_L$. Only corrections below the scale $M_S$ can change the mass and decay rate of $N_1$. The two-loop diagram shown in the left panel of Fig.~\ref{fig:correction_mass_coupling} dominantly corrects $M_1$, generating
\begin{align}
{\cal L} \; \simeq& \; \frac{1}{(16\pi^2)^2} \frac{M_{S,b} A^2}{m_{\Phi}^4} x_{1 a } x^*_{3 a} \lambda^*_{3b} \lambda^*_{3b} \;\; \bar{\ell}_1 \bar{\ell}_i H_RH_R \nonumber \\
\; \simeq & \;  \frac{0.25 \, y_\tau^2}{(16\pi^2)^2} M_{3} \; \frac{M_S^2}{v_R^4} \;\; \bar{\ell}_1 \bar{\ell}_i H_RH_R,
\end{align}
where we assume $M_S \ll m_{\Phi}$. In the second equality we use $x_{1a} x_{3a}^* A^2/ m_{\Phi}^4 = (U_{\tau I_1} U_{\tau I_i}y_\tau)$ $(U_{\tau I_3}U_{\tau I_i} y_\tau)^*/ v_R^2$ $\simeq (0.25 \, y_\tau^2/v_R^2)$, and $\lambda^2 /M_S \simeq M_{3}/ v_R^2$.
This term, after $H_R$ obtains a vev, gives a mass mixing between $N_1$ and $N_{3}$ resulting in a correction to the mass of $N_1$
\begin{align}
\label{eq:quadMSM1}
\Delta M_1 \simeq \; \left(\frac{ 0.25 \, y_{\tau}^2}{(16\pi^2)^2} \right)^2  \left(\frac{M_S}{v_R}\right)^4 M_3.
\end{align}
The mass mixing also induces a coupling of $N_1$ to $\ell H$,
\begin{align}
\label{eq:quadMSy1i}
\Delta y_{i1} \; \simeq \;  \left(\frac{ 0.25 \, y_{\tau}^2}{(16\pi^2)^2} \right)  \left(\frac{M_S}{v_R}\right)^2 y_{i3}.
\end{align}
The diagram in the right panel of Fig.~\ref{fig:correction_mass_coupling} also corrects $y_{i1}$ by a similar amount.

We next consider the case where the charged lepton yukawas arise from the exchange of heavy fermions $E$ and $\bar{E}$,
\begin{align}
\label{eq:yukawa_charge2}
 {\cal L} = z_{ia}^e \ell_i \bar{E}_a H^\dag_L + (z^e_{ia})^* \bar{\ell}_i E_a H^\dag_R + M_{E,a} E_a \bar{E}_a.
\end{align}
When $m_E > z^e v_R$, after integrating out $E$ and inserting the vev of $H_R$, we obtain the yukawa coupling
\begin{align}
    y_{ij}^e = z^e_{ia} \frac{v_R}{M_{Ea}}z_{aj}^{e\dag}
\end{align}
When $m_E < z^e v_R$, the SM right-handed charged leptons originate from $\bar{E}$, and the Yukawa coupling is $y^e \simeq z^e$.
\begin{figure}
\centering
	 \includegraphics[width=.5\linewidth]{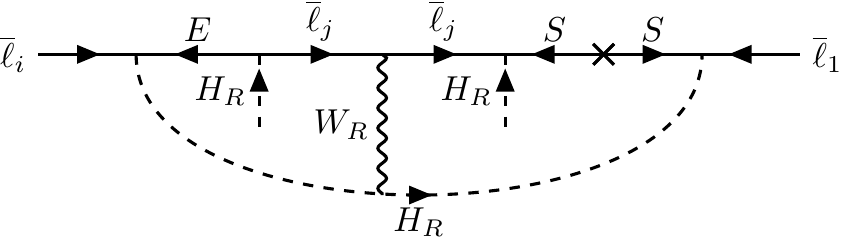}
	 \hfill
	 \includegraphics[width=.4\linewidth]{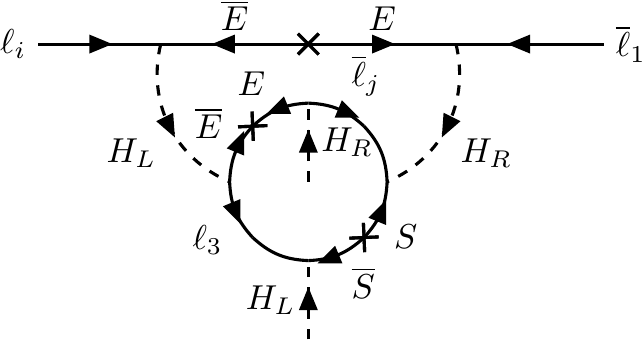}
	\caption{Two-loop diagrams correcting the decay rate and mass of the dark matter, $N_1$,  when the neutrino masses are generated by the exchange of a heavy singlet $S$, and when the charged lepton masses are generated by the exchange of a heavy fermion, $E$. The diagrams are UV completions to the EFT diagrams of Fig.~\ref{fig:octopus}.}
	\label{fig:correction_mass_coupling_Fermion}	
\end{figure}
The two-loop diagram with external $H_R$ and $\bar{\ell_3}$ in the left panel of Fig.~\ref{fig:correction_mass_coupling_Fermion} generates a mass-mixing between $N_3$ and $N_1$,%
\footnote{Without $W_R$ in the diagram, one of external $H_R$ must be charged.}
\begin{align}
 {\cal L} &\simeq \frac{1}{(16\pi^2)^2} \; g^2 z_{1 a } z^*_{3 a} \lambda^*_{3b} \lambda^*_{3b} \frac{M_{S,b}}{{\rm Max}\left\{M_{E,a}^2, m_{H_R}^2 \right\}} \; \bar{\ell}_1\bar{\ell}_3 H_RH_R \\
 & \simeq \frac{1}{\left(16 \pi^2\right)^2} \, \frac{M_S^2}{v_R^4} \, M_3 \;\;   \bar{\ell}_1\bar{\ell}_3 H_RH_R \times 
 \begin{cases}
    \left(0.5 y_{\tau}\right)^2 \left({z_{1a} z_{3a}^*}\right)^{-1} & \quad M_E \gtrsim v_R \\
    \left({z_{1a} z_{3a}^*}\right) & \quad M_E \lesssim v_R. 
 \end{cases}
\end{align}
In the second line, we use $z_{1a} z_{3a}^* =  U_{\tau I_1} U_{\tau I_3}y_\tau M_{E_a}/v_R$ $\simeq 0.25 y_\tau M_{E_a}/v_R$, and $\lambda^2 /M_S \simeq M_{3}/ v_R^2$. This term, after $H_R$ obtains a vev, gives a mass mixing between $N_1$ and $N_{3}$.
For $m_{E}\gtrsim v_R$, the correction is minimized for the largest $z = O(1)$. For $m_{E}\lesssim v_R$, the correction is minimized for the smallest $z \approx y_\tau$. The smallest quantum correction is then
\begin{align}
\label{eq:quadMSM1E}
\Delta M_1 \; \gtrsim \; \left(\frac{ 0.25 \, y_{\tau}^2}{(16\pi^2)^2} \right)^2  \left( \frac{M_S}{v_R}\right)^4 M_{3}.
\end{align}
Similarly, the mass mixing also induces a coupling of $N_1$ to $\ell H$,
\begin{align}
\label{eq:quadMSy1iE}
\Delta y_{i1} \; \gtrsim \; \left(\frac{ 0.25 \, y_{\tau}^2}{(16\pi^2)^2} \right)   \left( \frac{M_S}{ v_R}\right)^2 y_{i3}.
\end{align}
The two-loop diagram in the right panel of Fig.~\ref{fig:correction_mass_coupling_Fermion} with external $H_L$ and ${\ell_i}$ also corrects $y_{i1}$ by a similar amount. We see that Eqs.~\eqref{eq:quadMSM1} and \eqref{eq:quadMSy1i}, from a UV completion with $\Phi$, or Eqs.~\eqref{eq:quadMSM1E} and \eqref{eq:quadMSy1iE}, from a UV completion with $E$, are identical in form to Eqs.~\eqref{eq:M1rad} and \eqref{eq:y1rad} with $\Lambda$ replaced by $M_S$. Thus, with $M_S \ll v_R$ the naturalness of the theory is greatly improved.  When we take $\Lambda/v_R < 1$, $\Lambda$ should be interpreted as $M_S$.

\subsubsection{Corrections from charged fermion Yukawa couplings}
We next consider the quantum corrections from charged fermion Yukawa couplings. We introduce a UV completion for the up and down quark Yukawas by heavy fermions $U,\bar{U}$, and $D, \bar{D}$, with Lagrangian 
\begin{align}
\label{eq:yukawa_UV_ud}
 	{\cal L}_u &= z_{ia}^u q_i \bar{U}_a H_L + (z^u_{ia})^{*} \bar{q}_i U_a H_R + M_{U,a} U_a \bar{U}_a, \nonumber \\
	 {\cal L}_d &= z_{ia}^d q_i \bar{D}_a H^\dag_L + (z^d_{ia})^{*} \bar{q}_i D_a H^\dag_R + M_{D,a} D_a \bar{D}_a.
\end{align}
With $M_{U} > z^{u}v_R$, integrating out $U$ generates the up quark Yukawa couplings
\begin{align}
	    y_{ij}^{u}= z_{ia}^u \, \frac{v_R}{M_{U,a}} \, z_{aj}^{u\dagger}
\end{align}
via a seesaw, and similarly for the down quark Yukawas by integrating out $D$. When $m_U < z^u v_R$, on the other hand, the SM right-handed up quarks dominantly come from $\bar{U}$ rather than $\bar{q}$, so that the light fermion masses are ``flipped" rather than ``seesaw", with the Yukawa coupling $y^u \sim z^u$. In the up, down or charged lepton sectors, if $M > y v_R$ the light mass is seesawed, while it becomes flipped as $M$ drops below $y v_R$.

\begin{figure}
	\centering
	\includegraphics[width=0.41\linewidth]{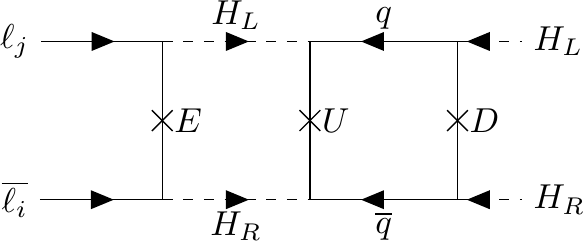}
	\hfill
	\includegraphics[width=0.5\linewidth]{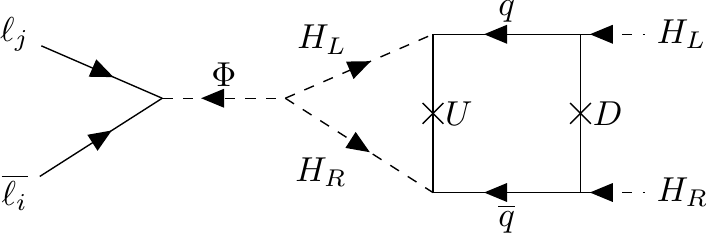}
	\caption{Two-loop diagrams correcting the decay rate of the dark matter, $N_1$, when the charged lepton masses are generated by the exchange of a heavy fermion, $E$ (left), or scalar, $\Phi$ (right), and the up-type quark and down-type quark masses are generated by the exchange of heavy fermions, $U$, $D$, respectively. Each diagram is a UV completion to the EFT diagram of Fig.~\ref{fig:eltonJohn}.}
	\label{fig:eltonJohnUV}	
\end{figure}

When the heavy fermion masses $M_U$, $M_D$, are less than $v_R$, the cutoff scale of the EFT generating the dimension-five quark masses is below $v_R$. As a result, the quadratically divergent radiative corrections to $y_{i1}$ as calculated in Eq.~\eqref{eq:y1radQuarks} and visualized in Fig.~\ref{fig:eltonJohn}, are absent. The radiative corrections to $y_{i1}$ in the UV complete theory are shown by the diagrams in Fig.~\ref{fig:eltonJohnUV}, which generate the operator
\begin{align}
    {\cal L} \simeq \frac{1}{(16\pi^2)^2} \frac{M_U M_D}{v_R m_{H_R}^2}\, (z^u_{k b} z^{u*}_{k b}) (z^d_{l c} z^{d*}_{l c}) \ell_i\bar{\ell}_1 H_L H_R \times \begin{cases} \frac{M_E v_R}{M_*^2} (z^e_{i a} z^{e*}_{1 a}) & : E~\text{exchange} \\
  y^e_{i1} & : \Phi~\text{exchange} 
  \end{cases}, \nonumber \\
  M_* = {\rm max}(M_{U},M_D, M_E, z^u v_R, z^d v_R, z^e v_R),
\end{align}
where we assume $M_{U,D,E} < m_{H_R}$. We consider the correction from the third generation fermions and their LR partners, since the smallest possible corrections are largest for the third generation.   
For $M_{U,D,E} > z^{u,d,e}v_R$, where we may integrate out the heavy fermions to obtain the dimension-5 operators, the quantum correction is bounded by
\begin{align}
    \Delta y_{i1} \gtrsim \frac{1}{(16\pi^2)^2} y_t^3 y_b^3 \times \begin{cases}
     y_\tau^3 \\
     y_\tau 
    \end{cases}
    \simeq
    \begin{cases}
      10^{-18} & :E~\text{exchange} \\
      10^{-14} & :\Phi~\text{exchange}.
    \end{cases},
    \label{eq:UDEcor1}
\end{align}
where we take $M_* \sim v_R$.
The correction is small enough for $M_1 < 10$ MeV/10 keV for $E/\Phi$ exchange.
For $M_{U,D,E} < z^{u,d,e}v_R$, where the SM right-handed fermions are dominantly $\bar{U},\bar{D},\bar{E}$, the quantum correction is bounded by
\begin{align}
\label{eq:UDEcor2}
 \Delta y_{i1} \gtrsim \frac{1}{(16\pi^2)^2}  y_{t}^3 y_{b}^3 \frac{M_U M_D}{y_t y_b v_R^2} \times
 \begin{cases}
  y_\tau^3\frac{M_E}{y_\tau v_R} \\
  y_\tau
 \end{cases},
\end{align}
which is even smaller than (\ref{eq:UDEcor1}).

In summary, these UV completions easily allow small $M_1$ to be natural throughout the allowed regions of Figs.~\ref{fig:thermal_FO} or \ref{fig:thermal_FI} for any $v_R$ consistent with Higgs Parity, $10^9 {\rm GeV} \lesssim v_R \lesssim 10^{13} {\rm GeV}$. The radiative correction of (\ref{eq:quadMSM1}), from the left panel of Fig.~\ref{fig:correction_mass_coupling_Fermion}, easily satisfies (\ref{eq:M1/M3approx}) for $M_S < v_R$. A possible tree-level contribution
from the Planck scale, (\ref{eq:UV correction result}), is natural if $M_S/v_R \lesssim (M_1/ \KEV)^{1/2} (3 \times 10^{11} \GEV/v_R)^2$.

Furthermore, corrections to the $N_1$ decay rate from Fig.~\ref{fig:correction_mass_coupling} or \ref{fig:correction_mass_coupling_Fermion} (Fig.~\ref{fig:eltonJohnUV}) involving lepton (charged fermion) yukawa couplings, can be made small enough in either cosmology by choosing $M_S$ ($M_U,M_D,M_E$) sufficiently less than $v_R$.  For (\ref{eq:quadMSy1i}) or (\ref{eq:quadMSy1iE}), the $N_1$ stability requirement (\ref{eq:y1iapprox}) is satisfied if $M_S/v_R < (30 \, \KEV / M_1)^{3/4}$, where we took $y_{33} = 10^{-6}$, typical for natural leptogenesis. For radiative corrections involving ``seesaw" charged fermions, (\ref{eq:UDEcor1}) shows that the $N_1$ lifetime is natural for $M_1 < 10$ MeV/10 keV for $E/\Phi$ exchange; for ``flipped" masses  (\ref{eq:UDEcor2}) shows that $M_1$ can naturally be much larger. Hence, the UV completion with the largest natural range for $M_1$ has charged lepton masses arising from $E$ exchange, rather than $\Phi$ exchange, and has ``flipped" rather than ``seesaw" charged fermion masses.  In such UV completions, the entire parameter of Figs. \ref{fig:thermal_FO} or \ref{fig:thermal_FI} can be made natural for $N_1$ DM. 

For sufficiently small Dirac masses $M_{U,D,E} \ll z^{u,d,e} v_R$, the SM fermion masses are ``flipped" with right-handed states dominantly $SU(2)_R$ singlets, $\bar{U}$, $\bar{D}$ and $\bar{E}$. This may suppress the decay of $N_2$ by $W_R$ exchange, relaxing the upper bound on $v_R$ in the cosmology with freeze-out and dilution by $N_2$. With ``flipped" masses, $\bar{q}$ and the charged component in $\bar{\ell}$ obtain large masses $z^{u,d,e} v_R = y^{u,d,e}v_R$. For $v_R$ around the upper bound, $N_2$ can decay only into the first generation of $\bar{q}$ and $\bar{\ell}$. The decay rate of $N_2$ via $W_R$ exchange is
\begin{align}
\Gamma_{N _2 \rightarrow  (\ell^+ \bar{u} d, \, \ell^- u \bar{d})} + \Gamma_{N _2 \rightarrow N _1 \ell^+ \ell^-}  = \frac{2}{1536\pi^3} \frac{M_2^5}{v_R^4} \; | U_{eI_2}|^2 \; (3 + |U_{eI_1}|^2)\,.
\label{eq:2beta}
\end{align}

\begin{figure}[]
    \centering
    \begin{minipage}{0.5\textwidth}
        \centering
        \includegraphics[width=1\textwidth]{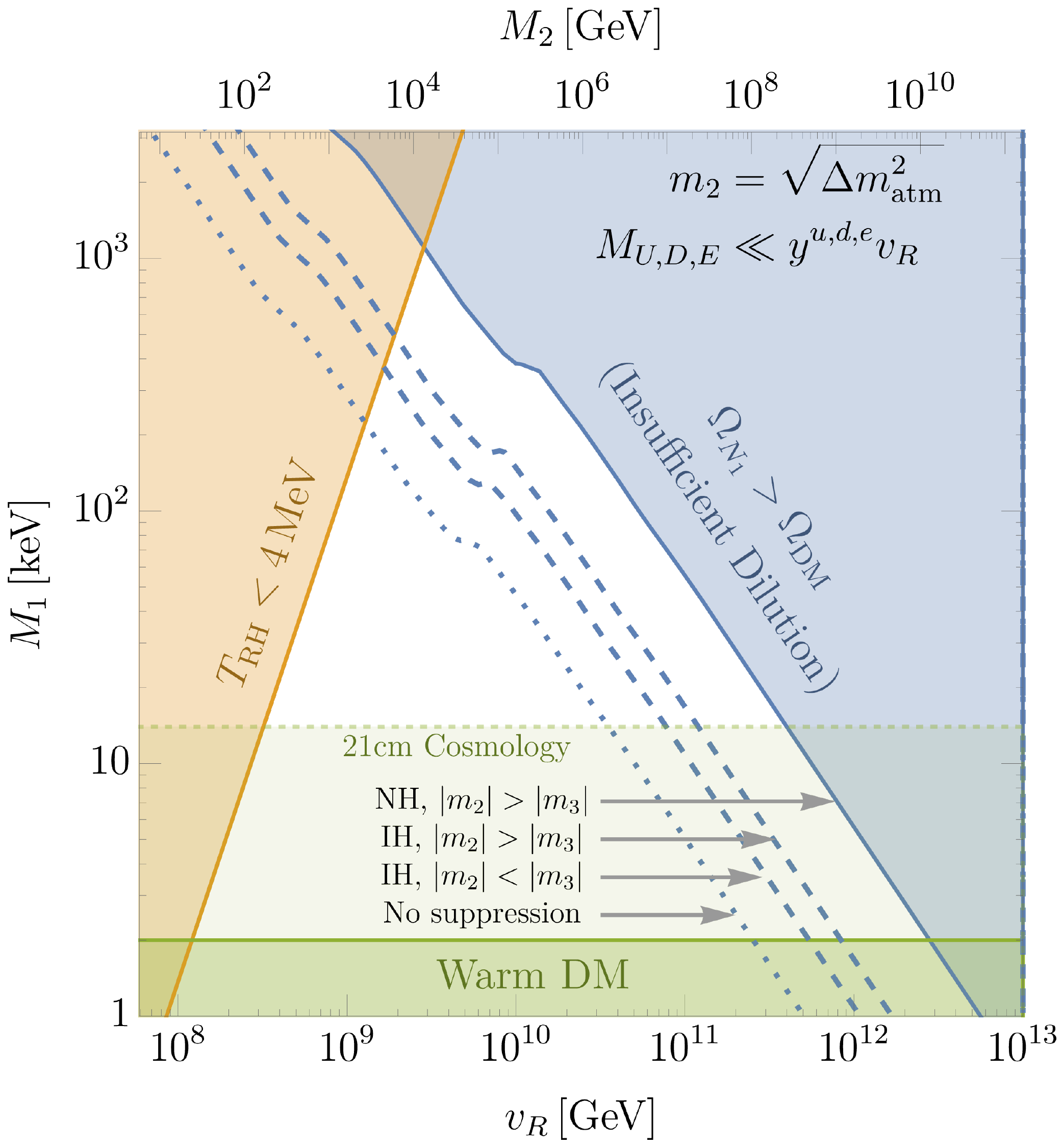} 
    \end{minipage}\hfill
    \begin{minipage}{0.5\textwidth}
        \centering
        \includegraphics[width=1\textwidth]{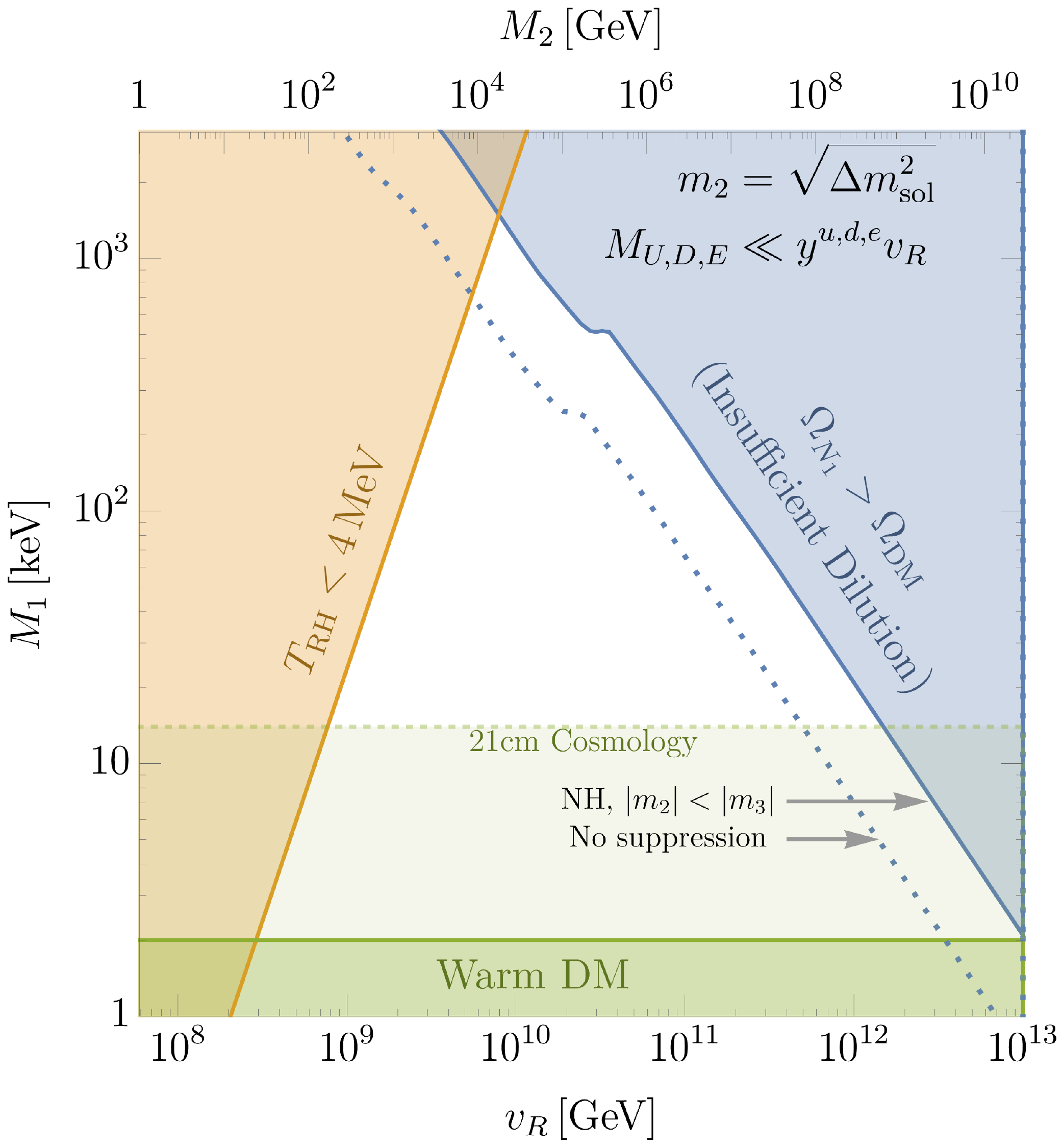} 
    \end{minipage}
    \caption{The parameter space of $N_1$ DM produced by relativistic freeze-out and dilution from $N_2$ decay when the masses of the heavy fermions, $M_{U,D,E}$, are far lighter than  $y^{u,d,e}v_R$. The shaded regions are identical to Fig.~\ref{fig:thermal_FO}, except that the beta decay rate of $N_2$ is suppressed, shifting the ({\color{c1} \bf blue}) insufficient dilution region to higher $v_R$. The $N_2$ beta decay rate decreases as the two heaviest generations of $\bar{q}$ and $\bar{\ell}$ becoming heavy, reducing the kinematically allowed decay channels and inducing suppressions from the PMNS matrix. We show the  allowed regions for $m_2 = \sqrt{ \smash[b]{ \Delta m_{\rm atm}  ^2}}$ ({\bf left}) and  $m_2 = \sqrt{ \smash[b]{ \Delta m_{\rm sol}  ^2}}$ ({\bf right}). The {\color{c1} \bf blue} contours show how the insufficient dilution boundary depends on whether $\nu_2$ and $\nu_3$ obey a normal (NH) or inverted hierarchy (IH).  Bounds from hot DM are discussed in the text.}
    \label{fig:thermal_FO_FN}
\end{figure}

The PMNS matrix elements are given by \cite{Esteban:2018azc}
\begin{align}
    |U_{e I_2}|^2 =&
    \begin{cases}
    |U_{e2}|^2 \simeq 0.30 &: {\rm NH}, |m_2| < |m_3| \\
    |U_{e3}|^2 \simeq 0.023 &: {\rm NH}, |m_2| > |m_3| \\
    |U_{e2}|^2 \simeq 0.30 &: {\rm IH}, |m_2| > |m_3| \\
    |U_{e1}|^2 \simeq 0.67 &: {\rm IH}, |m_2| < |m_3| 
    \end{cases}, \nonumber \\
    |U_{e I_1}|^2 =&
    \begin{cases}
    |U_{e1}|^2 \simeq 0.67 &: {\rm NH}\\
    |U_{e3}|^2 \simeq 0.023 &: {\rm IH}
    \end{cases}.
\end{align}
The suppression is most significant for NH with $|m_2| > |m_3|$. If the active neutrinos obey an IH, the suppression is also strongest when  $|m_2| > |m_3|$. The allowed parameter space of $N_1$ DM is shown in Fig.~\ref{fig:thermal_FO_FN} for all cases. The bounds from warmness and BBN are as in Fig.~\ref{fig:thermal_FO}; but the suppression of the $N_2$ beta decay rate relaxes the blue bound that arises from insufficient dilution, permitting the highest allowed $v_R$ to reach $10^{12-13} ~ \GEV$.
From (\ref{eq:2beta}), the fraction of $N_1$ DM that is hot is $|U_{e I_1}|^2/3 = 0.22 (\mbox{NH}), 0.007 (\mbox{IH})$. Thus, $N_2$ decaying dominantly via $W_R$ exchange is excluded for NH and allowed for IH.

\section{Natural leptogenesis}
\label{sec:natlepto}

In this section we study the extent to which successful leptogenesis can occur without the need for fine-tuning of parameters.  In Sections \ref{sec:RHnuDM}, \ref{sec:N1prod} and \ref{sec:lepto} we simply chose parameters of our theory to obtain a realistic light neutrino spectrum, decay rates, masses and interactions for $N_{1,2}$ that satisfy the constraints required for dark matter, and parameters that enhance leptogenesis to realistic values. While this is certainly possible, in this section we study the extra naturalness constraints imposed on the $(M_1,v_R)$ parameter space by requiring a natural theory without fine-tuning. We will use the UV completion described in the previous section that allows us to start with an understanding of why $N_1$ is light and sufficiently stable, and also limits the size of radiative corrections.

In Section~\ref{sec:lepto} we have seen that sufficient leptogenesis typically requires an enhancement of $\epsilon$ that can occur by near degeneracy of $N_2$ and $N_3$, or by increasing $y_{33}$ so that a cancellation between contributions to the light neutrino masses is required. Can these parameter choices be made natural by introducing approximate symmetries in the UV completion?  
In addition, in the last section we found a radiative correction to $y_{i1}$ proportional to $y_{33}$, leading to mixing between $N_1$ and $\nu_i$. Can a sufficiently long lifetime for $N_1$ be naturally maintained in the presence of an enhanced $y_{33}$ for leptogenesis?

\subsection{Models for enhanced asymmetry parameter}
\label{sec:tre}

Highly degenerate right-handed neutrinos, $M_{2} \simeq M_{3}$, can be explained by introducing an approximate flavor symmetry ensuring that $c_{22} \simeq c_{33}$ and $c_{23} \simeq 0$ in Eq.~(\ref{eq:yukawaNu}).
Such symmetries include an $SU(2)$ symmetry rotating $(\ell_2, \ell_3)$, or discrete symmetries $\ell_2 \leftrightarrow \ell_3$ and $\ell_{2} \rightarrow - \ell_{2}$. The symmetry is explicitly broken in the coupling $b_{ij}$ to explain the mass splitting of the two heaviest SM neutrinos. 

The symmetry is also explicitly broken by the charged lepton Yukawa couplings.  For example, when the charged lepton Yukawas arise from the exchange of a heavy scalar $\Phi$ of charge $(1,2,2,0)$, as in (\ref{eq:yukawa_charge}), one-loop quantum corrections from the coupling $x \Phi \ell \bar{\ell}$ give a wave-function renormalization,
\begin{align}
    {\cal L} = & ( 1 + \delta Z_{22})N_2^\dag \bar{\sigma}\partial N_2 +( 1 + \delta Z_{33})N_3^\dag \bar{\sigma}\partial N_3 + \left( \delta Z_{23} N_2^\dag \bar{\sigma}\partial N_3 + {\rm h.c.} \right), \nonumber \\
    \delta Z_{ij} \simeq & \frac{x_{ki}x^*_{kj}}{8\pi^2},
\end{align}
where we conservatively do not include a log-enhancement.
This generates a mass splitting
\begin{align}
 \frac{|M_2 - M_3|}{M_{2,3}} \simeq \sqrt{\left( \delta Z_{22} - \delta Z_{33} \right)^2 + \left( \delta Z_{23} + \delta Z_{23}^* \right)^2 } \gtrsim \frac{y_\tau^2}{8\pi^2} \simeq 10^{-6},
\end{align}
where we use $|x_{ki} x_{kj}^*| \gtrsim y_\tau^2$.  Near the resonance $x = 1$, $g(x) \simeq M_2/2(M_2 - M_3)$, so the maximum natural $g(x)$ is $ 5 \times 10^5$. We obtain the same bound for the case when charged lepton masses are generated by heavy fermion exchange, as in (\ref{eq:yukawa_charge2}). In summary, the maximum natural value for $g(x)$ is of order $10^6$.

Cancellation between the SM neutrino mass contributions from the see-saw of $N_3$ and the first dimension-5 operator of Eq.~\eqref{eq:m3}  can be explained in the following manner. Since we are interested in large $y_{33}$, we only consider $\ell_3$ and $\bar{\ell}_3$, and drop generation indices. Let us introduce only one singlet $S$ and couplings
\begin{align}
\label{eq:model_cancel}
{\cal L} = \lambda \ell  S H_L +  \lambda \bar{\ell} S H_R + \frac{1}{2} M_{S} S^2 + {\rm h.c.}
\end{align}
Integrating out $S$ gives the dimension-5 operator 
\begin{align}
{\cal L} = - \frac{\lambda^2}{2 M_S}\left( \ell H_L + \bar{\ell} H_R \right)^2 + {\rm h.c.},
\end{align}
corresponding to Eq.~(\ref{eq:yukawaNu}) with $b_{33} = c_{33}$. Only one linear combination of $\nu$ and $N$, which is dominantly $N$, obtains a Majorana mass and hence the SM neutrino remains massless.
This can be interpreted as a cancellation between $m^{(5)}$ and $m^{(ss)}$ in (\ref{eq:chi}), giving $|\chi| \ll 1$.

\begin{figure}
\centering
	\includegraphics[width=0.25\linewidth]{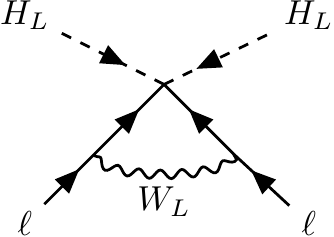}
	\caption{A diagram contributing to a non-zero neutrino mass for the case with tree-level cancellation between $m^{(5)}$ and $m^{(ss)}$. }
	\label{fig:cancel}	
\end{figure}

Since there is no symmetry forbidding the Majorana mass of $\nu$, it is generated by quantum corrections. Below the scale $v_R$, there is a quantum correction to $\ell\ell H_L H_L$ given by the diagram in Fig.~\ref{fig:cancel}, while there is no corresponding quantum correction to $\bar{\ell} \bar{\ell} H_R H_R$ and $\ell \bar{\ell}H_L H_R$. This quantum correction upsets the cancellation, giving a lower bound
\begin{align}
|\chi| > \frac{g^2}{16 \pi^2} \, \ln \left( \frac{{\rm min}(M_S,v_R)}{M_N} \right) \simeq 10^{-2}.
\end{align}

\subsection{Radiative corrections: $N_1$ lifetime}

Naturalness thus limits the maximum baryon asymmetry generated by $N_2$ in either cosmology,

\begin{align}
	Y_B \lesssim \frac{28}{79}\frac{1}{8\pi} y_{\rm max}^2 \; g(x)\sin^2 \alpha \sin 2\beta 
	\begin{dcases}
		\label{eq:naturalYB}
		\frac{\rho_{\rm DM}/s}{M_1} 
	&	\text{(Freeze-Out + Dilution)}  \\[3pt]
		\frac{\rho_{\rm DM}/s}{M_1} + 0.03 \, 
		\frac{m_2^{(ss)}}{10^{-4} \,\EV}
		Y_{\rm therm}
	&
		\left(\begin{array}{@{}c@{}}	
		\text{Freeze-In} \\
      	\text{Weak Washout}
    \end{array}\right) \\[3pt]
	0.03 \, Y_{\rm therm} \left(\frac{m_2^{(ss)}}{10^{-2} \,\EV} \right)^{-1.16}
	&
		\left(\begin{array}{@{}c@{}}	
		\text{Freeze-In} \\
      	\text{Strong Washout}
    \end{array}\right) \\
	\end{dcases},
\end{align}
where $y_{\rm max}$ is given in Eq.~\eqref{eq:ysqMax}.

\begin{figure}
	\center
	\includegraphics[width=0.9\linewidth]{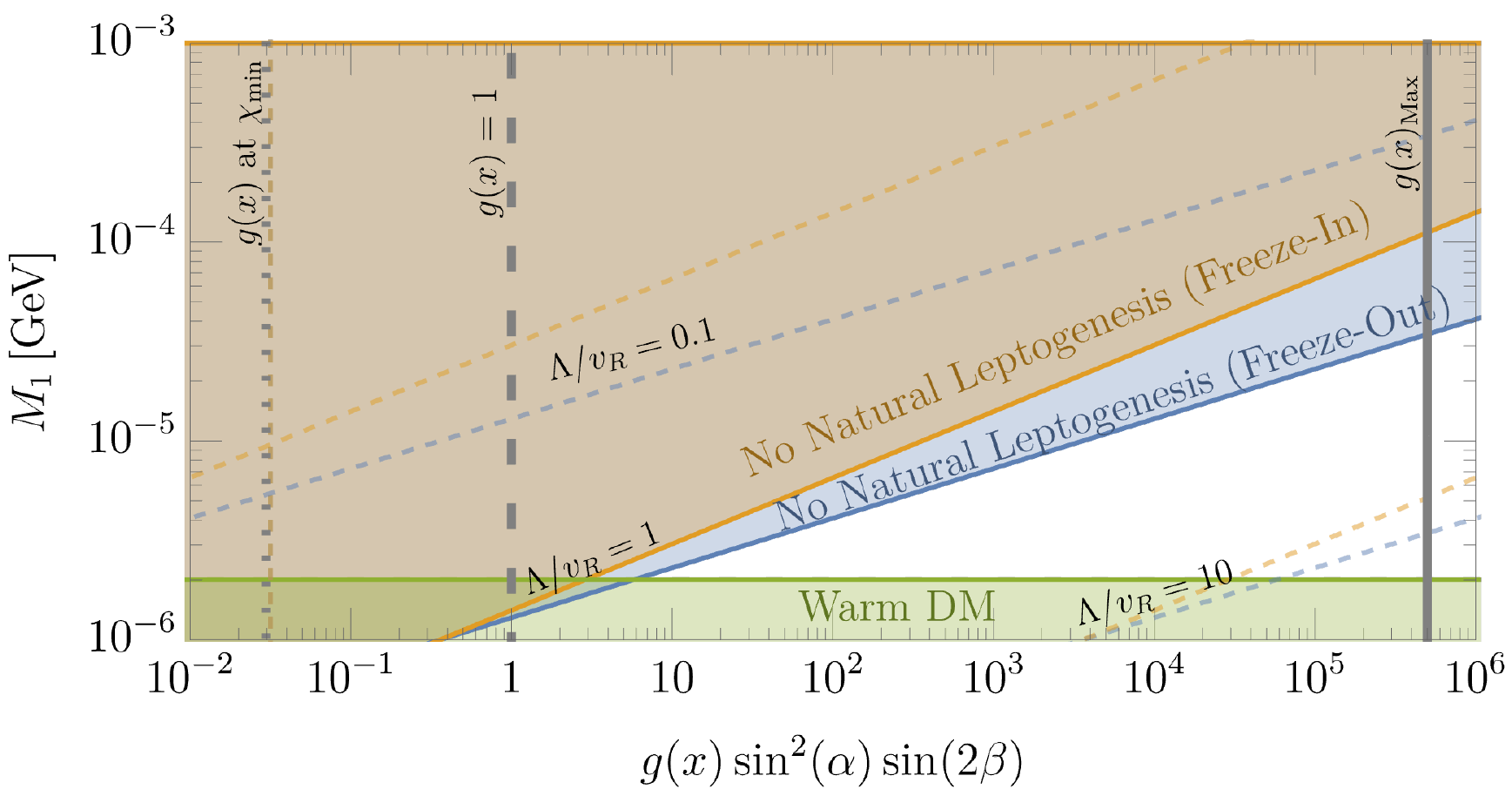}
	\caption{
	Parameter space for simultaneous $N_1$ DM and $N_2$ leptogenesis without fine-tuning.
	In the ({\color{c1} \bf blue}, {\color{c2} \bf orange}) shaded regions, the observed baryon asymmetry from $N_2$ decay, in the (freeze-out, freeze-in) cosmology, requires $y_{33}$ so large that fine-tuning is needed for sufficient stability of $N_1$,  when $\Lambda = v_R$. 
	The upper and lower dashed {\color{c1} \bf blue}  and {\color{c2} \bf orange}  contours show the analagous exclusion regions for $\Lambda/v_R = 0.1$ and $\Lambda/v_R = 10$, respectively. In the {\color{c3} \bf green} shaded region, $N_1$ DM is too warm. In both freeze-out or freeze-in cosmologies, successful $N_2$ leptogenesis requires $g(x) > 1$ for $\Lambda \gtrsim v_R$; the greater $\Lambda/v_R$ is, the more degenerate $M_2$ and $M_3$ must be to realize the observed baryon asymmetry. The vertical {\color{gray} \bf gray} solid, dashed, and dotted lines show representative values of $g(x)$ when $M_2$ and $M_3$ have the maximal natural degeneracy ($g(x)_{\rm max}$, solid), when $M_2$ and $M_3$ are comparable ($g(x) = 1$, dashed), and when $m_3^{(ss)}$ and $m_3^{(5)}$ are as naturally degenerate as can be ($g(x)$ at $\chi_{\rm min}$, dotted).}
	\label{fig:g(x)enhance}	
\end{figure}

The parameter space where $Y_B$ is unable to reach the observed baryon asymmetry without tuning is shown in Fig.~\ref{fig:g(x)enhance} in blue shading for the freeze-out cosmology and orange shading for the freeze-in cosmology, for $\Lambda/v_R = 1$. The dashed contours above and below show the analagous regions for $\Lambda/v_R = 0.1$ and $\Lambda/v_R = 10$, respectively. Because the radiative correction to the $N_1$ decay rate depends on the fourth power of $\Lambda/v_R$, the results are sensitive to this ratio; natural leptogenesis becomes implausible for $\Lambda \gg v_R$.  
The allowed parameter space within the freeze-in cosmology is greater than the freeze-out cosmology due to the additional contribution to $Y_B$ from $Y_{\ell H}$, which is assumed for the moment to saturate $0.1 Y_{\rm therm}$ for the purpose of showing the theoretical maximum allowed region of the freeze-in cosmology in Fig.~\ref{fig:g(x)enhance}. When $Y_{\ell H}$ is negligible compared to $Y_{W_R}$, the baryon asymmetry in the freeze-in cosmology is identical to the freeze-out cosmology and the orange region extends down to match the blue region.

The vertical gray lines show the asymmetry enhancement for three representative values of $g(x)$:  when $M_3$ and $M_2$ are as naturally degenerate as can be ($g(x)_{\rm Max}$, solid), when $M_3$ and $M_2$ are comparable ($g(x) = 1$, dashed), and when $m_3^{(ss)}$ and $m_3^{(5)}$ are as naturally degenerate as can be ($g(x)$ at $\chi_{\rm min}$, dotted).

A key result of Fig.~\ref{fig:g(x)enhance} is that, for a theory with $\Lambda/v_R > 1$, natural leptogenesis requires $g(x) \gg 1$ in either cosmology, which is only possible when $x \equiv (M_3/M_2)^2$ is close to unity.  Thus there are two ways to construct natural theories of leptogenesis. In the first, the structure of the theory below $v_R$ is modified to remove the quadratic divergence of (\ref{eq:y1rad}); such a theory is provided in Sec.~\ref{sec:model}. In the second, a symmetry is introduced to naturally yield near degeneracy of $N_2$ with $N_3$, as discussed in Sec.~\ref{sec:tre}.

The ratio $(\Lambda/v_R)$ can be less than one if the effective field theory described by \eqref{eq:yukawa} and \eqref{eq:yukawaNu} is generated by physics below the scale $v_R$.  In Sec \ref{sec:model} we construct an explicit model that generates \eqref{eq:yukawa} and \eqref{eq:yukawaNu} and show that in this theory the radiative corrections are given by (\ref{eq:quadMSM1}) and (\ref{eq:quadMSy1i}), which are identical to
\eqref{eq:M1rad} and \eqref{eq:y1rad} with $\Lambda$ replaced by $M_S$, the mass of the fermion which upon integrating out generates the operators of  \eqref{eq:yukawaNu}. Thus, when we take $\Lambda < v_R$, we understand it to be the mass $M_S$ of this fermion.

\subsection{Natural leptogenesis for freeze-out cosmology}
\label{sec:NatLeptoFO}

Although it appears the mass ratio $M_3/M_2$ can be freely adjusted to generate a large $g(x)$ independent of $y_{33}$, this is not the case as is shown in Section~\ref{sec:asymPerDecay}. This is because the neutrino mass matrix, \eqref{eq:numassmatrix}, relates $y_{33}, v_R, M_2,$ and $M_3$ together in a way that ensures the active neutrino masses, $m_2$ and $m_3$, remain $\mathcal{O}(0.1 \, \EV)$. 
In the freeze-out cosmology, the smallness of $y_{i1}$ and $y_{i2}$ together with \eqref{eq:numassmatrix} require that $m_2$ and $m_3$ satisfy Eqs.~(\ref{eq:M2m2}) and (\ref{eq:m3}),
so that $y_{33}^2$ must not only be less than $y^2_{\rm max}$, but equal to
\begin{align}
	  \label{eq:y33neutrino}
	y_{33}^2 &\simeq \left(\sqrt{x}\, m_2 - m_3 \right)\sqrt{x} \, m_2\frac{v_R^2}{v^4}. 		& \hspace{1.5cm} 
	\left(\begin{array}{@{}c@{}}	
		\text{Constraint from neutrino masses} \\
        \text{in freeze-out cosmology}
    \end{array}\right)
\end{align}
\begin{figure}[!]
    \centering
    \begin{minipage}{0.48\textwidth}
        \centering
        \includegraphics[width=1\textwidth]{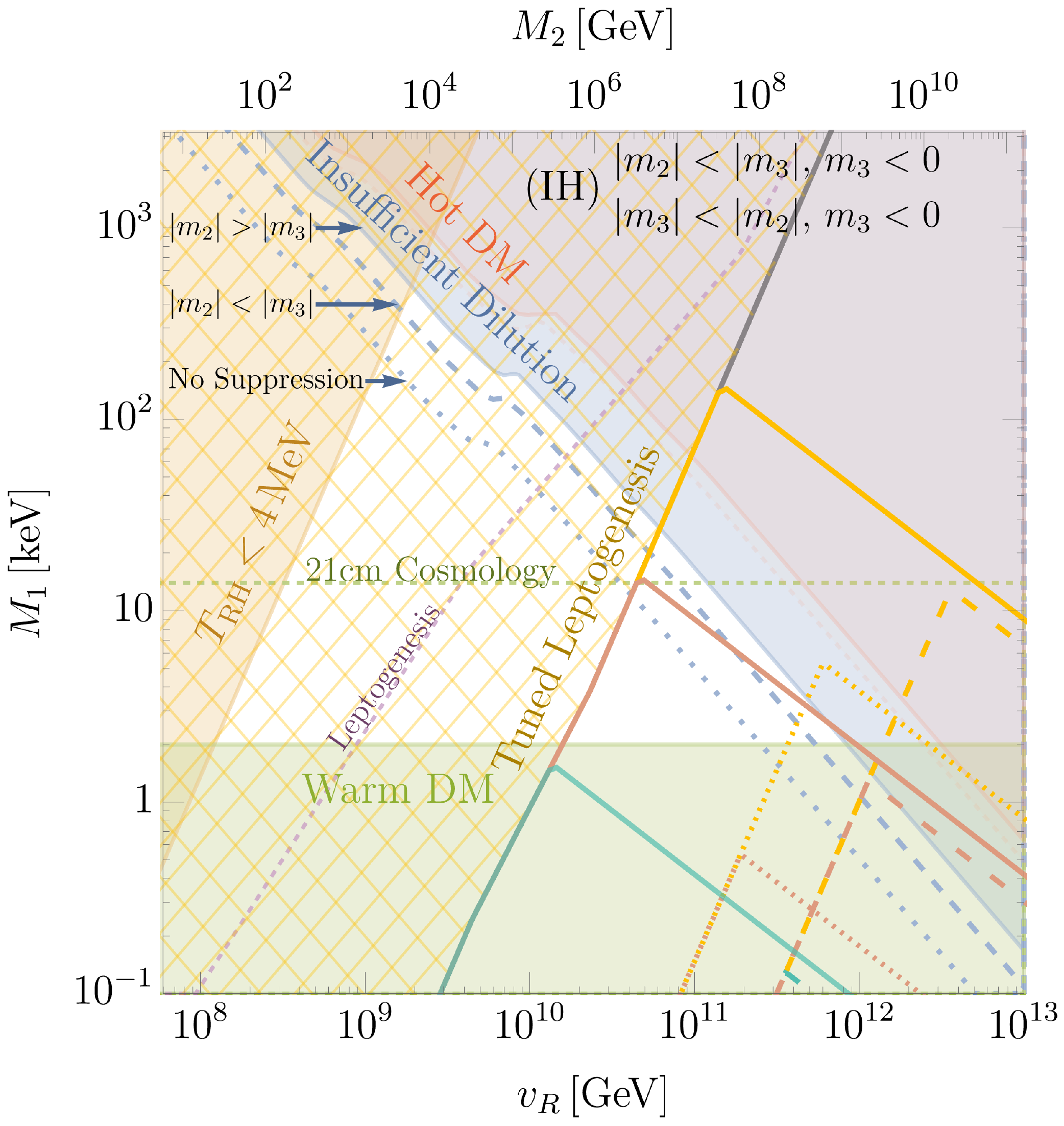} 
    \end{minipage}
    \begin{minipage}{0.48\textwidth}
        \centering
        \includegraphics[width=1\textwidth]{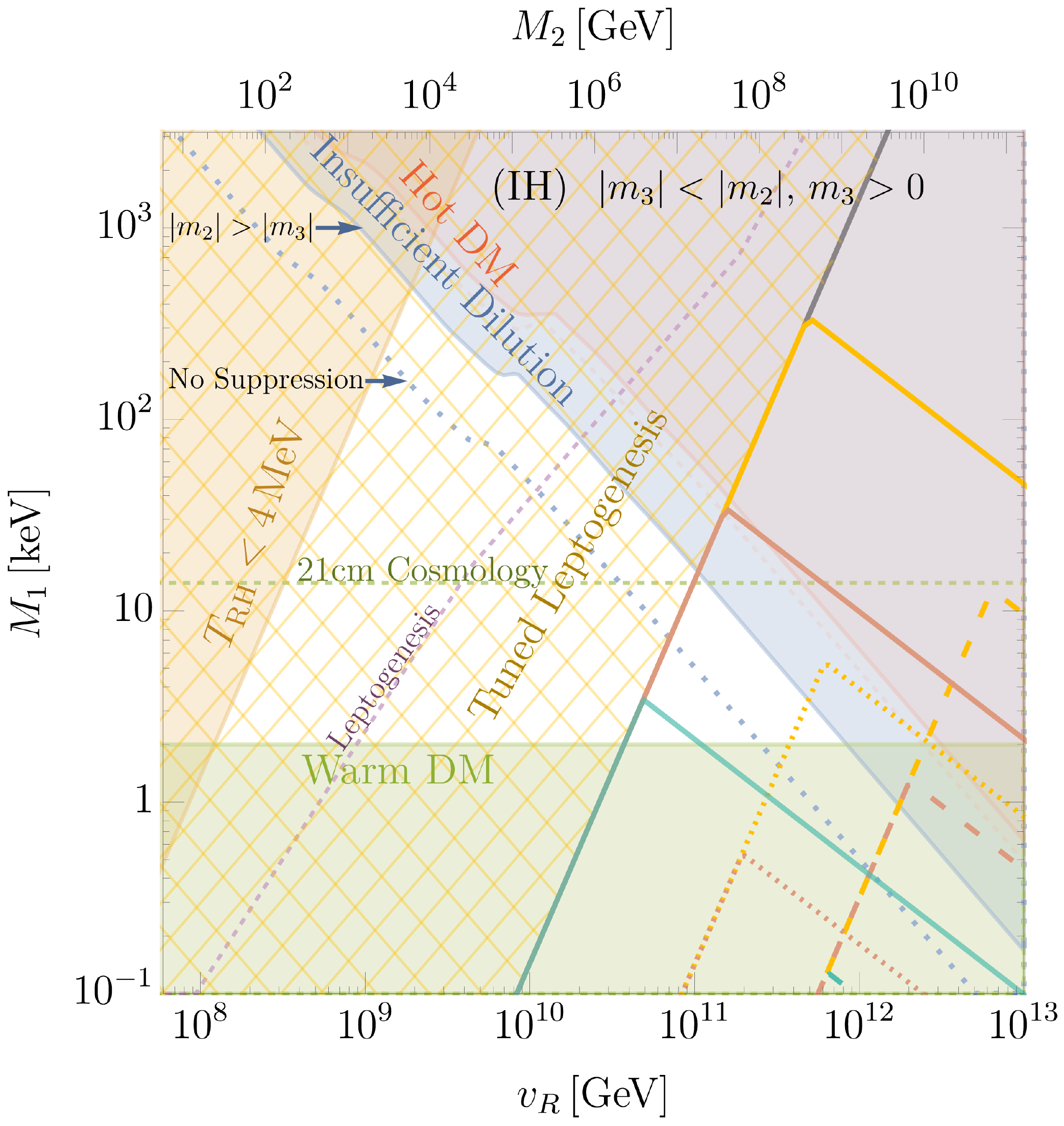} 
    \end{minipage}
    \begin{minipage}{0.48\textwidth}
        \centering
        \includegraphics[width=1\textwidth]{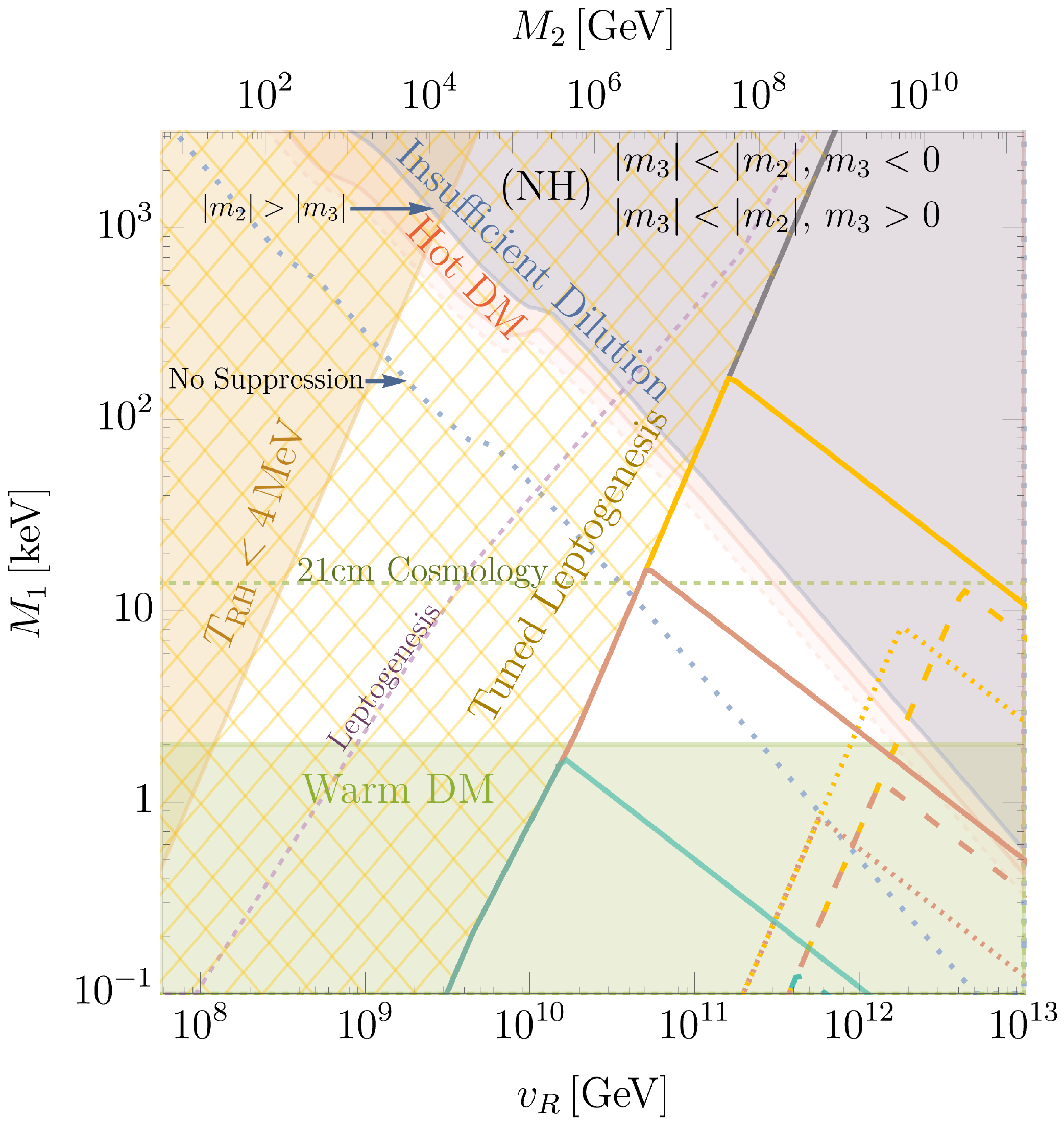} 
    \end{minipage}
     \begin{minipage}{0.48\textwidth}
        \centering
        \includegraphics[width=1\textwidth]{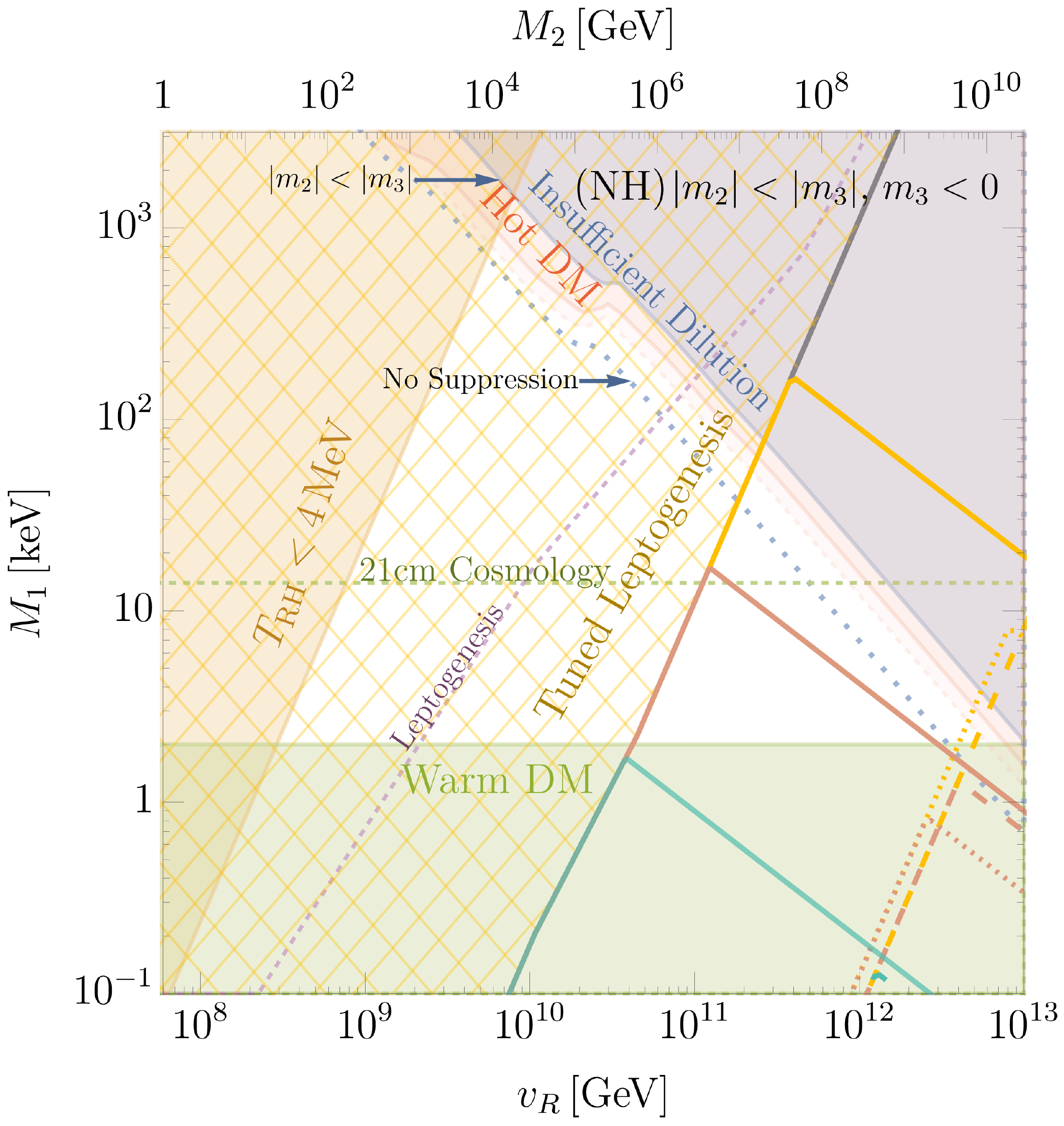} 
    \end{minipage}
     \begin{minipage}{0.7\textwidth}
        \centering
        \includegraphics[width=1\textwidth]{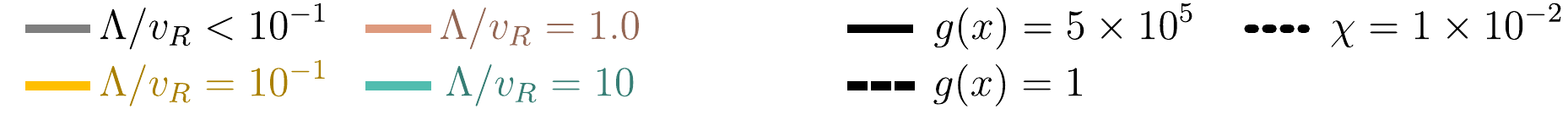} 
    \end{minipage}
    \caption{The parameter space where frozen-out $N_1$ DM and $N_2$ leptogenesis can naturally be realized without radiative corrections affecting the stability of $N_1$ DM and in accord with the active neutrino mass spectrum. The shaded (unhatched) regions solely constrain $N_1$ DM from freeze-out as in Fig.~\ref{fig:thermal_FO}. The hatched {\color{gold} \bf gold} region indicates where the baryon asymmetry generated by $N_2$ is unable to match the observed baryon asymmetry with $g(x)$ set to its largest, natural value, and $y_{33}$ set by consistent neutrino masses. The right, downward sloping contours mark where the radiative corrections to $y_{i1}$ are sufficiently large that they must be unnaturally tuned with tree contributions to keep $N_1$ DM stable when $g(x)$ is set to its largest, natural value, and $y_{33}$ set by consistent neutrino masses for $\Lambda/v_R = 0.1, 1, 10$ ({\color{gold} \bf gold}, {\color{c4} \bf red}, {\color{c3} \bf green}). The dashed and dotted contours show the same region when $M_2$ and $M_3$ are comparable, ($g(x) = 1$, dashed) and $m_3^{(ss)}$ and $m_3^{(5)}$ are as naturally degenerate as can be ($g(x)$ at $\chi_{\rm min}$, dotted). Naturalness and neutrino mass consistency excludes areas with too low or high values of $v_R$, and places a strong upper bound on the cutoff $\Lambda$. 
    We fix the $ \nu _2 $ and $\nu_3$ masses by the Inverted Hierarchy (IH, {\bf Top}) and Normal Hierarchy (NH, {\bf Bottom}).}
    \label{fig:FOLepto}
\end{figure}

In Fig.~\ref{fig:FOLepto}, we show the constraints on $(v_R, M_1)$ when incorporating leptogenesis naturally and consistently within the freeze-out $N_1$ DM cosmology. The shaded regions constraining $N_1$ DM remain from Fig.~\ref{fig:thermal_FO}, but newly added is a hatched gold region where natural leptogenesis is inconsistent with the observed neutrino masses. Within the allowed region reside three triangles with the same representative values of $M_3/M_2$ (equivalently, $g(x)$), shown in Fig.~\ref{fig:g(x)enhance}: when $M_3$ and $M_2$ are as naturally degenerate as can be ($g(x)_{\rm max}$, solid), when $M_3$ and $M_2$ are comparable ($g(x) = 1$, dashed), and when $m_3^{(ss)}$ and $m_3^{(5)}$ are as naturally degenerate as can be ($g(x)$ at $\chi_{\rm min}$, dotted), which occurs for $M_3 \gg M_2$. The right side of each triangle marks the region where $y_{33}$, as set by \eqref{eq:y33neutrino}, is greater than $y_{\rm max}$, \eqref{eq:ysqMax}; that is, where neutrino masses are incompatible with a natural $N_1$ lifetime. The left side of the triangle, i.e. the boundary of the hatched gold region, marks the region where $Y_B$ generated by $N_2$ (upper \eqref{eq:naturalYB}), is unable to match the observed baryon asymmetry with $y_{33}$ set by \eqref{eq:y33neutrino} and $\sin^2 \alpha \sin 2\beta = 1$; that is, where neutrino masses are incompatible with leptogenesis for the specified $x$. Within the unshaded region of each triangle, natural leptogenesis is possible for $\sin^2 \alpha \sin 2\beta < 1$. The gold, red, and green contours show the allowed regions when $\Lambda/v_R = 0.1$, $1$, and $10$, respectively.

Among the four panels of Fig.~\ref{fig:FOLepto}, the variation in location of the naturally allowed region can be understood by the differences in the values and relative signs of $m_2$ and $m_3$ taken in each panel. This is because the apex of each triangle is determined by the value of $y_{33}^2$ that satisfies the neutrino mass relations, \eqref{eq:y33neutrino}, and natural stability bounds for $N_1$ DM, \eqref{eq:ysqMax}. For the solid and dashed triangles, $x \approx 1$, and hence $y_{33}^2 \simeq m_2(m_2 - m_3)v_R^2/v^4$. When the active neutrinos obey an inverted hierarchy, as shown by the top two panels of Fig.~\ref{fig:FOLepto}, $m_2 \approx |m_3| \approx \sqrt{m_{\rm atm}^2}$, so that if $m_3 < 0$, (top left panel), $m_2(m_2 - m_3) \simeq (0.1 \, \EV)^2$, and if $m_3 > 0$, (top right panel), $m_2(m_2 - m_3) \ll (0.1 \, \EV)^2$.%
\footnote{$m_2 = |m_2|$ since it is determined solely by the positive-definite dimension five mass contribution, $m_2^{(5)}$. $m_3$ is not necessarily positive because it may have a non-negligible see-saw contribution with a negative sign.}
The first scenario gives a relatively larger value of $y_{33}^2$ compared to the second, meaning leptogenesis can be realized at slightly lower values of $v_R$ in the top left panel compared to the top right panel. However, a lower value of $y_{33}^2$ means radiative corrections to $y_{i1}$ are smaller, so that slightly higher values of $M_1$ can be reached in the top right panel compared to the top left. Identical reasoning explains the slight variation in the bottom two panels when the active neutrinos obey a normal hierarchy.%
\footnote{A consistent neutrino mass spectrum requires $m_2 > m_3 $ when $x \sim 1$, otherwise $y_{33}^2$, a positive definite quantity, would be negative (see \eqref{eq:y33neutrino}). This is violated if $|m_2| < |m_3|$ and $m_3 > 0$, which is why this case is absent in Fig.~\ref{fig:FOLepto}.}

Last, Fig.~\ref{fig:FOLepto} does not show the parameter region  where radiative corrections to the mass of $N_1$, \eqref{eq:M1rad}, exceed $M_1$. This is because the radiative corrections to $M_1$ are far less constraining than the radiative corrections to $y_{i1}$ affecting the stability of $N_1$.  For example, when  $\Lambda/v_R \leq 1$, $\Delta M_1 > M_1$ only when $v_R > 10^{13} \, \GEV$, which is not visible on Fig.~\ref{fig:FOLepto}. For larger values of $\Lambda/v_R$, the constraints from $\Delta M_1$ do affect regions of parameter space for $v_R < 10^{13}$ GeV, but only for parameter space already excluded by the constraints from $\Delta y_{i1}$.

\subsection{Natural leptogenesis for freeze-in cosmology}
\label{sec:NatLeptoFI}

\begin{figure}[!]
    \centering
    \begin{minipage}{0.48\textwidth}
        \centering
        \includegraphics[width=1\textwidth]{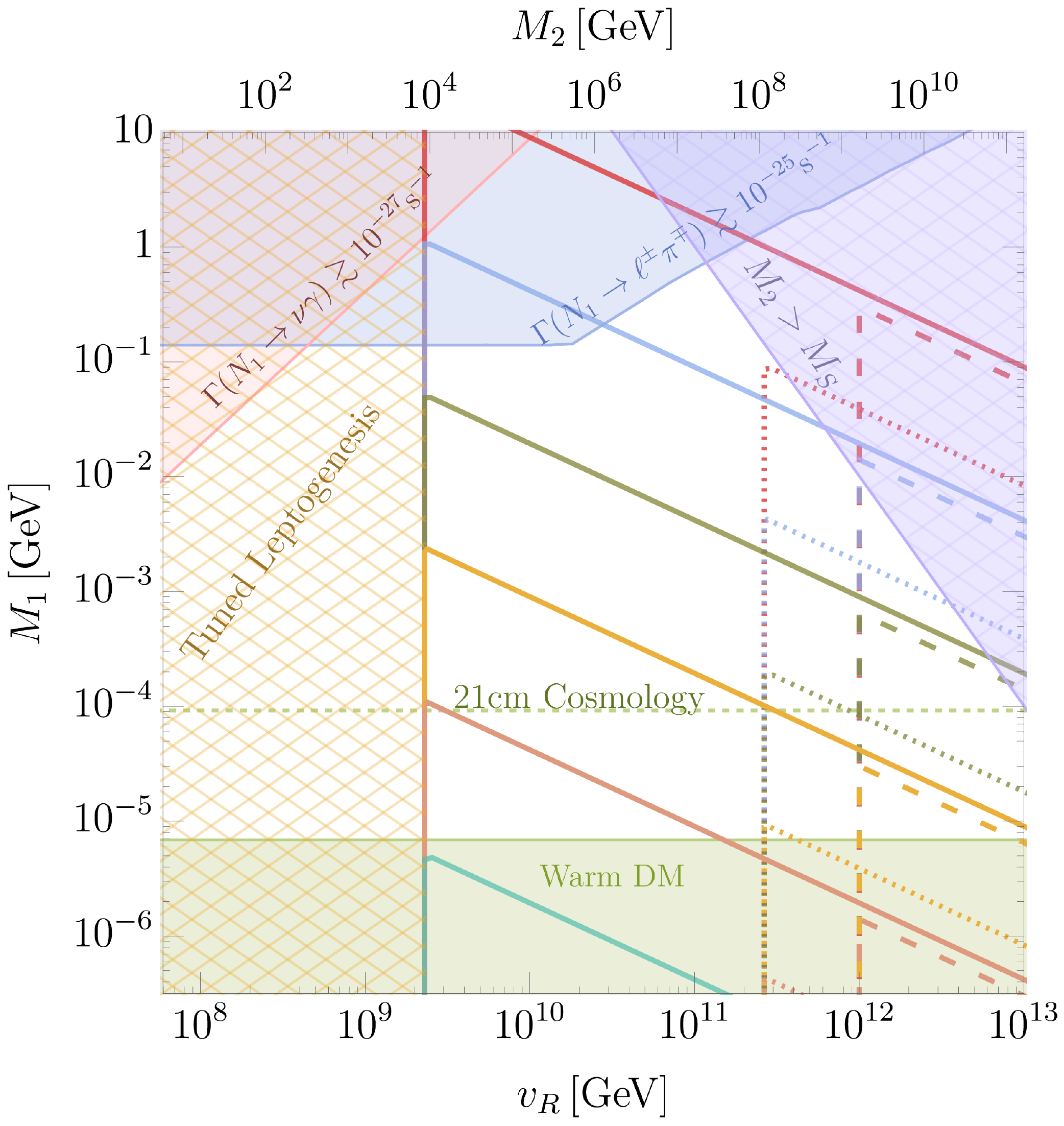}
    \end{minipage}
    \begin{minipage}{0.48\textwidth}
        \centering
        \includegraphics[width=1\textwidth]{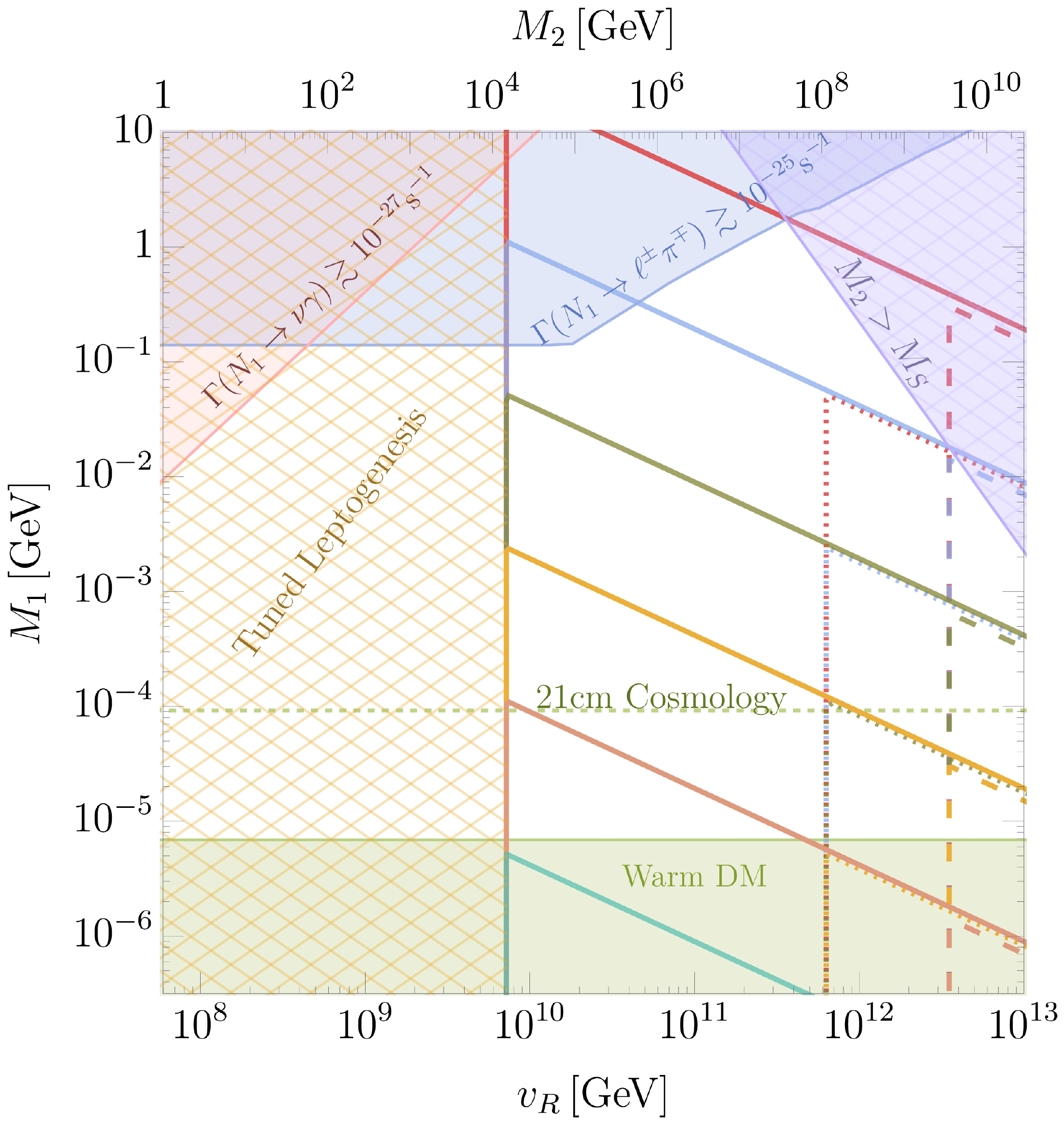}
    \end{minipage}
    \begin{minipage}{0.6\textwidth}
        \centering
        \includegraphics[width=1\textwidth]{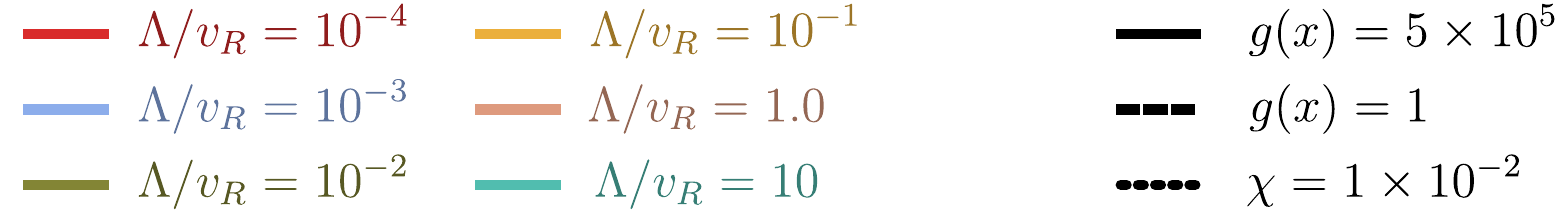} 
    \end{minipage}
    \caption{The parameter space where $N_1$ DM from freeze-in and $N_2$ leptogenesis can naturally be realized without radiative corrections affecting the stability of $N_1$ DM and in accord with the active neutrino mass spectrum. The unhatched shaded regions are constraints solely on $N_1$ DM from freeze-in as in Fig.~\ref{fig:thermal_FI}. In the hatched {\color{gold} \bf gold} region, the baryon asymmetry generated by $N_2$, at the maximum possible $\eta Y_2 \simeq 0.1 Y_{\rm therm}$, is unable to match the observed baryon asymmetry with $g(x)$ set to its largest natural value, and $y_{33}$ constrained by neutrino masses. The right, downward sloping contours indicate where the radiative corrections to $y_{i1}$ are sufficiently large that they must be unnaturally tuned with tree contributions to keep $N_1$ DM stable when $g(x)$ is set to its largest natural value, and $y_{33}$ set by consistent neutrino masses. Each contour corresponds to a specific $\Lambda/v_R$, as shown by the legend at the bottom. The dashed and dotted contours show the same region when $M_2$ and $M_3$ are comparable, ($g(x) = 1$, dashed) and $m_3^{(ss)}$ and $m_3^{(5)}$ are as naturally degenerate as can be ($g(x)$ at $\chi_{\rm min}$ , dotted).  Naturalness and neutrino mass consistency excludes areas with too low or high values of $v_R$, and places a strong upper bound on the cutoff $\Lambda$. Regions with larger $M_1$ are only allowed if $\Lambda < v_R$, as occurs for the model of Sec.~\ref{sec:model}. The hatched {\color{violet} \bf violet} region shows the inconsistent region where the mass of $N_2$ is greater than the mass of the heavy fermion that generates it.
    {\bf Left:} We fix $m_{22} =  \sqrt{\smash[b]{ \Delta m_{\rm atm}  ^2}}$ and $m_{33} =  -\sqrt{\smash[b]{ \Delta m_{\rm atm}  ^2 + \Delta m_{\rm sol}  ^2}}$ resembling the Inverted Hierarchy. Consequently, $m_{22}(m_{22} - m_{33}) \simeq (0.1 \, \EV)^2$ and $y_{33}^2$ is relatively large at $x = 1$. {\bf Right:}  We fix $m_{22} =  \sqrt{\smash[b]{ \Delta m_{\rm sol}  ^2}}$ and $m_{33} =  -\sqrt{\smash[b]{ \Delta m_{\rm atm}  ^2 }}$, resembling the Normal Hierarchy. Consequently, $m_{22}(m_{22} - m_{33}) \ll (0.1 \, \EV)^2$ and $y_{33}^2$ is relatively small at $x = 1$.}
   \label{fig:FILepto}
\end{figure}
Just as neutrino mass relations tie together $g(x)$ and $y_{33}$ in the freeze-out cosmology, so too do they tie $g(x)$ and $y_{33}$ in the freeze-in cosmology, as is shown in Section~\ref{sec:numass_FI}.
After requiring $\tilde{m}_2 < 0.001$ eV to avoid strong wash-out, a similar relationship to \eqref{eq:y33neutrino} occurs:
\begin{align}
	y_{33}^2 \simeq \left(\sqrt{x} \, m_{22} - m_{33} \right)\sqrt{x}\, m_{22}\frac{v_R^2}{v^4}, & \hspace{1.5cm} 
	\left(\begin{array}{@{}c@{}}	
		\text{Constraint from neutrino masses} \\
        \text{in freeze-in cosmology}
    \end{array}\right)
    \label{eq:y33neutrinoFI}
\end{align}
where $|m_{33}| \lesssim 0.05 \, \EV$ and $m_{22} = M_2(v/v_R)^2 = 0.01-0.05$ eV.

In Fig.~\ref{fig:FILepto}, we show the constraints on $(v_R, M_1)$ when leptogenesis is incorporated naturally and consistently in the cosmology with $N_1$ DM from freeze-in. The shaded regions constraining $N_1$ DM remain from Fig.~\ref{fig:thermal_FI}, but newly added is a hatched gold region where natural and consistent leptogenesis is inconsistent with the observed neutrino masses. Within the allowed region reside three triangles associated with the three familiar values of $M_3/M_2$: $g(x)_{\rm max}$, solid; $g(x) = 1$, dashed; $g(x)$ at $\chi_{\rm min}$, dotted. The right side of each triangle marks the region where $y_{33}$, as set by \eqref{eq:y33neutrinoFI}, is greater than $y_{\rm max}$, \eqref{eq:ysqMax}. The left side of the triangle, i.e. the boundary of the hatched gold region, marks the region where $Y_B$ generated by $N_2$, at the maximum possible $\eta Y_2 \simeq 0.1 Y_{\rm therm}$, is unable to match the observed baryon asymmetry with $y_{33}$ set by \eqref{eq:y33neutrinoFI} and $\sin^2 \alpha \sin 2\beta = 1$. Within the unshaded region of each triangle, natural leptognesis is possible for $\sin^2 \alpha \sin 2\beta <1$. Each contour color corresponds to a different $\Lambda/v_R$ spanning six decades from $10^{-4}-10$, as shown by the legend at the bottom of the figure. Fig.~\ref{fig:FILepto} demonstrates that naturally reaching the highest masses of $N_1$ DM allowed in the freeze-in cosmology requires $\Lambda/v_R \ll 10^{-1}$.

The left side of the triangle in Fig.~\ref{fig:FILepto} is vertical unlike Fig.~\ref{fig:FOLepto} because $\eta Y_2$ at its maximum is independent of $M_1$ due to the additional contribution from $Y_{\ell H}$.  When $\widetilde{m}_2 \not \sim 10^{-3} \, \EV$, $\eta Y_2 \leq 0.1 Y_{\rm therm}$, and the triangular region shrinks ($\widetilde{m}_2$ is defined in (\ref{eq:m2tilde})). If $\widetilde{m}_2 \ll 10^{-3}$, $Y_{\ell H} \ll Y_{W_R}$, and the left side of the triangular regions of Fig.~\ref{fig:FILepto} contract to match those of Fig.~\ref{fig:FOLepto} for freeze-out. 

Since $m_{22}$ and $m_{33}$ are unknown quantities generally misaligned with the active neutrino masses, it is impossible to know the exact parameter space associated with the normal and inverted hierarchies. Nevertheless, since $m_{22}$ and $|m_{33}|$ remain of order the observed neutrino masses, the variations in the allowed parameter space do not change dramatically when scanning over possible values of $m_{22}$ and $m_{33}$. For example, in the left panel of Fig.~\ref{fig:FILepto}, $m_{22}(m_{22} - m_{33})\simeq (0.1 \, \EV)^2$ so that $y_{33}^2$ is at its largest when $x \sim 1$ for the same reasons discussed in Sec.~\ref{sec:NatLeptoFO} for freeze-out. In this case, leptogenesis can probe lower $v_R$ due to the slight enhancement in $y_{33}$. In the right panel, $m_{22}(m_{22} - m_{33}) \ll (0.1 \, \EV)^2$ so that $y_{33}^2$ is much smaller, and larger $v_R$ is required to realize the observed baryon asymmetry. The right panel of Fig.~\ref{fig:FILepto} assumes $m_{33}$ and $m_{22}$ are not more degenerate than the observed neutrino mass spectrum. If they are significantly more degenerate, $y_{33}^2$ decreases and large $v_R$ is required to generate the observed baryon asymmetry. Consequently, the naturally allowed triangular region shifts to higher $v_R$. Finally, the allowed region where $m_{22}(m_{22} - m_{33}) \lesssim (0.1 \, \EV)^2$  lies between the triangular regions in the left and right panels of Fig.~\ref{fig:FILepto}.

Within the hatched violet region, the mass of $N_2$ is greater than the mass of the heavy fermion, $M_S$, that generates it, which is inconsistent. This region is always more constraining than the region where the reheat temperature after inflation, $T_{\rm RH}^{\rm inf}$, is below $M_2$ and leptogenesis becomes challenging. We do not analyze this region in this work.

Last, Fig.~\ref{fig:FILepto} does not show the region of parameter space where radiative corrections to the mass of $N_1$, \eqref{eq:M1rad}, are greater than $M_1$ for the same reasons discussed for the freeze-out cosmology: the radiative corrections to $M_1$ are weaker than the radiative corrections to $y_{1i}$ and either do not show up on Fig.~\ref{fig:FILepto}, or are already excluded by other means. 

\section{Conclusions and discussion}
\label{sec:conclusion}

The discovery of the Higgs with a mass of 125 GeV has revealed that the Higgs quartic coupling nearly vanishes at a high energy scale $(10^9-10^{13})$ GeV. In extensions of the SM with a $Z_2$ symmetry called Higgs Parity, the spontaneous breaking of Higgs Parity yields the SM as a low energy effective theory. The SM Higgs quartic coupling is predicted to vanish at the $Z_2$ symmetry breaking scale, and hence precise measurements of SM parameters can narrow down the symmetry breaking scale. Observable quantities correlated with the symmetry breaking scale are correlated with SM parameters.

In this paper, we identified Higgs Parity with Left-Right symmetry, which is broken at scale $v_R$. By combining Left-Right Higgs Parity with space-time parity, the absence of CP violation in strong interactions is explained. Left-Right symmetry predicts three right-handed neutrinos. The lightest, $N_1$, may be dark matter and the decay of a heavier one, $N_2$, may create the baryon asymmetry of the universe through leptogenesis.

We studied two cosmological histories of the universe. In the freeze-out cosmology, the reheating temperature of the universe is high enough that right-handed neutrinos are initially thermalized via exchange of additional gauge bosons required by Left-Right symmetry. $N_i$ later decouple from the thermal bath; $N_1$ are overproduced, but are diluted by the late-time decay of $N_2$. $N_2$ decays also create the baryon asymmetry.  In the freeze-in cosmology, the reheating temperature is low, so that the right-handed neutrinos are not thermalized, but an appropriate amount of $N_1$ is produced via new gauge boson exchange around the completion of reheating. $N_2$ are produced by the new gauge boson exchange and by Yukawa couplings to SM particles. The $N_2$ decays again produce the baryon asymmetry.

\begin{figure}
\centering
    \begin{minipage}{0.48\textwidth}
        \centering
        \includegraphics[width=1\textwidth]{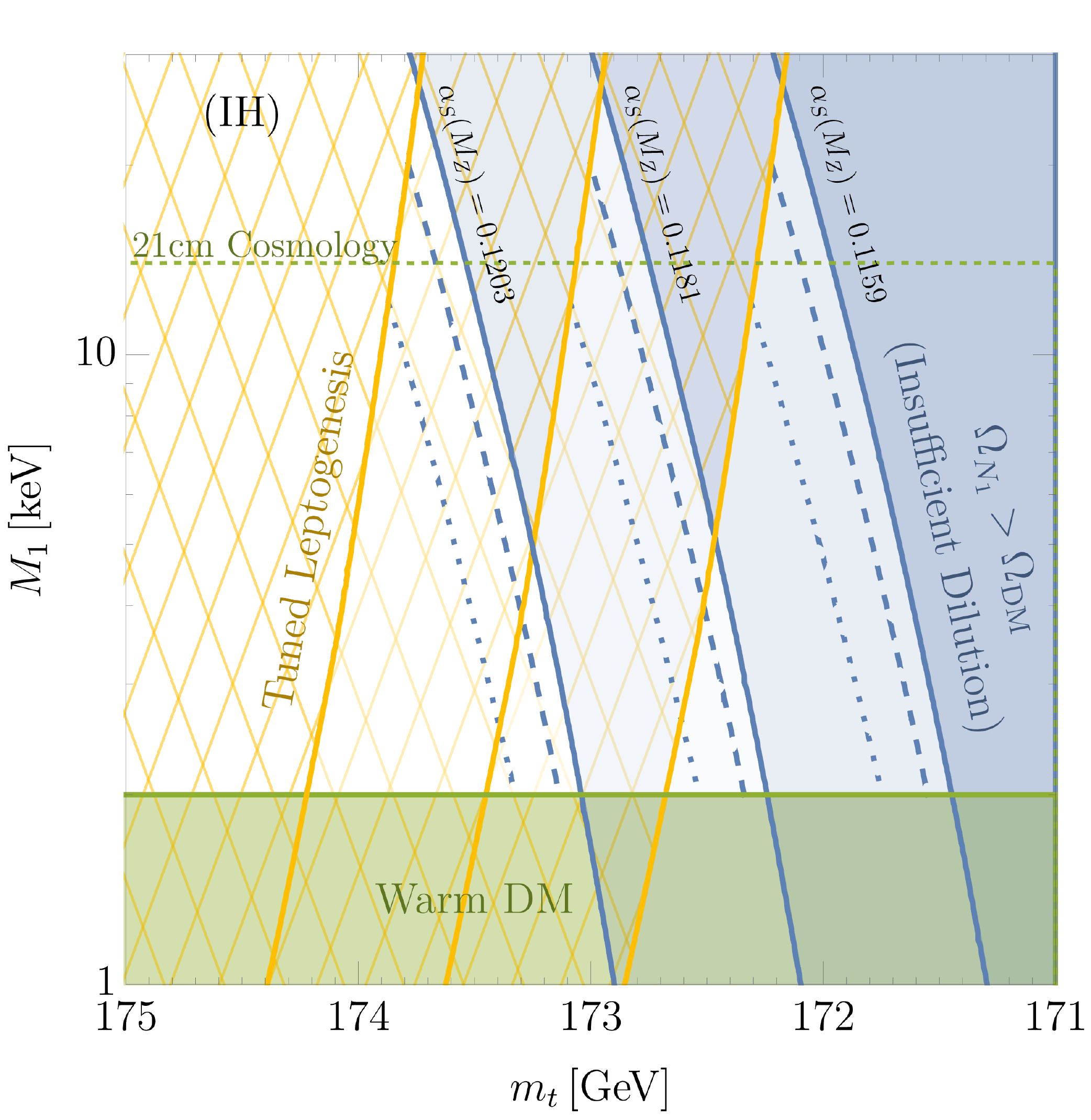}
    \end{minipage}
    \begin{minipage}{0.48\textwidth}
        \centering
        \includegraphics[width=1\textwidth]{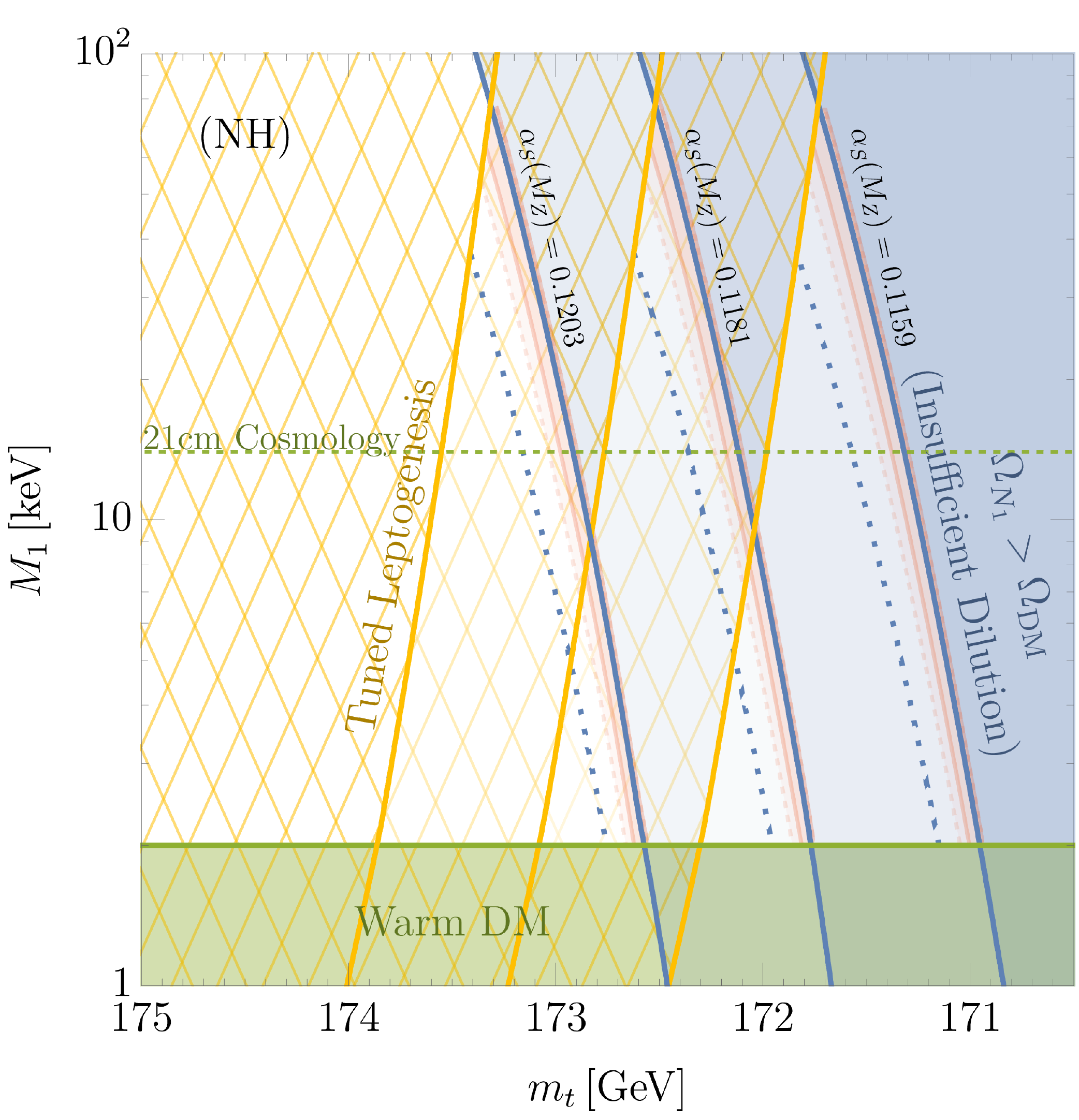}
    \end{minipage}
	\caption{The parameter space of $N_1$ DM from freeze-out, natural leptogenesis, and consistent neutrino masses in terms of the mass of $N_1$, $M_1$, and the mass of the top quark, $m_t$.   Remarkably, $N_1$ DM, natural leptogenesis, and the observed neutrino masses are consistent with the current measurement of $m_t = 173.0 \pm 0.4 \, \GEV$. 
	The center triangle fixes $\alpha_S(M_Z)$ at its central value, and the triangles to the left and right at $\pm 2\sigma$ values. We fix $m_h$ at its central value throughout, since variations in $m_h$ within its uncertainty do not appreciably change the parameter space.
	The $\nu_2$ and $\nu_3$ masses are fixed by: {\bf Left} the Inverted Hierarchy (IH) in accordance with the top left panel of Fig.~\ref{fig:FOLepto} and {\bf Right} by the Normal Hierarchy (NH), in accordance with the bottom right panel of Fig.~\ref{fig:FOLepto}.}
	\label{fig:mtop1181}	
\end{figure}

The freeze-out cosmology is tightly constrained.  With quark and lepton masses generated by the effective theory of (\ref{eq:yukawa}) and (\ref{eq:yukawaNu}), successful dark matter and baryogenesis can be achieved simultaneously in the unshaded regions of the $(v_R,M_1)$ plane of Fig.~\ref{fig:epsilonFO}. The symmetry breaking scale is predicted to be $v_R= 10^8-10^{13}$ GeV; remarkably, this coincides with the window predicted from SM parameters and Higgs Parity. The parameter space can be probed by 21cm line cosmology and by precise measurements of SM parameters. If the effective theory has a UV completion below $v_R$, the allowed region is slightly enlarged, as shown in Fig.~\ref{fig:thermal_FO_FN}. The freeze-in cosmology, on the other hand, is consistent with simultaneous dark matter and baryogenesis over a wide range of $(v_R,M_1)$, including the entire unshaded region of Fig.~\ref{fig:thermal_FI}.

Naturalness of the scheme further constraints the parameter space as well as the origin of the fermion masses in the model. The stability of $N_1$ DM is not protected by any symmetry. Quantum corrections may induce Yukawa couplings of $N_1$ to the SM lepton doublets and Higgs, making $N_1$ decay too fast. We identified two types of quantum corrections. First, $N_3$ must have significant Yukawa couplings for efficient leptogenesis, while the tau Yukawa coupling explicitly breaks any symmetry that distinguishes $N_3$ from $N_1$; quantum corrections involving $N_3$ and tau Yukawa couplings destabilize $N_1$. In some of the parameter space, to suppress these quantum corrections, the neutrino mass operators of (\ref{eq:yukawaNu}) should be UV-completed by fields with a mass below $v_R$. Second, the $SU(2)_R$ doublet to which $N_1$ is embedded, $\bar{\ell}_1$, has Yukawa couplings to generate the charged lepton Yukawa couplings. The chiral symmetry of $N_1$ which can forbid its decay is explicitly broken by a combination of this Yukawa and the quark Yukawas. To suppress the resulting quantum corrections, the UV completion of the operators of (\ref{eq:yukawa}), that generate charged fermion masses, requires fields with masses below $v_R$.

In most of parameter space, sufficient baryon asymmetry requires either the two heavier right-handed neutrinos $N_{2,3}$ are nearly degenerate, or the see-saw contribution to the SM neutrino masses from $N_3$ is nearly cancelled by a contribution from dimension-5 operators. These two features can be explained naturally by UV models of the neutrino sector presented in Sec.~\ref{sec:tre}. However, the near degeneracy or cancellation may be destabilized by quantum corrections, limiting the enhancement. This excludes lower values of $v_R$, where the masses of $N_{2,3}$ are small and significant enhancement of the CP asymmetry is required.

Constraints on the freeze-out cosmology, summarised in Fig.~\ref{fig:FOLepto}, allow $v_R \sim 10^{10}-10^{13} (10^{12})$ GeV and $M_1 \sim 2-100 (30)$ keV for the normal (inverted) hierarchy of SM neutrinos, respectively. Measurements of SM parameters, the warmness of DM, and the hierarchy of SM neutrinos can probe this parameter space.
For example, if an inverted hierarchy is confirmed, $v_R<10^{12}$ GeV is required, giving precise predictions for future measurements of $m_{t}$ and $\alpha_s$. Also, observations of cosmic 21cm line radiation  will discover DM to be warm, unless $v_R \sim 10^{11}$ GeV.
For a normal hierarchy, a wider range of $v_R$ is allowed, but discovery or constraints on the warmness of DM will narrow down $v_R$, and hence SM parameters.
If the CP asymmetry of leptogenesis is not enhanced by either degeneracy or cancellation, $v_R$ and $M_1$ are required to be above $10^{12}$ GeV and around a few keV, respectively. This parameter region can be probed by measurements of SM parameters and the warmness of DM.

In Fig.~\ref{fig:mtop1181}, we recast the constraints on the $(m_{t}, M_1)$ plane for a fixed Higgs mass and several values of a strong coupling constant. In Higgs Parity, the scale $v_R$ depends dominantly on $m_{t}$, and to a lesser extent, $\alpha_S(M_Z)$ and the Higgs mass, $m_h$ (see e.g.~Fig.~\ref{fig:vpPrediction}). Consequently, for fixed $\alpha_S(M_Z)$ and $m_h$, $m_{t}$ acts as a direct substitute for the scale $v_R$. The allowed parameter space is in remarkable agreement with the observed top quark mass.  Future measurements of $m_{t}$, $\alpha_S(M_Z)$, and $m_h$ will hone in on the scale $v_R$ and, together with determination of the neutrino mass hierarchy, will narrow the allowed range of $M_1$. This can then be confirmed or excluded by 21 cm line cosmology.
Here we assume that the running of gauge coupling constants is that of the SM up to the scale $v_R$. If the Dirac mass terms in Eqs.~(\ref{eq:yukawa_charge2}) and (\ref{eq:yukawa_UV_ud}) are smaller than $v_R$, the running is slightly altered. If all of the Dirac masses are smaller than $ y^{u,d,e} v_R$, there exists a set of new particles with masses $ y^{u,d,e} v_R$. Even for this extreme case, the prediction for $v_R$ for given SM parameters is increased only by a factor of two. For fixed $v_R$, this corresponds to an increase in the prediction for the top quark mass by $150$ MeV. If the Dirac masses of fermions generating the first generation  Yukawas are above $v_R$, the increase in $v_R$ is at most only $10 \%$. The corresponding increase in the top quark mass is $20$ MeV, which is smaller than the expected uncertainty of top quark mass measurements at future lepton colliders~\cite{Seidel:2013sqa,Horiguchi:2013wra,Kiyo:2015ooa,Beneke:2015kwa}.

In the freeze-out cosmology, if $N_2$ decays dominantly via $W_R$, a component of hot dark matter is predicted due to the subdominant decay mode $N_2 \rightarrow N_1 \ell^+ \ell^- $.  This is a very natural possibility, occurring whenever the $N_2$ Yukawa couplings are sufficiently small. In this case the prediction for $v_R$, or equivalently $m_{t}$, is sharpened, corresponding to the right-hand blue side of the allowed regions in Fig.~\ref{fig:mtop1181}. The branching ratio of the decay into $\ell H$, which creates lepton asymmetry, is less than unity, but this can be compensated by the enhancement of the CP asymmetry. When charged fermion masses arise from the effective theory of (\ref{eq:yukawa}), this hot component provides 10\% of dark matter. However, in the case of UV completions discussed in Sec.~\ref{sec:model}, for a normal neutrino mass hierarchy too much hot dark matter is produced if $N_2$ decays dominantly via $W_R$, while for the inverted hierarchy the hot fraction is only $0.7\%$.  The relevant $N_2$ branching ratios can be computed because the lepton flavor mixing matrix for $W_R$ is the complex conjugate of the PMNS matrix.

The freeze-in cosmology is also constrained, as shown in Fig.~\ref{fig:FILepto}; $v_R$ must be above $10^9$ GeV. If the CP asymmetry of leptogenesis is not enhanced by degeneracy or cancellation, $v_R$ is required to be above $10^{12}$ GeV, constraining the parameters.

Theories of Higgs Parity suffer from the domain wall problem~\cite{Zeldovich:1974uw} if the Higgs Parity symmetry breaking occurs after inflation. To avoid the problem requires that the reheating temperature is at most $v_R$; the constraint is typically stronger since the maximal temperature of the universe is in general higher than the reheating temperature~\cite{Kolb:1990vq,Harigaya:2013vwa,Mukaida:2015ria} (see, however,~\cite{Co:2020xaf}). As we have shown in this paper, the baryon asymmetry can be produced naturally via leptogenesis with the reheating temperature much smaller than $v_R$, especially in the freeze-in cosmology, safely avoiding the domain wall problem.

We conclude the paper by stressing the importance of cosmology and precise measurements for Higgs Parity.
New physics scales in theories of Higgs Parity are high. New particles are heavy and/or very weakly coupled to SM particles. Direct confirmation of these theories by discovery of new particles or deviation from SM predictions at collider experiments will be difficult in the near future. In testing such theories, theoretical considerations on the early universe, cosmological observations, and predictions of SM parameters (including those of neutrinos) play key roles. In this paper, we investigated the production of dark matter and baryon densities in a Left-Right symmetric Higgs Parity theory. The theory can be in fact probed by the warmness of DM, precise determination of SM parameters by future colliders and lattice computations, and by the measurement of the neutrino hierarchy.

\section*{Acknowledgement}
This work was supported in part by the Director, Office of Science, Office of High Energy and Nuclear Physics, of the US Department of Energy under Contracts DE-AC02-05CH11231 (LJH) and DE-SC0009988 (KH), by the National Science Foundation under grants PHY-1316783 and PHY-1521446 (LJH), as well as by the Raymond and Beverly Sackler Foundation Fund (KH).

\bibliographystyle{JHEP}
\bibliography{LeptoBib}

\end{document}